\documentclass{jfm}
\usepackage[hyperindex,breaklinks]{hyperref}
\usepackage{pslatex,graphicx,amssymb}
\usepackage{natbib}
\usepackage{multirow}
\usepackage[ansinew]{inputenc}

\usepackage[T1]{fontenc}      
\usepackage{pstricks}

\newcommand{\etal}[0]{{\itshape et al.}}
\newcommand{\bu}{\mathbf u}

\title[Standing and travelling waves in cylindrical Rayleigh--B\'enard convection]
{Standing and travelling waves in cylindrical Rayleigh--B\'enard convection}

\author[K. Boro\'nska and L.~S. Tuckerman]
{Katarzyna Boro\'nska\thanks
{email: {\tt kasia@limsi.fr} -- Web page: 
{\tt http://www.limsi.fr/Individu/kasia/en}}
and Laurette S. Tuckerman\thanks
{email: {\tt laurette@limsi.fr} -- Web page:
{\tt http://www.limsi.fr/Individu/laurette}}
}

\affiliation{Laboratoire d'Informatique pour la M\'ecanique et les Sciences de 
l'Ing\'enieur (LIMSI--CNRS)\\
B.P.~133, 91403 Orsay, France}

\begin{document}
\maketitle

\begin{abstract}
The Boussinesq equations for Rayleigh--B\'enard convection are simulated for a
cylindrical container with an aspect ratio near 1.5. The transition from an
axisymmetric stationary flow to time-dependent flows is studied using
nonlinear simulations, linear stability analysis and bifurcation theory.  
At a Rayleigh number near
$25\,000$, the axisymmetric flow becomes unstable to standing or
travelling azimuthal waves.  The standing waves are slightly unstable to
travelling waves.  This scenario is identified as a Hopf bifurcation in 
a system with O(2) symmetry.
\end{abstract}
\section{Introduction}
Rayleigh--B\'enard instability in a fluid layer heated from below
in the presence of gravity 
is the classic prototype of pattern
formation. A new chapter in its investigation began with the increase of
computer performance that made feasible three-dimensional, nonlinear,
high-resolution simulations of the Boussinesq equations governing this system.

We are interested in a fluid layer confined in a vertical cylinder whose upper and lower bounding surfaces are maintained at a temperature difference measured by the Rayleigh number. The conductive solution for this
system is a motionless state with 
a uniform vertical temperature
gradient. This solution is stable up to a critical Rayleigh number $Ra_c$,
whose value depends on the aspect ratio $\Gamma\equiv $ radius/height. 
Above $Ra_c$, convective motions appear and form various roll structures.

A summary covering the developments since the mid 1980s for convective systems
with large aspect ratio ($\Gamma \gg 1$) can be found in \cite*{BodPesAhl}. In
such domains a rich variety of patterns was reported: ``Pan Am'' patterns
\cite*[arches with several centres of curvature, see][]{AhlCanSte}, straight
parallel rolls \cite*[]{Cro,CroGalPoc}, concentric rolls \cite*[targets,
see][]{KosPal,CroMorSch}, one- and several- armed rotating spirals
\cite*[][]{PlaEgoBodPes}, targets with dislocated centre \cite[][]{Cro},
hexagonal cells \cite*[][]{CilPamPer} and spiral-defect chaos
\cite*[][]{MorBodCanAhl}. A large overview on convective phenomena observed
experimentally before this time can also be found in \cite{Kos}.

We focus here on cylinders with moderate aspect ratio $\Gamma \sim 1 $. 
The flow structure then depends strongly on system geometry. 
For this regime, the 
stability of the conductive state was well established in the
1970s--1980s  by \cite{ChaSan70}, \cite{StoMul} and \cite{BueCat}.
Critical Rayleigh numbers $Ra_c$ are about $2000$ for
$\Gamma\geq 1$, increasing steeply for lower $\Gamma$ and decreasing
asymptotically towards $Ra_{c}=1708$ for $\Gamma\rightarrow\infty$.
\cite{ChaSan70} estimated by a numerical variational technique the onset of
axisymmetric convection in cylinders of aspect ratios between 0.5 and 8, with
insulating and conducting sidewalls. They found the critical Rayleigh numbers
($Ra_c=2545$ for $\Gamma=1$, decreasing for higher $\Gamma$) and the
corresponding number of rolls. They then generalised this analysis
\cite[][]{ChaSan71}, including non-axisymmetric modes and predicting $Ra_c$
and corresponding critical azimuthal wavenumbers.
\cite{StoMul} observed experimentally convective patterns in annuli and
cylinders of aspect ratio $0.7 \leq \Gamma \leq 3.2 $, varying the sidewall
insulation. Their critical Rayleigh numbers were in good agreement with those
predicted by Charlson and Sani.
\cite{Ros} investigated convective instabilities numerically for free-slip
boundary conditions, using a severely truncated expansion in a small number of
eigenmodes. He described non-axisymmetric motions existing just above onset
for aspect ratios between 0.5 and 2.0. 
Finally, \cite{BueCat} described how the onset of convection is influenced by
the ratio of the fluid conductivity to that of the wall, by performing linear
analysis for the aspect ratio range $0 < \Gamma \leq 4$.  
They determined the critical Rayleigh number and azimuthal wavenumber as a 
function of both aspect ratio and sidewall conductivity, thus completing the 
results of the previous investigations, which considered either perfectly 
insulating or perfectly conducting walls. 
These results were confirmed by \cite*{Marques}.
The flow succeeding the conductive 
state is three-dimensional over large ranges of aspect ratios, contrary to the 
expectations of \cite{Kos}.

The stability of the first convective state, depending on both aspect ratio
and Prandtl number, has been investigated mainly for situations in which
the primary flow is axisymmetric. \cite{ChaSan75}
attempted to predict numerically the stability of the primary axisymmetric
flow, but the resolution available at that time was inadequate to the task.
\cite*{MulNeuWeb} investigated convective flows experimentally and
theoretically. They observed
axisymmetric flows for $\Gamma=1$ and non-axisymmetric flows for $0.1\leq
\Gamma \leq 0.5$.  
\cite{HarSan}
calculated weakly nonlinear solutions to the Boussinesq equations for several
moderate and small aspect ratios. They found a bifurcation from the
axisymmetric state towards a mode with azimuthal wavenumber $m=2$ for
$\Gamma=1$, $Pr=6.7$ and $Ra_{c2}=2430$.

The most complete numerical study of secondary convective instabilities for
moderate aspect ratio cylinders was performed by \cite*{WanKuhRat}. 
For cylinders with insulating sidewalls and $0.9 < \Gamma < 1.57$, 
the primary bifurcation to convection occurs at $Ra_c \approx 2000$ 
and leads to an axisymmetric flow whose stability was 
investigated for Prandtl numbers 0.02 and 1.
Wanschura et al. predicted the succeeding flows to be steady, 
except over a narrow aspect
ratio range $1.45\leq\Gamma \leq 1.57$ at $Pr=1$, where they found oscillatory
instabilities at $Ra_{c2}\approx 25 000$
towards flows with azimuthal wavenumbers $m=3$ and $m=4$.
The primary aim of this paper is to provide a more detailed description of
these bifurcations.

\cite*{TouHadHen}
numerically investigated the stability of the conductive state 
for aspect ratios $\Gamma=0.5$ and $\Gamma=1$. They
described the main critical modes and established a diagram of primary
bifurcations, including unstable branches. They also found a secondary
bifurcation point $Ra_{c2}$, at which
the axisymmetric flow becomes unstable
towards a two-roll flow and calculated $Ra_{c2}$ for 
$\Gamma=1$ and $0<Pr<1$.

An interesting experimental study was carried out by \cite*{HofLucMul}.
Varying the Rayleigh number through different sequences of values, for fixed
parameters $\Gamma=2.0$ and $Pr=6.7$, they obtained several 
different stable
patterns for the same final Rayleigh number. They also reported a transition
from an axisymmetric steady state towards azimuthal waves.
Our numerical simulations of this phenomenon are the subject of a
separate investigation.

More recently convective patterns were numerically investigated by 
\cite{RudFeu} and by \cite{Leo}. R{\"u}diger and Feudel found
stability ranges for multi-roll
patterns, targets and spirals for $\Gamma=4$, $Pr=1$. 
Leong observed several steady convective 
patterns for aspect ratios 2 and 4 and Prandtl number $Pr=7$, 
all of which were stable in the range $6250\leq Ra\leq 37\,500$,
and calculated the heat transfer for each pattern.

Convective systems often display oscillatory behavior.
In binary fluid or rotating convection, the primary
bifurcation is usually to periodic states, while
in Rayleigh-B\'enard convection, periodic behavior
occurs as a secondary bifurcation.
The oscillatory and skew-varicose instabilities of long straight 
parallel rolls calculated in, e.g. \cite{Busse74} and \cite{Busse79},
are manifested as travelling waves along rolls 
and as periodic defect nucleation \cite[]{Cro,CroGalPoc,RudFeu};
rotating spirals were observed by the same investigators; and
radially propagating patterns of concentric rolls 
were observed by \cite{Tuc88}.
However, none of these manifestations of oscillatory behavior
resemble the azimuthal waves we describe in this study.

Competition between standing and rotating azimuthal waves
has been extensively studied in thermocapillary convection, 
driven by surface-tension gradients.
For example, competition between rotating and standing waves is
observed on the upper free surface of an open cylindrical container 
by \cite{Zebib} and in the midplane of a cylindrical liquid bridge 
with free outer surface by \cite*{Leypoldt}, both of aspect ratio 1.
These azimuthal waves are very similar to those we describe
in this study; however, such flows are uncommon in the 
Rayleigh--B\'e\-nard (buoyancy-driven) convection literature.

We wished to study in detail the time-periodic non-axisymmetric states
in cylindrical Ray\-leigh--B\'enard convection
resulting from the bifurcation found by \cite{WanKuhRat}.
Hence we have simulated numerically the loss of stability
of the first convective axisymmetric solution undergoing an oscillatory
bifurcation for $1.45\leq\Gamma \leq 1.57$ and $Pr=1$. 
In this paper we describe the results of nonlinear simulations and 
linear stability analysis, which identify the scenario in terms of 
bifurcation theory in systems with symmetries.

In addition to obtaining results particular to cylindrical 
Rayleigh--B\'enard convection with these parameter combinations, 
our purpose is to demonstrate how numerical 
and theoretical techniques can be combined in order
to obtain a complete bifurcation-theoretic understanding of the
oscillatory states produced by this secondary bifurcation.
Such an approach can be applied to analyse transitions in 
a wide variety of other physical systems, ranging from flows driven
by differentially rotating boundaries \cite*[]{Nore} to 
Bose--Einstein condensation \cite*[]{Huepe}.

\section{Method}
\subsection{Governing equations}
\begin{figure}
\begin{center}
	\includegraphics[width=0.2\textwidth]{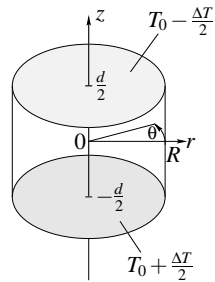}
\end{center}
	\caption{Geometry and coordinate system.}
	\label{fig:cylinder}
\end{figure}

We consider a fluid confined in a cylinder of depth $d$ and radius $R$ 
(figure~\ref{fig:cylinder}).  The aspect ratio is defined as $\Gamma\equiv R/d$.  The
fluid has kinematic viscosity $\nu$, density $\rho$, thermal diffusivity $\kappa$ and
thermal expansion coefficient (at constant pressure) $\gamma$. The top and
bottom temperatures of the cylinder are kept constant, at $T_0-\Delta T /2$
and $T_0+\Delta T /2$, respectively, 
leading to the linear conductive temperature profile $T(z)=T_0-z\Delta T/d$. 
The lateral walls are insulating.
The Rayleigh number $Ra$ and the Prandtl number $Pr$ are defined by
\begin{subeqnarray}
Ra & \equiv & \frac{\Delta T g \gamma d^3}{\kappa\nu}, \\ 
Pr & \equiv & \frac{\nu}{\kappa} .
\end{subeqnarray}
Using the units $d^2/\kappa$, $d$, $\kappa/d$ and $\nu \kappa /\gamma g d^3$
for time, distance, velocity and temperature, 
we define $\bu$ and $h$ to be the nondimensionalised velocity and 
deviation of the temperature from the basic vertical profile, respectively.
We obtain the Boussinesq equations governing the system:
\begin{subeqnarray}
\label{eq:N-S}
Pr^{-1}\left({\partial _t}\bu +
\left(\bu\cdot{\mathbf\nabla}\right)\bu
\right)
 &=& - {\mathbf\nabla}p + \Delta\bu
+h{\mathbf e_z} \\
\partial _t h
+\left(\mathbf u\cdot\nabla\right)h &=& Ra\;u_z+\Delta h \\
{\mathbf\nabla}\cdot\bu&=&0.
\end{subeqnarray}
The boundary conditions for velocity are no-slip and no-penetration
\begin{equation}
{\bf u} = 0 \qquad \quad {\rm for} \quad r=\Gamma \quad {\rm or} 
\quad z=\pm 1/2 . 
\label{eq:bcu}\end{equation}
Since the horizontal plates are assumed to be perfectly conducting (Dirichlet
condition for $h$) and the vertical walls are insulating (Neumann condition),
the boundary conditions for the temperature are
\begin{subequations}
\label{eq:bch}
\begin{eqnarray}
	h= 0 &&{\rm for}\quad z=\pm 1/2, \label{eq:bchz}\\
\frac{\partial h}{\partial r} = 0 &&{\rm for}\quad r=\Gamma .\label{eq:bchr}
\end{eqnarray}
\end{subequations}
\subsection{Symmetries}
\label{Symmetries}

Symmetries play an important role in the possible transitions
undergone by this system.
The Boussinesq equations (\ref{eq:N-S}) with boundary conditions
(\ref{eq:bcu})--(\ref{eq:bch}) have reflection symmetry in
the vertical direction $z$, and rotational and reflection symmetry in the
azimuthal direction $\theta$.
The reflection symmetry in $z$\ is broken by the first bifurcation
to a convective state.
If the first convective state consists of axisymmetric convective rolls,
then its remaining symmetries are reflection and rotation in $\theta$,
which together comprise the symmetry group $O(2)$.
Bifurcations in the presence of $O(2)$ symmetry were studied
and classified during the 1980s by a large number of researchers, 
e.g.~\cite{Bajaj,GolSte,Knobloch,GilMal,Kuznetsov,CoulletIooss}.
We give a brief summary of their results.

First, the critical eigenvector may be axisymmetric.
This case may be further subdivided
according to whether the eigenvector is reflection-symmetric or
antisymmetric in $\theta$ 
and whether the eigenvalue is real or complex.
A reflection-symmetric eigenvector can lead to a target pattern
of radially propagating rolls, e.g.~\cite{Tuc88}.
The breaking of reflection symmetry is associated with azimuthal flow.

Secondly, the critical eigenvector may be non-axisymmetric.
If the critical eigenvalue is real, then the resulting bifurcation is a circle
pitchfork, leading to a ``circle'' of steady states parametrised by phase.
Each steady state is reflection symmetric in $\theta$ (about some value
$\theta_0$). If reflection symmetry is broken by a subsequent bifurcation, the
scenario is that of a drift pitchfork, leading to slow motion (``drift'')
along the circle.
A complex eigenvalue corresponding to a non-axisymmetric eigenvector, 
like that found by \cite{WanKuhRat} for parameters
$1.45\leq\Gamma \leq 1.57$, $Pr=1$, $Ra > 23\,000$, leads to a Hopf
bifurcation which engenders three nonlinear branches: standing waves,
counterclockwise travelling waves, and clockwise travelling waves.
The standing waves are
reflection-symmetric in $\theta$ (again about some value $\theta_0$), while
the travelling waves break this symmetry. Our aim is to determine
which of these types of waves is realised by our physical system.

\subsection{Numerical integration}
We integrated the equations by a classical pseudospectral method
\cite[]{GotOrs}, in which each scalar $f$ of the fields {\bf u} and $h$ is
represented using Chebyshev polynomials in the radial and vertical direction
and Fourier series in the azimuthal direction
\begin{equation}
f\left(r,z,\theta,t\right) \quad = \quad
\sum _{j,k,m=0}^{N_r,N_z,N_ \theta }
\hat{f}_{jkm}\left(t\right)C_j(r/\Gamma)C_k(2z)e^{im\theta}+{\rm~ c.c.},
\end{equation}
where the permitted combinations of $(j,m)$ are restricted by the
parity and regularity conditions described in \cite{Tuc} for
$u_r$, $u_\theta$, $u_z$ and $h$.  The nonlinear (advective)
terms were calculated in physical space and integrated via the Adams--Bashforth
formula, while the linear (diffusive) terms were calculated in spectral space
and integrated via the Crank--Nicolson formula.  
An influence matrix method was used to impose incompressibility
\cite[]{Tuc}.  A resolution of $N_r+1=36$,
$2(N_\theta+1)=80$, $N_z+1=18$ gridpoints or modes was found to be 
sufficient for nonlinear simulations.  
All computations were performed on the NEC SX-5 vector
supercomputer, with time step $2\times 10^{-4}$ or $4\times 10^{-4}$,
depending on $Ra$, with CPU time per time step per grid point of $10^{-6}$.

\subsection{Linear stability analysis}
\label{linstab}
An important additional element in understanding the phenomena undergone
by the system is linear stability analysis. The procedure, which we summarise
below, is described in more detail in \cite{MamTuc,TucBar} and references therein.
We linearise the equations about a steady state 
$({\mathbf U},H)$:
\begin{subeqnarray}
\label{eq:linN-S}
Pr^{-1}\left({\partial _t}\bu +
\left({\mathbf U}\cdot{\mathbf\nabla}\right)\bu
+\left(\bu\cdot{\mathbf\nabla}\right){\mathbf U}
\right)
 &=& - {\mathbf\nabla}p + \Delta\bu+h{\mathbf e_z} \\
\partial _t h
+\left(\mathbf U\cdot\nabla\right)h 
+\left(\mathbf u\cdot\nabla\right)H
&=& Ra\;u_z+\Delta h \\
{\mathbf\nabla}\cdot\bu&=&0.
\end{subeqnarray}
Equations (\ref{eq:linN-S}) with boundary conditions 
(\ref{eq:bcu})-(\ref{eq:bch}) are then integrated in time in the same way
as the nonlinear equations (\ref{eq:N-S}).
We abbreviate the linear evolution problem (\ref{eq:linN-S}) by
\begin{equation}
\partial_t 
\left(\begin{array}{c}\bu \\ h \end{array}\right)=
L \left(\begin{array}{c}\bu \\ h \end{array}\right)
.
\label{eq:abbrev}
\end{equation}
Temporal integration is equivalent to carrying out the power method on
the approximate exponential operator, since
\begin{equation}
\label{eq:expEv}
\left(\begin{array}{c}\bu \\ h \end{array}\right)(t+\Delta t) 
= e^{L\Delta t} \left(\begin{array}{c}\bu \\ h \end{array}\right)(t)
.
\end{equation}

In order to extract the leading real or complex eigenvalues
(those of largest real part) and corresponding eigenvectors, 
we postprocess the results of integrating (\ref{eq:linN-S})
as follows.
A small number of fields 
\begin{equation}
\left(\begin{array}{c}\bu \\ h \end{array}\right)(0),
\left(\begin{array}{c}\bu \\ h \end{array}\right)(T),
\left(\begin{array}{c}\bu \\ h \end{array}\right)(2T),\dots ,
\left(\begin{array}{c}\bu \\ h \end{array}\right)((K-1)T)
\label{eq:Kvecs}
\end{equation}
are calculated, by carrying out $T/\Delta t$ linearised timesteps.
The {\it Krylov space} corresponding to initial vector $(\bu,h)^T$ and
matrix $e^{LT}$ is the $K$-dimensional linear subspace consisting of
all linear combinations of vectors in (\ref{eq:Kvecs}).
These vectors are 
orthonormalised to one another to generate a set of vectors 
$v_1, v_2, v_3,~\ldots v_K$ which form a basis for the Krylov space.  
The action of the operator on the Krylov space is
represented by a small ($K\times K$) matrix $M$ whose elements are
\begin{equation}
M_{jk}\equiv \langle v_j,e^{LT} v_k\rangle.
\end{equation}
The small matrix $M$ can be directly diagonalised. 
Its eigenvalues $\lambda$ approximate a small number of the eigenvalues of the large matrix $e^{LT}$:
this is the essence of {\it Arnoldi's method}.
The procedure of generating the Krylov space via repeated action of $e^{LT}$
selects preferentially the $K$ dominant values (those of largest magnitude) of $e^{LT}$,
i.e.~the $K$ leading eigenvalues (those of largest real part) of $L$.

The eigenvectors of $M$ prescribe coefficients of the vectors $v_j$
which can be combined to form approximate eigenvectors 
$\phi$ of $e^{LT}$.
The accuracy of these approximate eigenpairs $(\lambda,\phi)$ is measured by the residue
$||e^{LT}\phi - \lambda \phi||$ in the case of real eigenvalues or by
the residues
$||e^{LT}\phi^R - (\lambda^R \phi^R -\lambda^I \phi^I)||$,
$||e^{LT}\phi^I - (\lambda^R \phi^I +\lambda^I \phi^R)||$
in the case of complex eigenvalues.
If the desired eigenvalues have sufficiently small residues,
they are accepted; otherwise
we continue integration of (\ref{eq:linN-S}), replacing (\ref{eq:Kvecs}) by
\begin{equation}
\left(\begin{array}{c}\bu \\ h \end{array}\right)(T),
\left(\begin{array}{c}\bu \\ h \end{array}\right)(2T),
\left(\begin{array}{c}\bu \\ h \end{array}\right)(3T),\dots ,
\left(\begin{array}{c}\bu \\ h \end{array}\right)(KT)
\end{equation}
and so on, until the residue is below the acceptance criterion.

After integrating the axisymmetric version of the nonlinear equations 
(\ref{eq:N-S}) at a given Rayleigh number to create the nonlinear 
axisymmetric solution $({\mathbf U}, H)$, we 
integrated the non-axisymmetric linearised equations (\ref{eq:linN-S}) 
to evolve $({\mathbf u},h)$ from an arbitrary initial condition.
To integrate (\ref{eq:linN-S}), we used a timestep 
of $\Delta t = 10^{-4}$ and a spatial resolution of $N_r=47, N_z=29$ 
for each azimuthal mode.
To construct the Krylov space (\ref{eq:Kvecs}) and approximate eigenpairs,
we used $K=10$ vectors, a time interval of $T=100\Delta t= 10^{-2}$, 
and an acceptance criterion of $10^{-5}$.

\subsection{Complex eigenvectors and their representations}
\label{Complex eigenvectors and their representations}

The linear problem (\ref{eq:linN-S}) for perturbations
$(\bu,h)$ about an axisymmetric convective state $({\mathbf U},H)$ 
can be divided into decoupled subproblems, each corresponding 
to a single azimuthal wavenumber $m$.
The problem for wavenumber $m$ can in turn be divided
into two identical decoupled subproblems, corresponding
to fields of the form
\begin{subequations}
\begin{eqnarray}
\label{eq:linreal}
\hat{u}_r(r,z) \cos(m\theta),\;\;\hat{u}_\theta (r,z) \sin(m\theta),\;\;&&
\hat{u}_z(r,z)\cos(m\theta),\;\; \hat{h}(r,z)\cos(m\theta),\label{eq:costype}\\
&&\mbox{and} \nonumber\\
\hat{u}_r(r,z) \sin(m\theta),\;\;\hat{u}_\theta (r,z) \cos(m\theta),\;\;&&
\hat{u}_z(r,z)\sin(m\theta),\;\; \hat{h}(r,z)\sin(m\theta). \label{eq:sintype}
\end{eqnarray}
\end{subequations}
For simplicity, we will represent each of these types of vector fields 
by its temperature component $\hat{h}(r,z)$ and leave the dependence
on $\theta$ and on $t$ to be written explicitly.
We may write the linear evolution problem (\ref{eq:abbrev}) 
restricted to fields with trigonometric dependence on $m\theta$ such as 
(\ref{eq:costype})--(\ref{eq:sintype}) as
\begin{equation}
\partial_t \hat{h} = \hat{L}_m \hat{h}. \label{eq:abbrev2}
\end{equation}

A real eigenvalue breaking azimuthal symmetry in an $O(2)$ symmetric
situation is associated with a two-dimensional eigenspace,
consisting of linear combinations of vectors of type
(\ref{eq:costype}) and (\ref{eq:sintype}).
Since
\begin{subequations}
\begin{equation}
\alpha\:\hat{h}(r,z)\cos(m\theta)+\beta\:\hat{h}(r,z)\sin(m\theta)=
C\: \hat{h}(r,z) \cos(m(\theta-\theta_0)),
\label{eq:combinetrig}\end{equation}
where
\begin{equation}
C=\sqrt{\alpha^2+\beta^2},\;\;\;\;\;m\theta_0={\rm atan}(\beta/\alpha),
\end{equation}
\end{subequations}
all real eigenvectors have $m$ nodal lines and reflection symmetry
about some $\theta_0$. If we take $C\propto\sqrt{Ra-Ra_{c2}}$ and
add (\ref{eq:combinetrig}) to the basic axisymmetric state,
we obtain the ``circle'' of steady states resulting from
a circle pitchfork mentioned in~\S\,\ref{Symmetries}.

A complex eigenvalue in the $O(2)$ symmetric situation is associated with a 
four-dimensional eigenspace.
Within each eigenvector class (\ref{eq:costype}) and (\ref{eq:sintype}),
the eigenspace is two-dimensional, spanned by two linearly independent
eigenvectors $\hat{h}^R$ and $\hat{h}^I$, which are transformed by $\hat{L}_m$ as
\begin{equation}
\hat{L}_m \left(\begin{array}{c}\hat{h}^R \\ \hat{h}^I \end{array}\right)
= \left( \begin{array}{cr}\mu & -\omega \\ \omega & \mu \end{array} \right)
\left(\begin{array}{c}\hat{h}^R \\ \hat{h}^I \end{array}\right)
\label{eq:howcomplexacts}\end{equation}
In (\ref{eq:howcomplexacts}), $\hat{h}^R$ can be replaced
by any linear combination of $\hat{h}^R$ and $\hat{h}^I$, but
once $\hat{h}^R$ is selected, the choice of $\hat{h}^I$ follows from 
(\ref{eq:howcomplexacts}).
Although the components of equation (\ref{eq:howcomplexacts}) are the real and imaginary parts of the complex equation
\begin{equation}
\hat{L}_m (\hat{h}^R + i \hat{h}^I) = (\mu + i\omega)(\hat{h}^R + i \hat{h}^I),
\end{equation}
the customary designation of $\hat{h}^R$ and $\hat{h}^I$
as the real and the imaginary part of the eigenvector is arbitrary,
as reflected by the fact that an eigenvector can be multiplied 
by any complex number.

To form eigenvectors of the full cylindrical problem belonging to
the four-dimensional eigenspace,
each of $\hat{h}^R$ and $\hat{h}^I$ is multiplied by a trigonometric function.
This yields as a basis for the four-dimensional eigenspace:
\begin{subequations}
\begin{eqnarray}
\label{eq:fourforms}
\hat{h}^R(r,z) \cos(m\theta), \label{eq:RC}\\
\hat{h}^I(r,z) \cos(m\theta), \label{eq:IC}\\
\hat{h}^R(r,z) \sin(m\theta), \label{eq:RS}\\
\hat{h}^I(r,z) \sin(m\theta).\;  \label{eq:IS}
\end{eqnarray}
\end{subequations}
One choice for a complex eigenvector pair is
(\ref{eq:RC})-(\ref{eq:IC}), since
\begin{equation}
\hat{L}_m 
\left(\begin{array}{c}\hat{h}^R \cos(m\theta)\\ \hat{h}^I \cos(m\theta)\end{array}\right)
= \left( \begin{array}{cr}\mu & -\omega \\ \omega & \mu \end{array} \right)
\left(\begin{array}{c}\hat{h}^R \cos(m\theta)\\ \hat{h}^I \cos(m\theta)\end{array}\right)
\end{equation}
More generally, the trigonometric dependence can be taken
as in (\ref{eq:combinetrig}), with the same trigonometric dependence for
each of $\hat{h}^R$ and $\hat{h}^I$, to form a complex conjugate eigenvector
pair each of whose members has $m$ nodal lines and $m$ axes of reflection 
symmetry, including $\theta=\theta_0$.
The evolution in time under (\ref{eq:abbrev2}) for a field with an
initial condition of this form is
\begin{equation}
h(r,\theta,z,t)=\alpha e^{\mu t}\left[\hat{h}^R(r,z) \cos(\omega t)
-\hat{h}^I(r,z) \sin(\omega t) \right] \cos(m(\theta-\theta_0)).
\label{eq:standingwave}\end{equation}
The subspace of fields with azimuthal dependence
$\cos(m(\theta-\theta_0))$ is invariant under linearised time evolution.
(There also exists an invariant subspace under the nonlinear time evolution,
which includes harmonics $\cos(km(\theta-\theta_0))$, with the same $m$
axes of reflection symmetry.)
If we take $\mu=0$ and $\alpha \propto \sqrt{Ra-Ra_{c2}}$
in (\ref{eq:standingwave}), and add this to the basic axisymmetric 
solution, then we obtain to first order
the standing wave solution mentioned in~\S\,\ref{Symmetries}.

Any combination of (\ref{eq:RC})-(\ref{eq:IS}) is also
a member of a complex eigenvector pair.
The calculation
\begin{eqnarray}
&&\hat{L}_m \left( \begin{array}{c}
\alpha \hat{h}^R(r,z) \cos(m\theta) + \beta \hat{h}^I(r,z) \sin(m\theta) \\
\alpha \hat{h}^I(r,z) \cos(m\theta)-\beta \hat{h}^R(r,z)\sin(m\theta)
\end{array}\right) \nonumber\\
&&= \left( \begin{array}{cr}\mu & -\omega \\ \omega & \mu \end{array} \right)
\left( \begin{array}{c}
\alpha \hat{h}^R(r,z) \cos(m\theta) + \beta \hat{h}^I(r,z) \sin(m\theta) \\
\alpha \hat{h}^I(r,z) \cos(m\theta)-\beta \hat{h}^R(r,z)\sin(m\theta)
\end{array}\right),
\label{eq:asym}\end{eqnarray}
when compared with (\ref{eq:howcomplexacts}), shows that 
the two components of the vector in (\ref{eq:asym})
form a complex conjugate pair of eigenvectors 
for the full cylindrical problem, as in (\ref{eq:howcomplexacts}).
Because $\hat{h}^R(r,z)$ and $\hat{h}^I(r,z)$ have different functional
forms in $(r,z)$, these vectors, unlike those of (\ref{eq:combinetrig}),
cannot be combined into a single trigonometric function.
Neither of the two components of (\ref{eq:asym}) has nodal lines
or reflection symmetry about any axis if both $\alpha$ and $\beta$
are non-zero.
The evolution in time under (\ref{eq:abbrev2}) for a field whose
initial condition is the first component of (\ref{eq:asym}) is
\begin{eqnarray}
h(r,\theta,z,t)
=e^{\mu t} 
[\hat{h}^R(r,z) (\alpha \cos(m\theta) \cos(\omega t)
-\beta \sin(m\theta)\sin(\omega t)) \nonumber\\
+ \hat{h}^I(r,z) (\alpha \cos(m\theta)\sin(\omega t))
+\beta \sin(m\theta)\cos(\omega t))].
\label{eq:complexevol}\end{eqnarray}
If $\beta=\pm\alpha$, then (\ref{eq:complexevol}) becomes
\begin{equation}
h(r,\theta,z,t)=
e^{\mu t} \alpha
[\hat{h}^R(r,z) \cos(m\theta \pm \omega t)
+ \hat{h}^I(r,z)  \sin(m\theta \pm \omega t)],
\label{eq:travellingwave}\end{equation}
where $t$ or $\theta$ may be replaced by $(t-t_0)$ or $(\theta-\theta_0)$.
If we take $\mu=0$ and $\alpha \propto \sqrt{Ra-Ra_{c2}}$ 
in (\ref{eq:travellingwave}) and add the basic axisymmetric solution,
then we obtain, to first order, the expression for 
clockwise $(m\theta+\omega t)$
or counterclockwise $(m\theta-\omega t)$ travelling waves
mentioned in~\S\,\ref{Symmetries}.

\subsection{Amplitude equations and normal form}

The linearised evolution treated in the previous section permits any
combinations of (\ref{eq:RC})--(\ref{eq:IS}).  The mathematical analysis of
Hopf bifurcation in the presence of $O(2)$ symmetry carried out by
e.g. \cite{Bajaj,GolSte,Knobloch,GilMal,Kuznetsov} describes the
effect of including generic nonlinear terms compatible with the symmetries.
Following the formulation of these authors, we decompose the field
into a sum of clockwise and counterclockwise travelling waves
with complex amplitudes $\zeta_- = \rho_- e^{i\phi_-}$ and 
$\zeta_+= \rho_+ e^{i\phi_+}$, respectively.
The four variables $\rho_\pm, \phi_\pm$ form another description
of the four-dimensional space described in the previous section.
The nonlinear evolution of $\zeta_\pm$ near the bifurcation
can be described by the following amplitude equations or {\it normal form}:
\begin{subequations}
\label{nfz}
\begin{eqnarray}
\dot{\zeta}_+ &=& \left(\mu + i \omega + a |\zeta_-|^2 
+ b (|\zeta_+|^2 + |\zeta_-|^2) \right) \zeta_+\label{nfz1},\\
\dot{\zeta}_- &=& \left(\mu + i \omega + a |\zeta_+|^2 
+ b (|\zeta_+|^2 + |\zeta_-|^2) \right) \zeta_- \label{nfz2}.
\end{eqnarray}
\end{subequations}
We use the normal form to interpret the results of our
full numerical simulations.

\begin{figure}
\begin{center}
	\includegraphics[width=0.5\textwidth,clip=true]{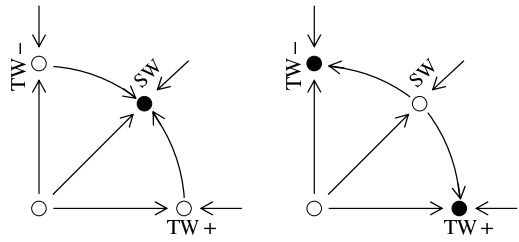}
\end{center}
	\caption{Phase diagram illustrating stability of standing waves (left)
	or travelling waves (right). The origin is the basic state and
the axes represent amplitudes
of counterclockwise and clockwise travelling waves $\rho_+$ and $\rho_-$. 
Standing waves can be constructed as an equal superposition of the two.}
	\label{fig:phase}
\end{figure}

Separating (\ref{nfz}) into equations for real amplitudes $\rho\pm$ and 
phases $\phi_\pm$ leads to
\begin{subequations}
\label{nf}
\begin{eqnarray}
\dot{\rho}_+ &=& \left(\mu + a_r \rho_-^2 + b_r (\rho_+^2 + \rho_-^2) \right) \rho_+\label{nf1},\\
\dot{\rho}_- &=& \left(\mu + a_r \rho_+^2 + b_r (\rho_+^2 + \rho_-^2) \right) \rho_- \label{nf2},\\
\dot{\phi}_+ &=& \omega + a_i \rho_-^2 + b_i (\rho_+^2 + \rho_-^2) \label{nf3},\\
\dot{\phi}_- &=& -\omega - a_i \rho_+^2 - b_i (\rho_+^2 + \rho_-^2) \label{nf4}.
\end{eqnarray}
\end{subequations}
Periodic solutions to (\ref{nf}) must be either standing or 
travelling waves.
Solutions to (\ref{nf}) and their properties are given in 
Table \ref{tab:nfsols}.
This table shows that both standing and travelling wave solutions 
exist for $\mu>0$ if $b_r$ and $a_r+2b_r$ are both negative.  A positive growth rate from a solution
indicates instability.  Thus, the stability of the solutions
depends on the sign of $a_r$: if $a_r>0$, then standing waves are stable and
travelling waves unstable, and vice versa for $a_r<0$.
Figure~\ref{fig:phase} shows phase portraits for the amplitudes 
$(\rho_+,\rho_-)$, for the cases in which all three branches co-exist
and either the standing or the travelling waves are stable.  
\begin{table}
\begin{eqnarray*}
\begin{array}{llcc} 
\mbox{{\hfill name \hfill}} & \mbox{solution} & \mbox{growth rates} & \mbox{frequencies} \\
\hline 
\mbox{Basic state} & \rho_+=\rho_- = 0 & \mu,\; \mu &\\
\mbox{Counterclockwise wave} & \rho_+ = \sqrt{\frac{-\mu}{b_r}},\; 
\rho_-=0 & -2\mu,\; -\frac{a_r}{b_r}\mu &
\omega-\frac{b_i}{b_r} \mu \\
\mbox{Clockwise wave} & \rho_- = \sqrt{\frac{-\mu}{b_r}},\; 
\rho_+=0 & -2\mu,\; -\frac{a_r}{b_r}\mu  &
-\left(\omega-\frac{b_i}{b_r}\mu\right) \\
\mbox{Standing wave} & \rho_+ = \rho_-=\sqrt{\frac{-\mu}{a_r+2b_r}} 
& -2\mu,\; \frac{2a_r}{a_r+2b_r}\mu &
\pm\left(\omega-\frac{a_i+2b_i}{a_r+2b_r}\mu\right) 
\end{array}
\end{eqnarray*}
\caption{Solutions to (\ref{nf}) and their properties.}
\label{tab:nfsols}
\end{table}

\section{Results}

\subsection{Conductive state}
\begin{figure}
	\begin{center}
	\includegraphics[width=10cm]{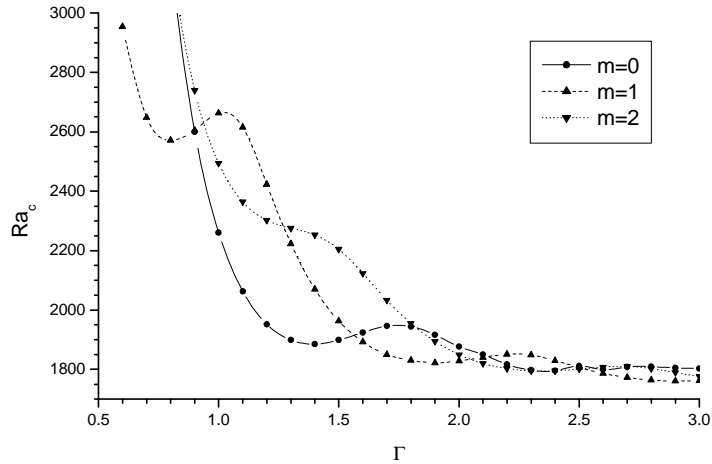}
	\end{center}
	\caption{Linear stability of the conductive state}
	\label{fig:Ra_c1}
\end{figure}

Figure~\ref{fig:Ra_c1} shows the linear stability limits of the conductive
state to perturbations with azimuthal wavenumbers $m=0$, 1, and 2 
\cite[][]{BorBor}.  These results, obtained with the linearised version of our
code, agree very closely with those presented by \cite{WanKuhRat}.  Note that
in the range $0.9<\Gamma<1.57$, the primary instability is axisymmetric.
Immediately below and above this range of aspect ratio, the first instability
is to an eigenvector with azimuthal wavenumber $m=1$.
Instability of the conductive state is independent of $Pr$.
However, the resulting nonlinear states and their stability depend on $Pr$;
in the remainder of the study we fix $Pr=1$.

\subsection{Steady axisymmetric state}
\begin{figure}
\begin{center}
\begin{tabular}{cc}
\includegraphics[angle=180,width=6cm]{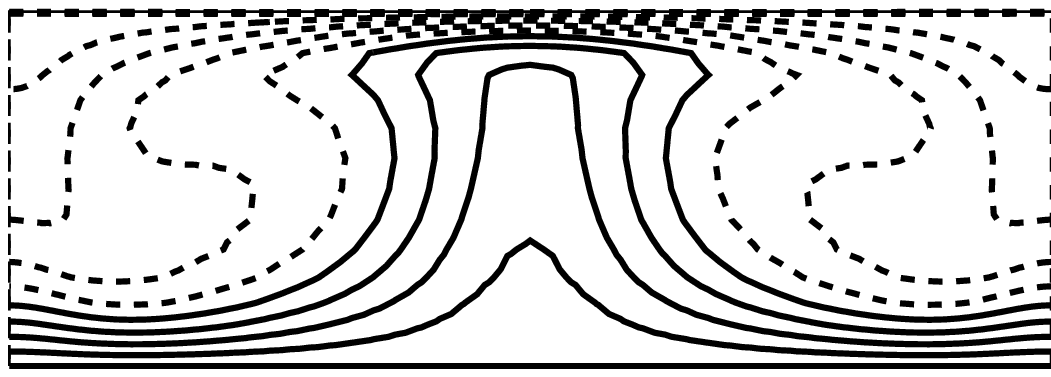}&  
\includegraphics[angle=180,width=6cm]{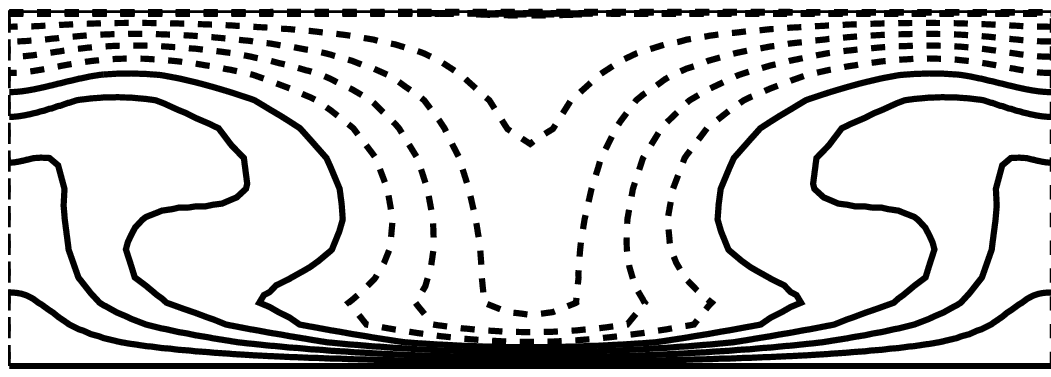}  
\\
(\textit{a})&(\textit{b})
\end{tabular}
\end{center}
\caption{Temperature contours for axisymmetric solutions 
at $\Gamma=1.47$ and $Ra=1950$ with upward (left) 
and downward (right) flow at the centre.
Solid (dashed) curves correspond to positive (negative) values,
here and in subsequent visualisations.}
\label{fig:axitemp}
\end{figure}

We reproduced the primary flow for $\Gamma = 1.47$ and $Ra=1950$,
parameters for which, according to \cite{WanKuhRat} and figure~\ref{fig:Ra_c1},
the conductive state is
unstable only to axisymmetric perturbations.  In a fully three-dimensional
simulation, starting the evolution from an arbitrary non-axisymmetric
perturbation about the conductive state, we obtained a flow 
consisting of one toroidal roll. While axisymmetric, this flow breaks
the reflection symmetry in $z$ and thus two such states exist, with
either upflow or downflow at the centre; these
are illustrated in figure \ref{fig:axitemp}.
We used the state with downflow at the centre as the initial condition 
for higher Rayleigh numbers.
According to the calculations of \cite{WanKuhRat}, the axisymmetric state 
first bifurcates towards a flow with azimuthal wavenumber $m=3$ for 
$1.45\leq\Gamma <1.53$ and with wavenumber
$m=4$ for $1.53 \leq \Gamma\leq 1.57$.  
The critical Rayleigh numbers $Ra_{c2}$ at which this loss of stability
occurs are given in table \ref{tab:Ra}.

\subsection{Eigenvalues and eigenvectors}
\label{Eigenvalues and eigenvectors}
Using the methods described in~\S\,\ref{linstab}, 
we integrated the evolution equations (\ref{eq:linN-S})
linearised about axisymmetric solutions
for aspect ratios $1.45 \leq \Gamma \leq 1.57$ and 
several different Rayleigh numbers.
The leading eigenpairs calculated for $Ra=24\,000$, $\Gamma=1.57$ are given in 
Table \ref{tab:lintable}.  For these parameter values, the 
critical eigenvectors are (in order of decreasing growth rate): 
two conjugate pairs with azimuthal wavelengths $m=4$ and $m=3$, 
a real eigenvector with $m=1$, and another conjugate pair with $m=5$.

\begin{figure}
\begin{center}
	\begin{tabular}{cc}
		  \includegraphics{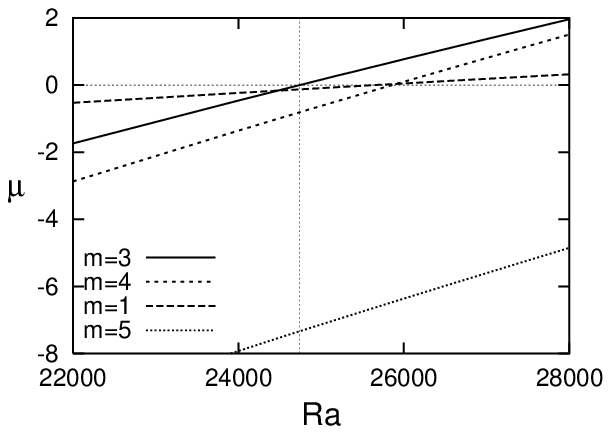} &
		  \includegraphics{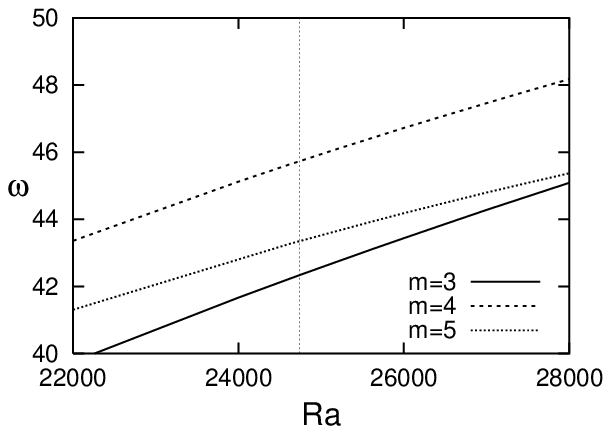}\\
		(\textit{a}) & (\textit{b})\\ 
		\includegraphics{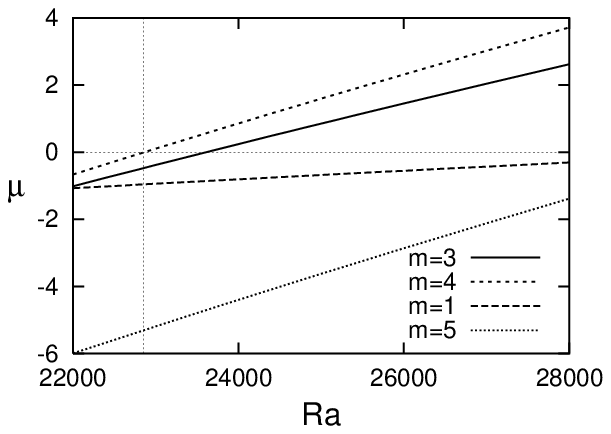}&
		\includegraphics{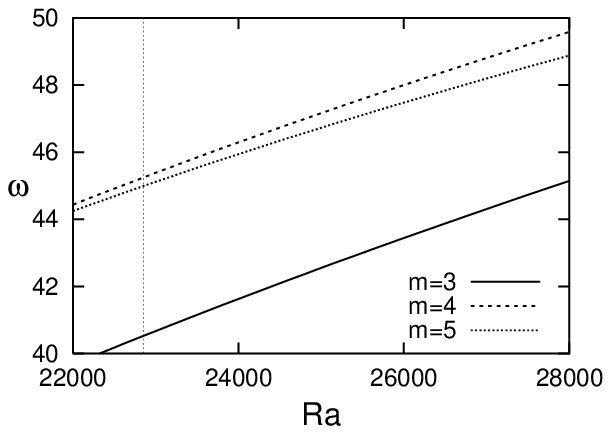}\\
		(\textit{c}) & (\textit{d})
	\end{tabular}
\end{center}
 	\caption{Leading eigenvalues as a function of Rayleigh number 
	for aspect ratio $\Gamma=1.47$: (\textit{a})  real part,  (\textit{b}) imaginary part and for
	aspect ratio $\Gamma=1.57$: (\textit{c}) real part, (\textit{d}) imaginary part. 
	Vertical thin dashed line marks $Ra_{c2}=24738$ for $\Gamma=1.47$ and $Ra_{c2}=22849$ for $\Gamma=1.57$.}
	\label{fig:eigenvalues}
\end{figure}

Figure~\ref{fig:eigenvalues} represents the dependence of the 
leading eigenvalues on Rayleigh number 
for aspect ratios $\Gamma=1.47$ and $\Gamma=1.57$,
along with the azimuthal wavenumbers of the corresponding eigenvectors.
$Ra_{c2}$ was calculated by determining
the zero crossing of $\mu\left(Ra\right)$, the
growth rate of the leading eigenvalue (that of largest real part), by linear
interpolation. 
(Critical Rayleigh numbers calculated by introducing perturbations 
into nonlinear simulations at various values of $Ra$, 
and fitting the initial evolution to an 
exponential to calculate growth or decay rates $\mu(Ra)$ gave similar results.)
We then calculated $\omega_{c2}\equiv \omega(Ra_{c2})$, also by
linear interpolation.  The values we obtained for two aspect ratios
$\Gamma=1.47$ and $\Gamma=1.57$, and the corresponding values published by
\cite{WanKuhRat} are those given in Table \ref{tab:Ra}.
The critical wavenumbers are the same, and
the errors in $Ra_{c2}$ and in $\omega_{c2}$ are less than 1\%.
In what follows, we will focus 
on the $m=3$ instability, since the $m=4$ transition is similar; the
aspect ratio is $\Gamma=1.47$ unless otherwise specified.

\begin{table}
	\begin{center}
	\begin{tabular}[]{ccccc}
$\Gamma$	&  & present study&Wanschura \etal & error \\\hline
	\multirow{3}{*}{$1.47$} &	
$Ra_{c2}$	& $24\,738$			& $24\,928$ 	& 0.76\% \\
& $\omega_{c2}$	& 42.33	& 42.54 & 0.48\% \\
& $m_{c2}$	& 3	& 3\\ \hline
	\multirow{3}{*}{$1.57$} 
& $Ra_{c2}$	& $22\,849$	& $23\,011$ & 0.70\% \\
& $\omega_{c2}$	& 45.26	& 45.47 & 0.45\%\\
& $m_{c2}$	& 4	& 4\\
\end{tabular}
	\end{center}
	\caption{The parameters of the oscillatory bifurcations found by linear 
	analysis: critical Rayleigh numbers $Ra_{c2}$, critical frequencies 
	$\omega_{c2}$ and azimuthal 
wavenumbers of critical eigenvectors for two aspect ratios. }
	\label{tab:Ra}
\end{table}
\begin{table}
\begin{center}
\begin{tabular}{ccccc}
	 \multirow{2}{*}{eigenvalue}			&	\multicolumn{2}{c}{eigenvector visualisation}				&	 \multirow{2}{*}{wavenumber}	&	 \multirow{2}{*}{error}	\\	
				&	real part		&	$\pm$ imaginary part	&		&		\\	\hline
$	0.86	\pm	46.3	i$ & 	\parbox[c]{1.7cm}{\includegraphics[width=1.7cm]{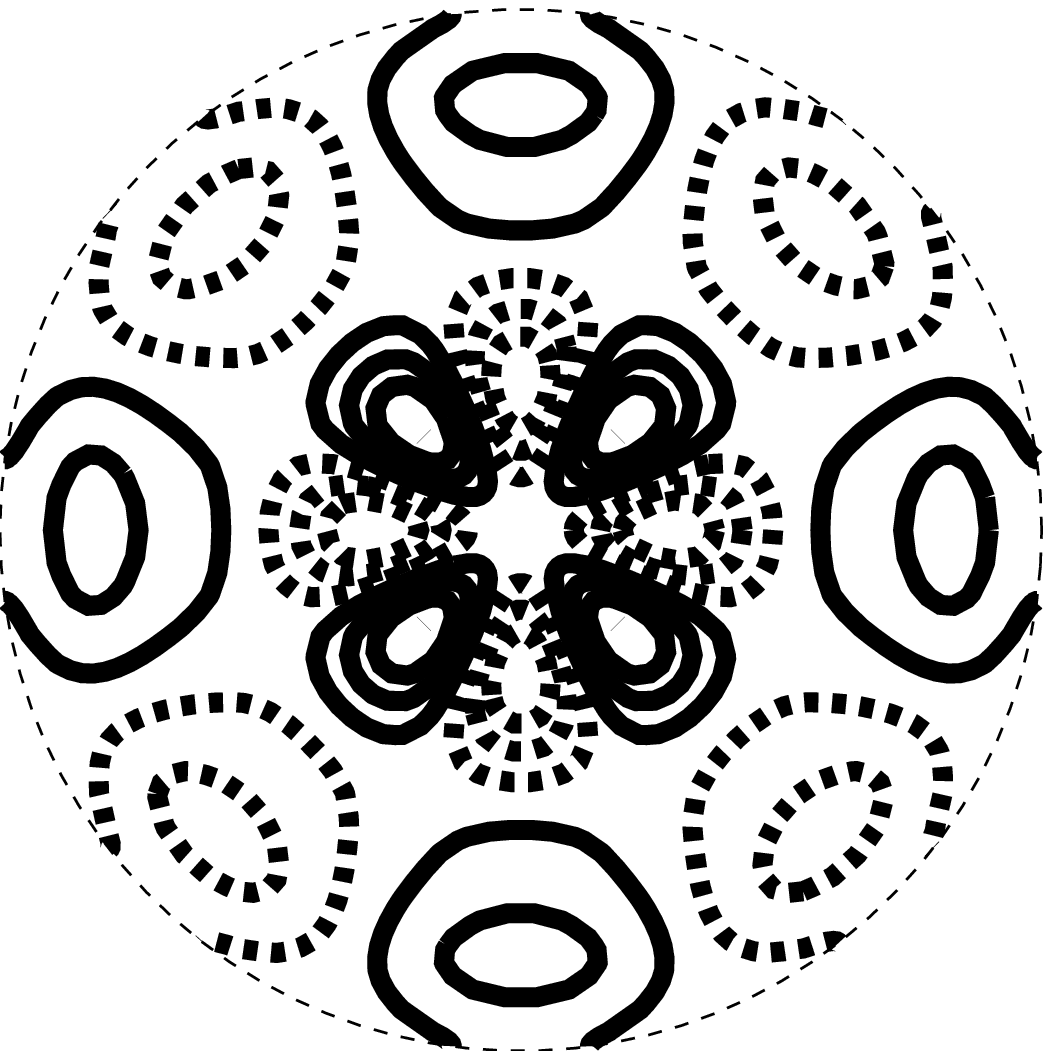}}		&	\parbox[c]{1.7cm}{\includegraphics[width=1.7cm]{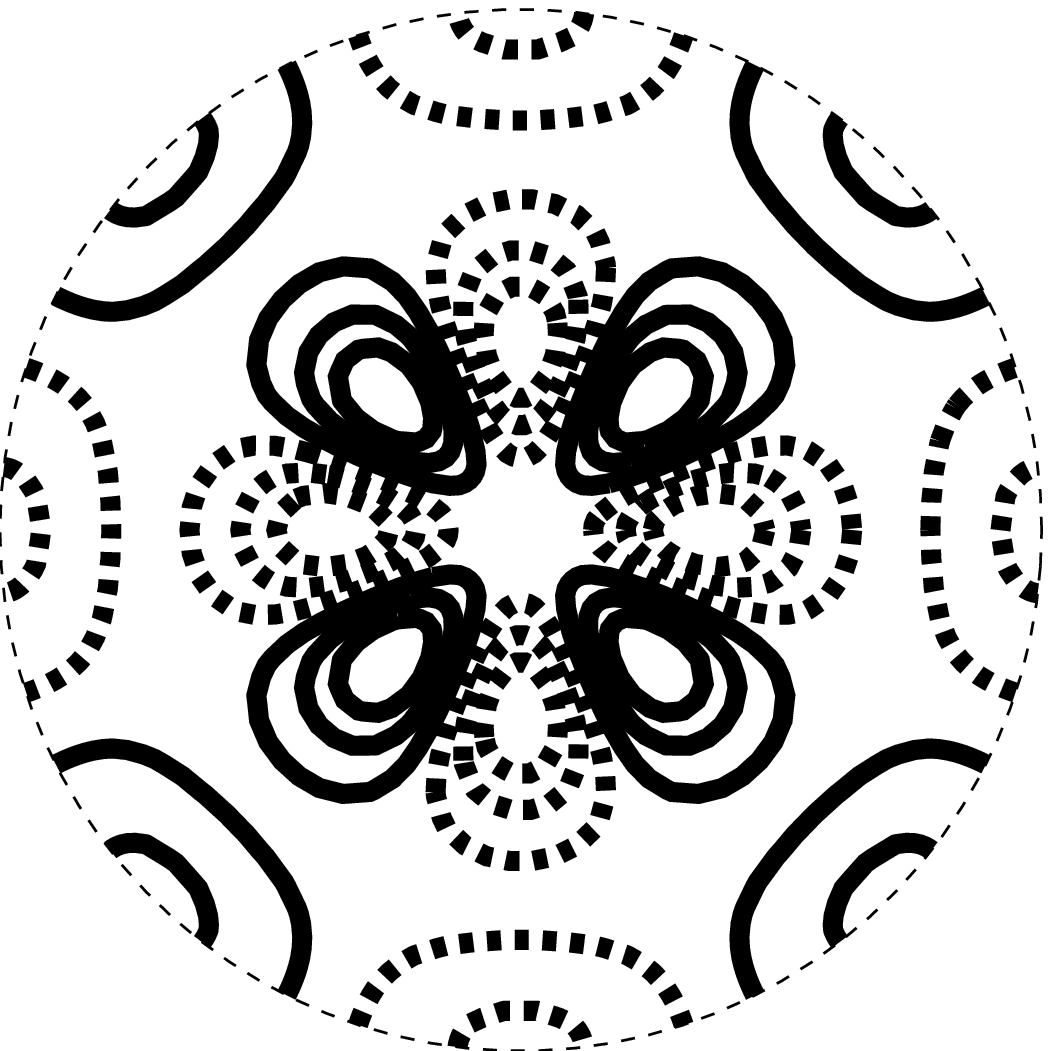}}	&$	4	$&$	 10 ^ {-10}	$\\\hline
$	0.24	\pm	41.6	i$ & 	\parbox[c]{1.7cm}{\includegraphics[width=1.7cm]{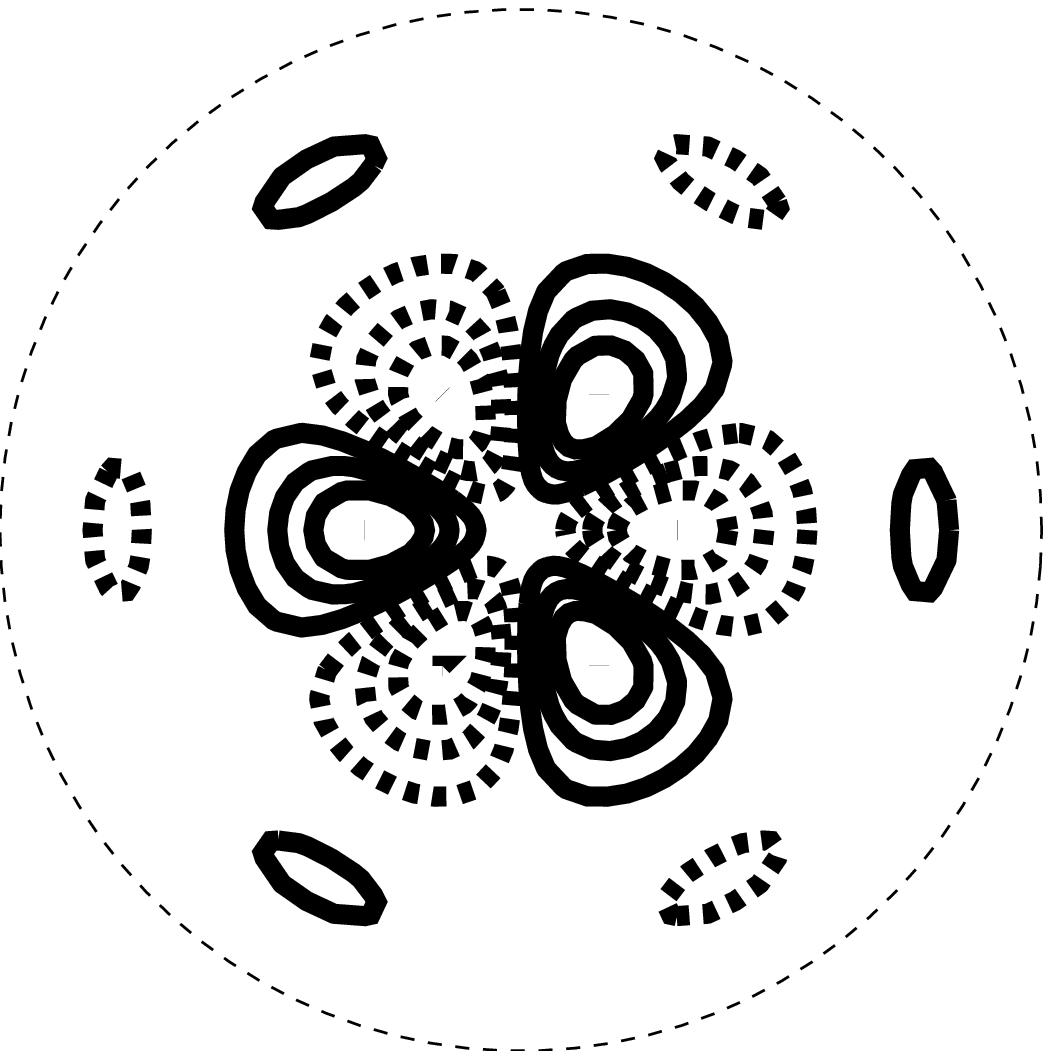}}		&	\parbox[c]{1.7cm}{\includegraphics[width=1.7cm]{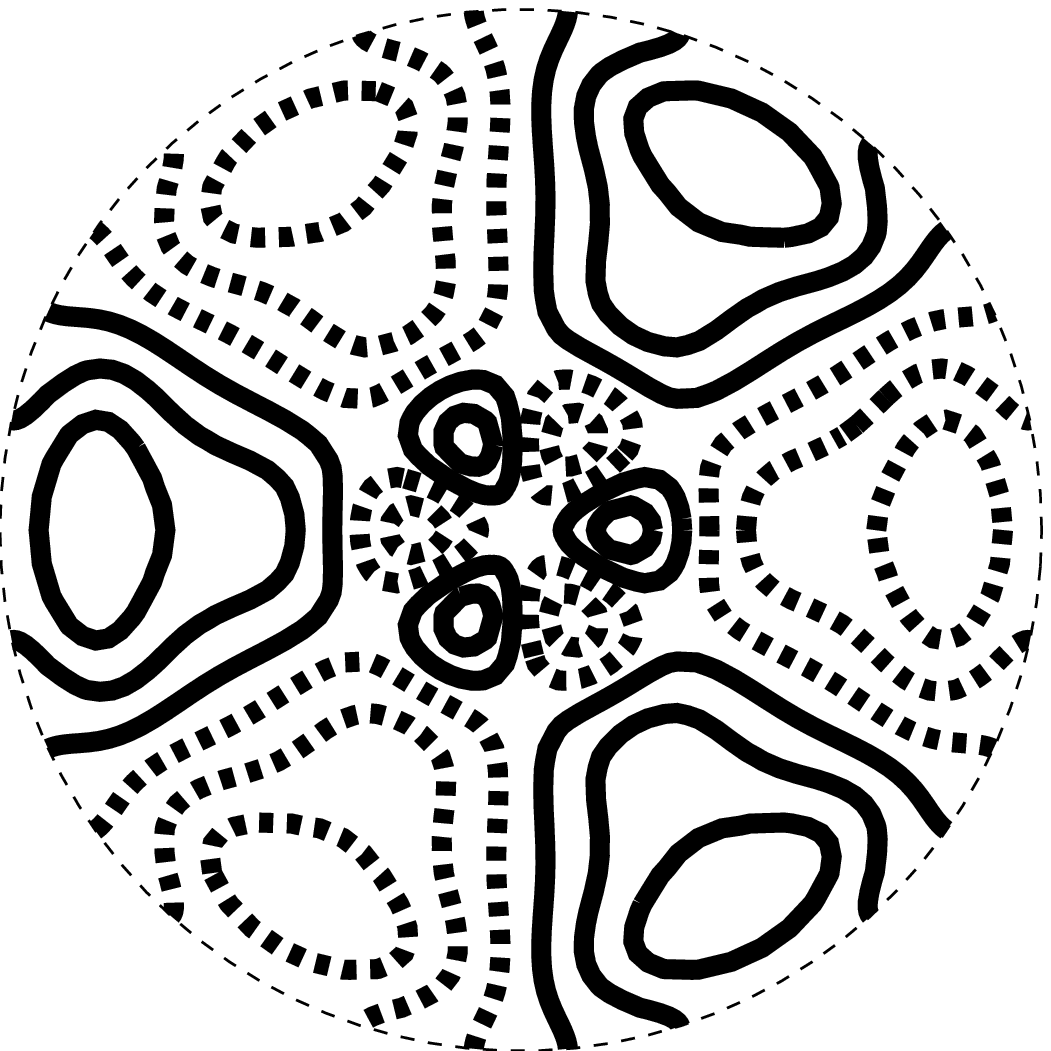}}	&$	3	$&$	2\times 10 ^ {-10}	$\\\hline
$	-0.81			$ & 	\parbox[c]{1.7cm}{\includegraphics[width=1.7cm]{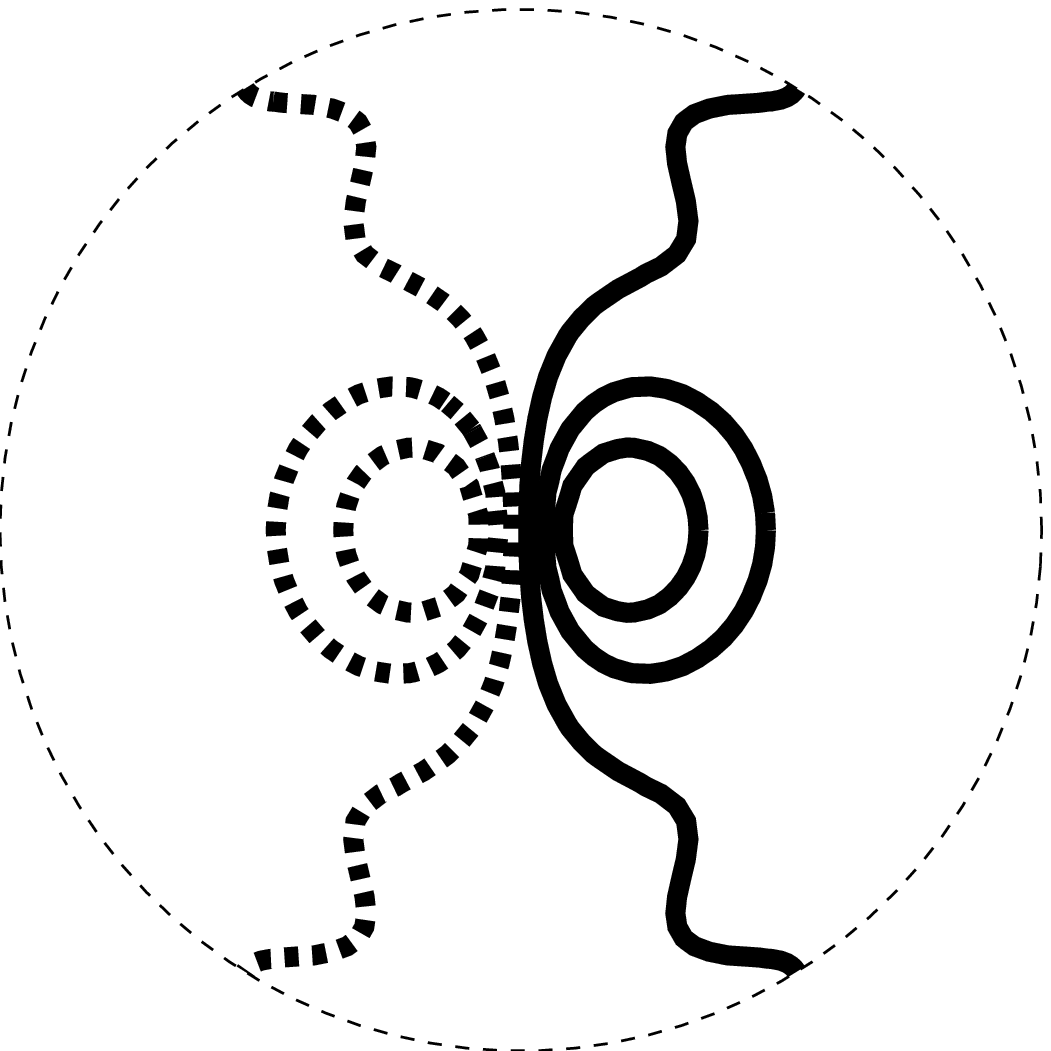}}		&	--	&$	1	$&$	6\times 10 ^ {-10}	$\\\hline
$	-4.40	\pm	45.9	i$ & 	\parbox[c]{1.7cm}{\includegraphics[width=1.7cm]{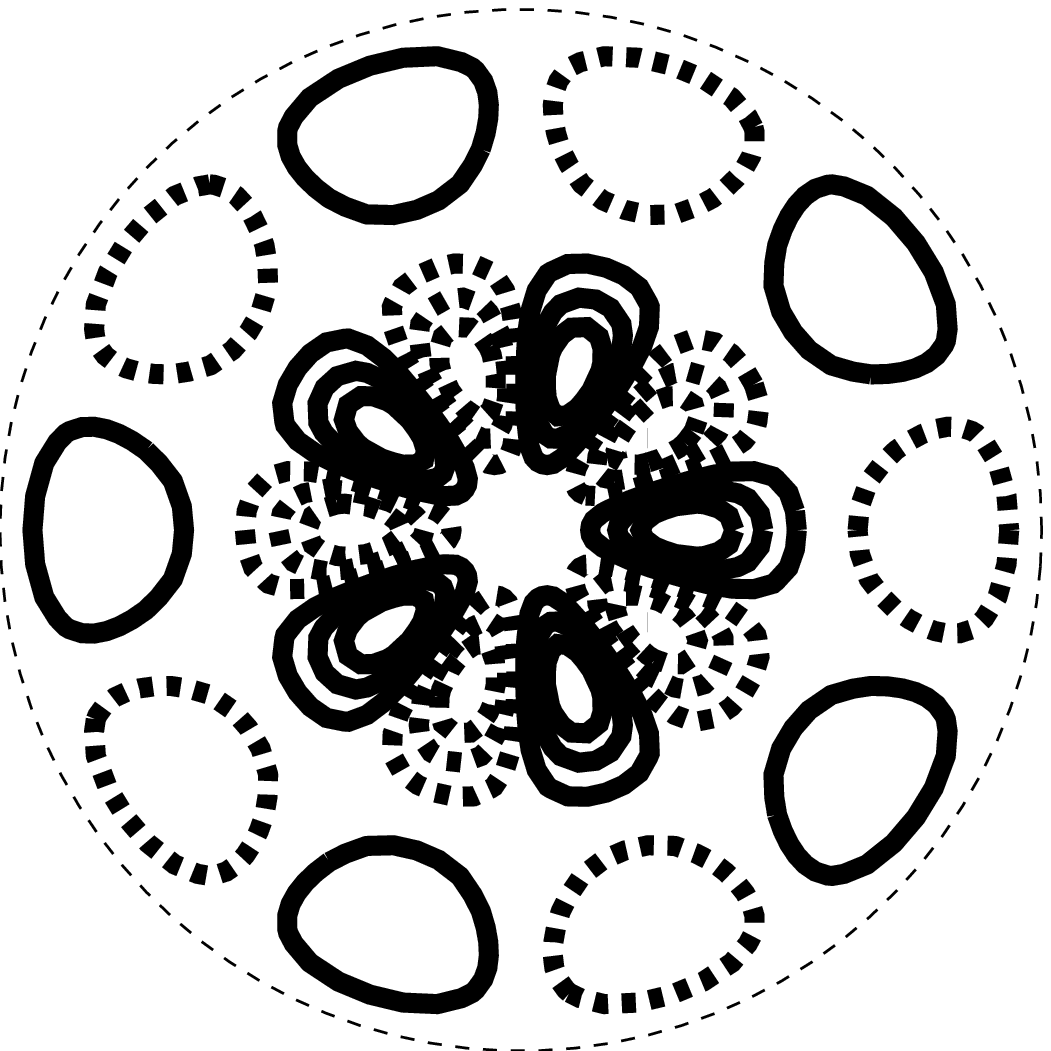}}		&	\parbox[c]{1.7cm}{\includegraphics[width=1.7cm]{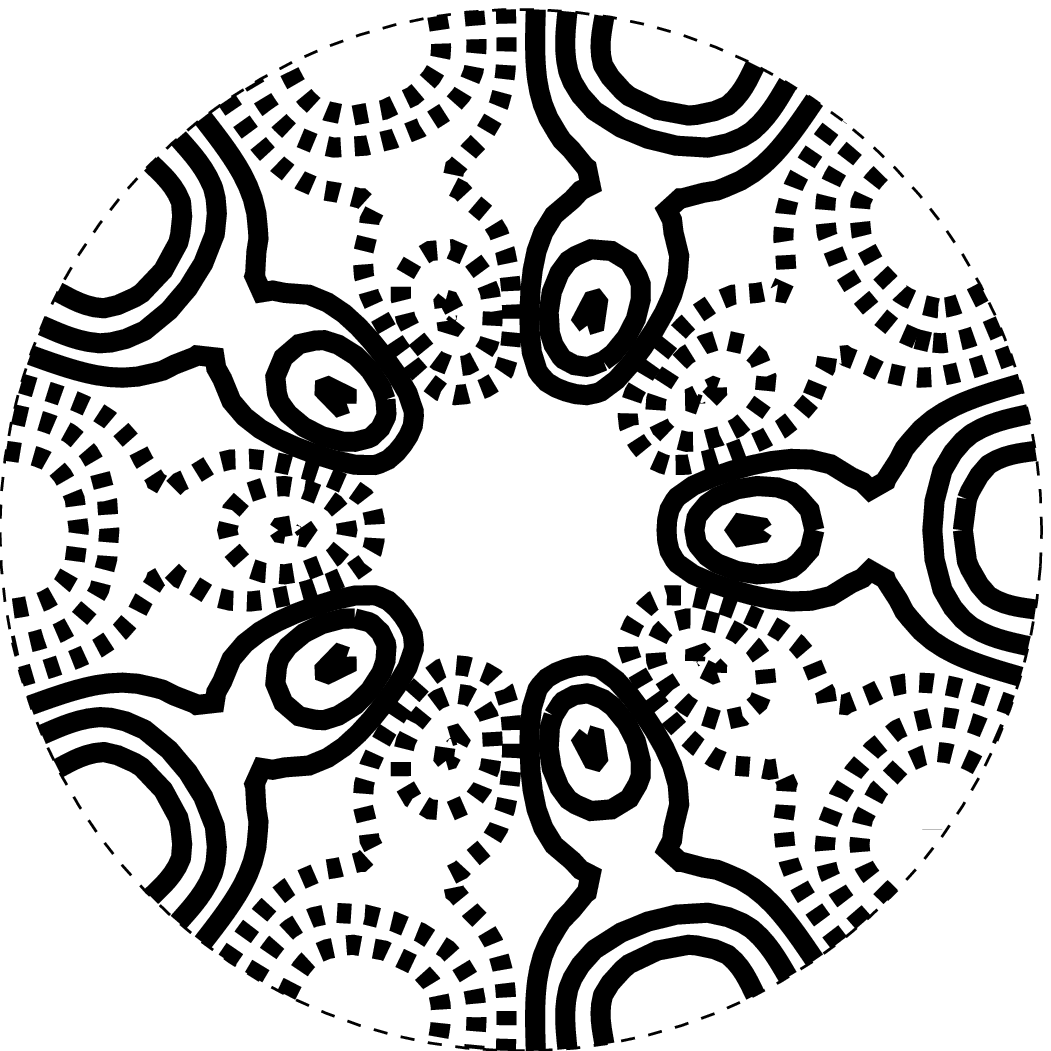}}	&$	5	$&$	9\times 10 ^ {-07}	$\\	\hline

\end{tabular}
\end{center}
	\caption{For $Ra=24\,000$, $\Gamma=1.57$: 
	eigenvalues, visualisation 
	of corresponding eigenvectors, azimuthal wavenumber and residual error. The visualised field is the temperature at the 
	midplane; for complex conjugate 
	eigenpairs the real and imaginary parts of the eigenvector are depicted.
	}
	\label{tab:lintable}
\end{table}
\begin{figure}
\begin{center}
\begin{tabular}{cc}
\includegraphics[width=2cm]{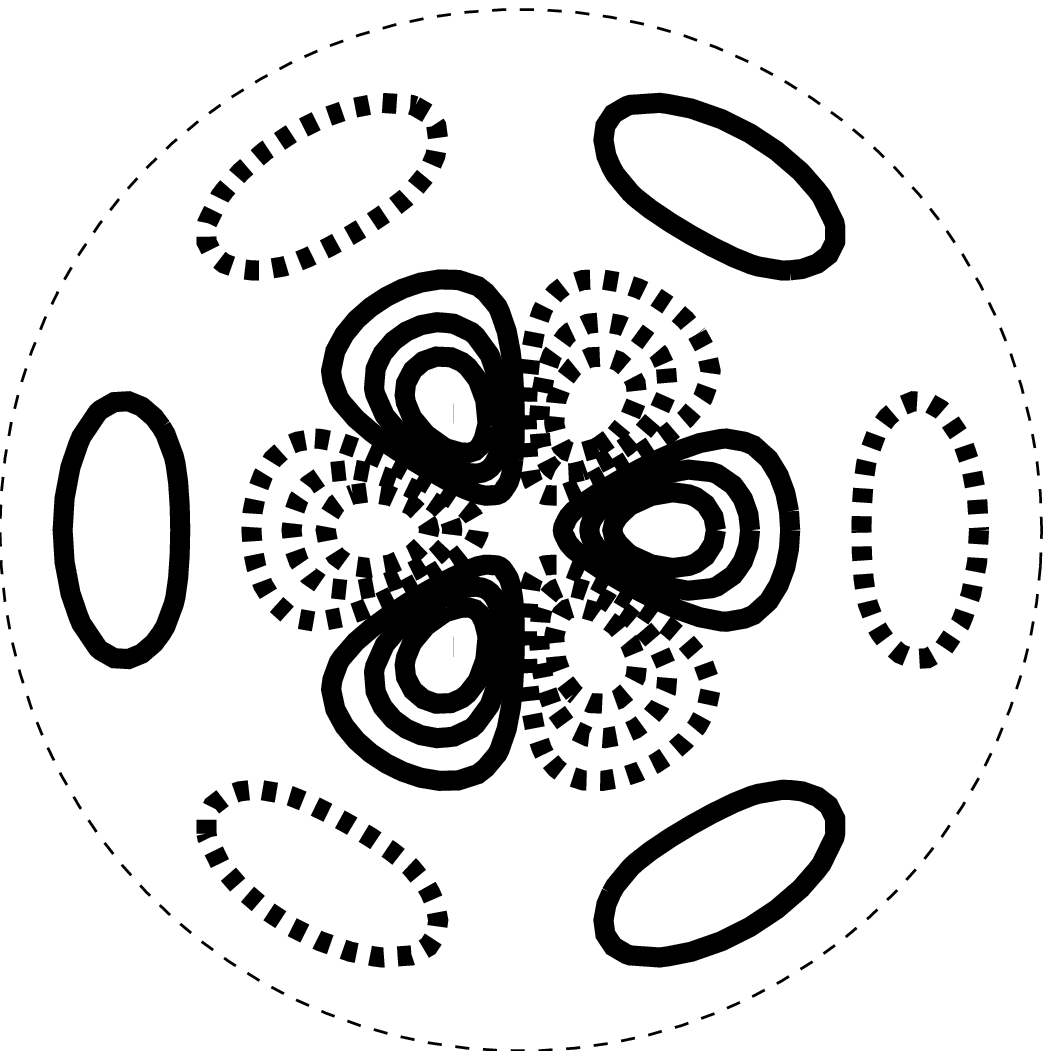}&
\includegraphics[width=2cm]{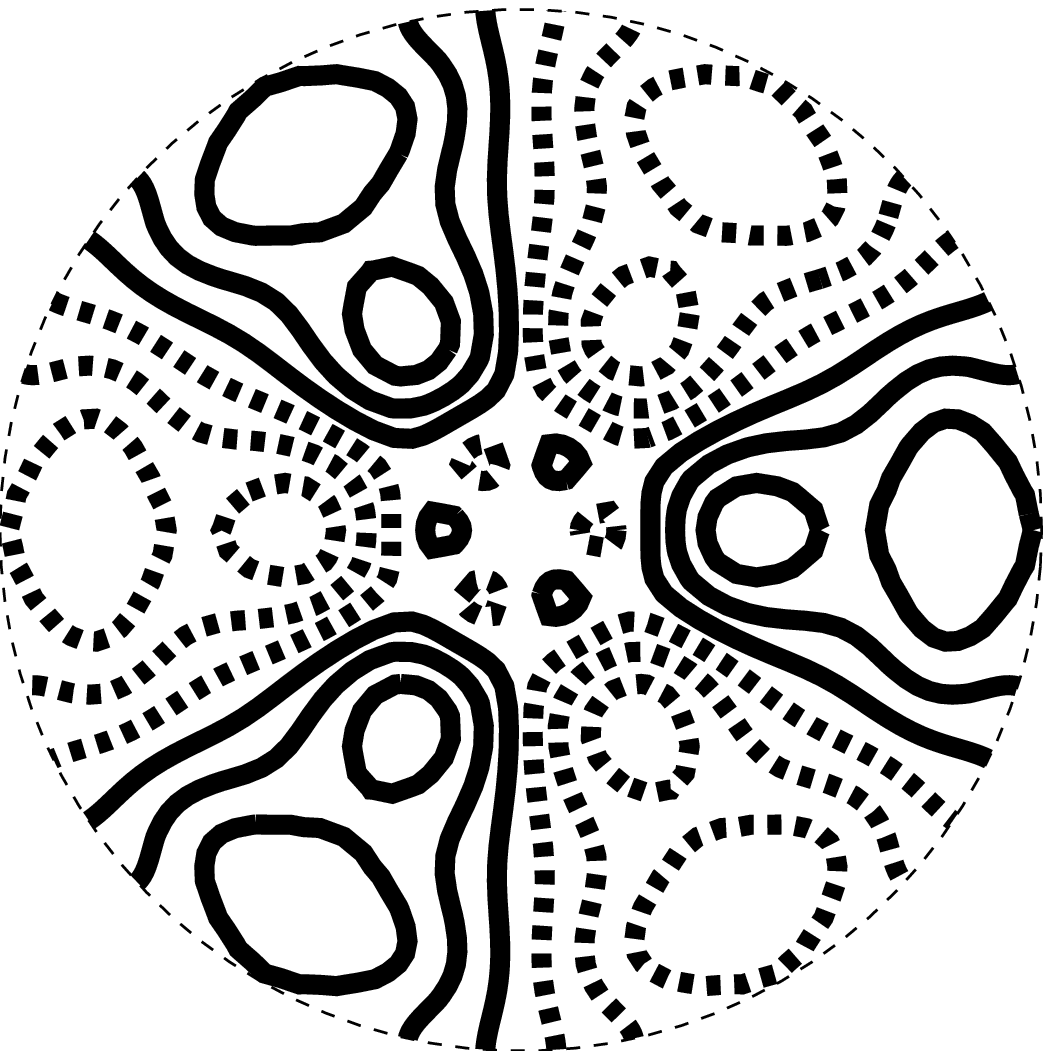}\\ 
(\textit{a}) & (\textit{b})
	\end{tabular}
\begin{tabular}{ccccc}
\includegraphics[width=2cm]{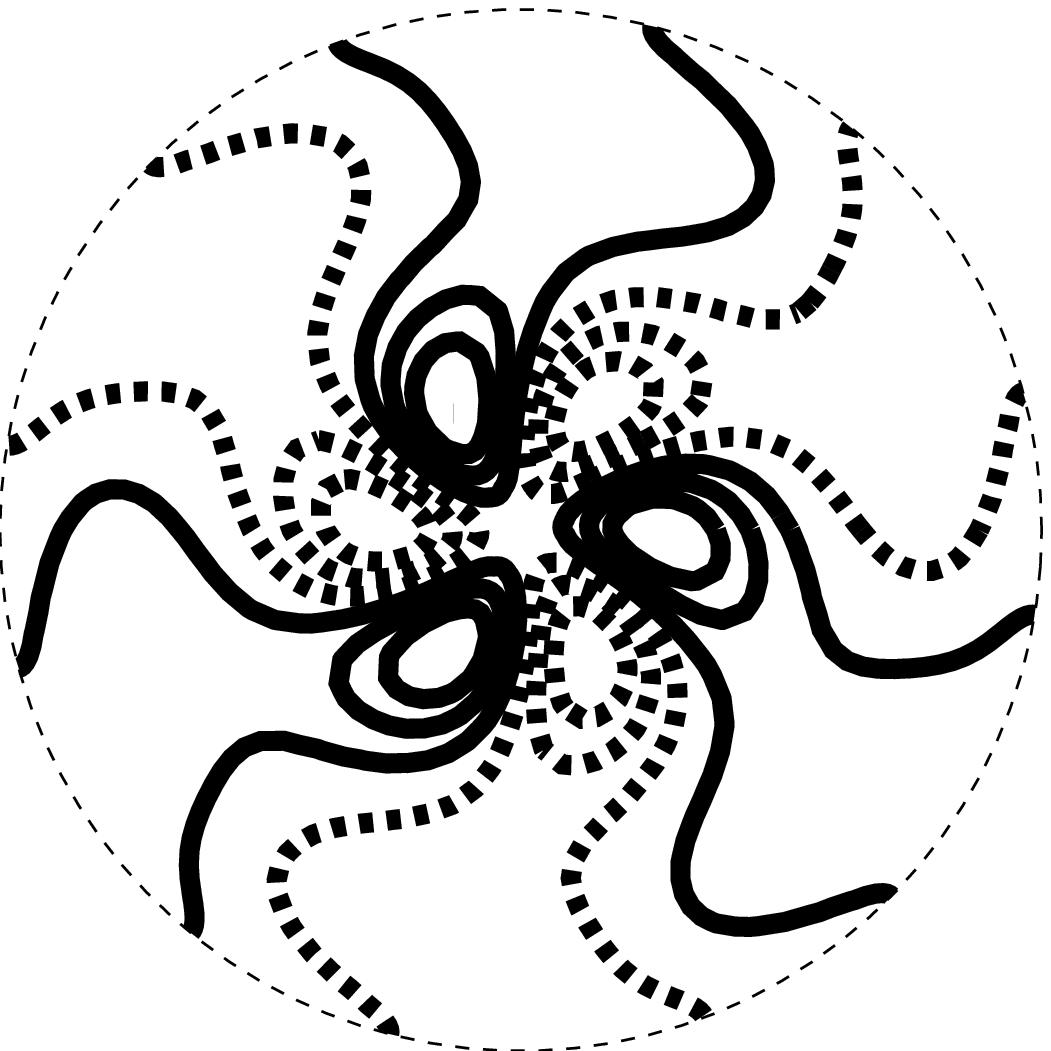}&
\includegraphics[width=2cm]{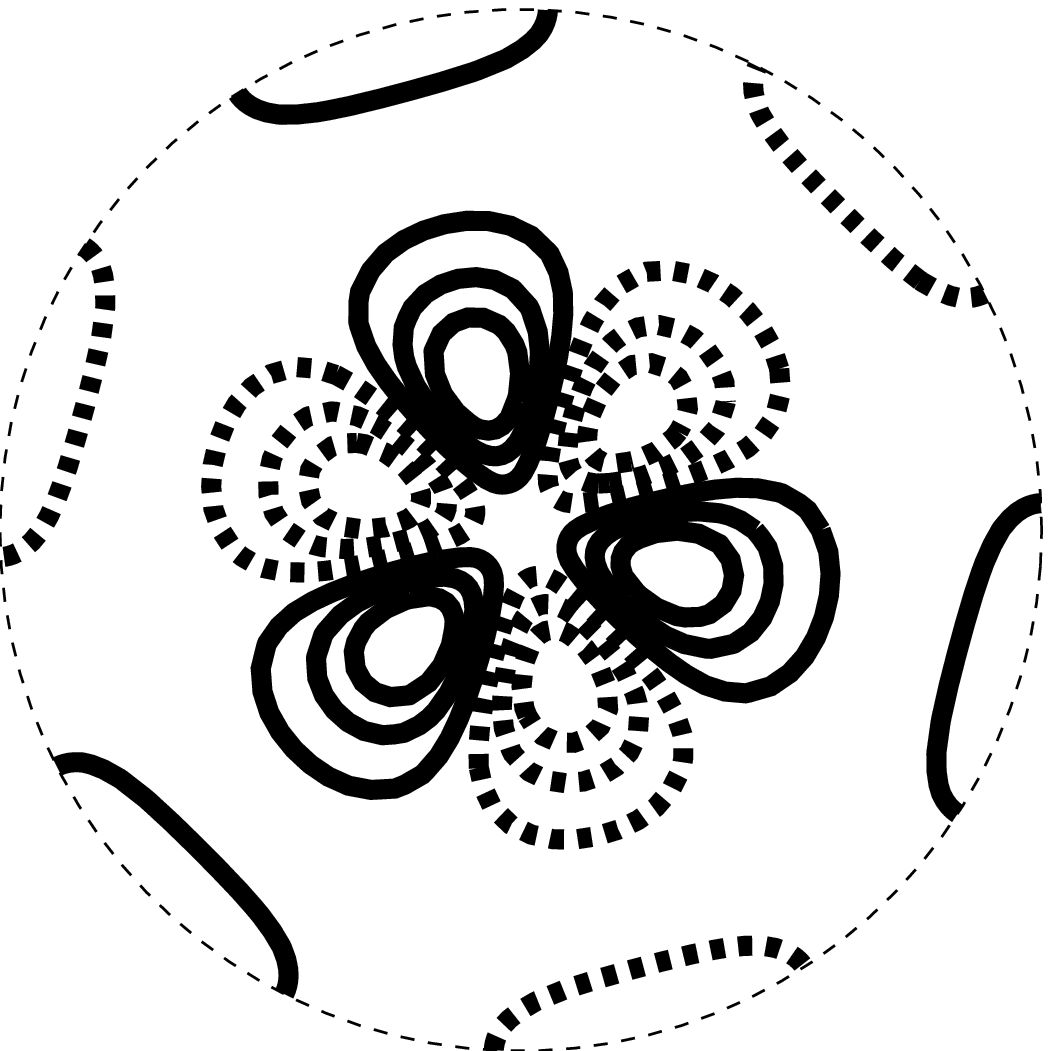}&
\includegraphics[width=2cm]{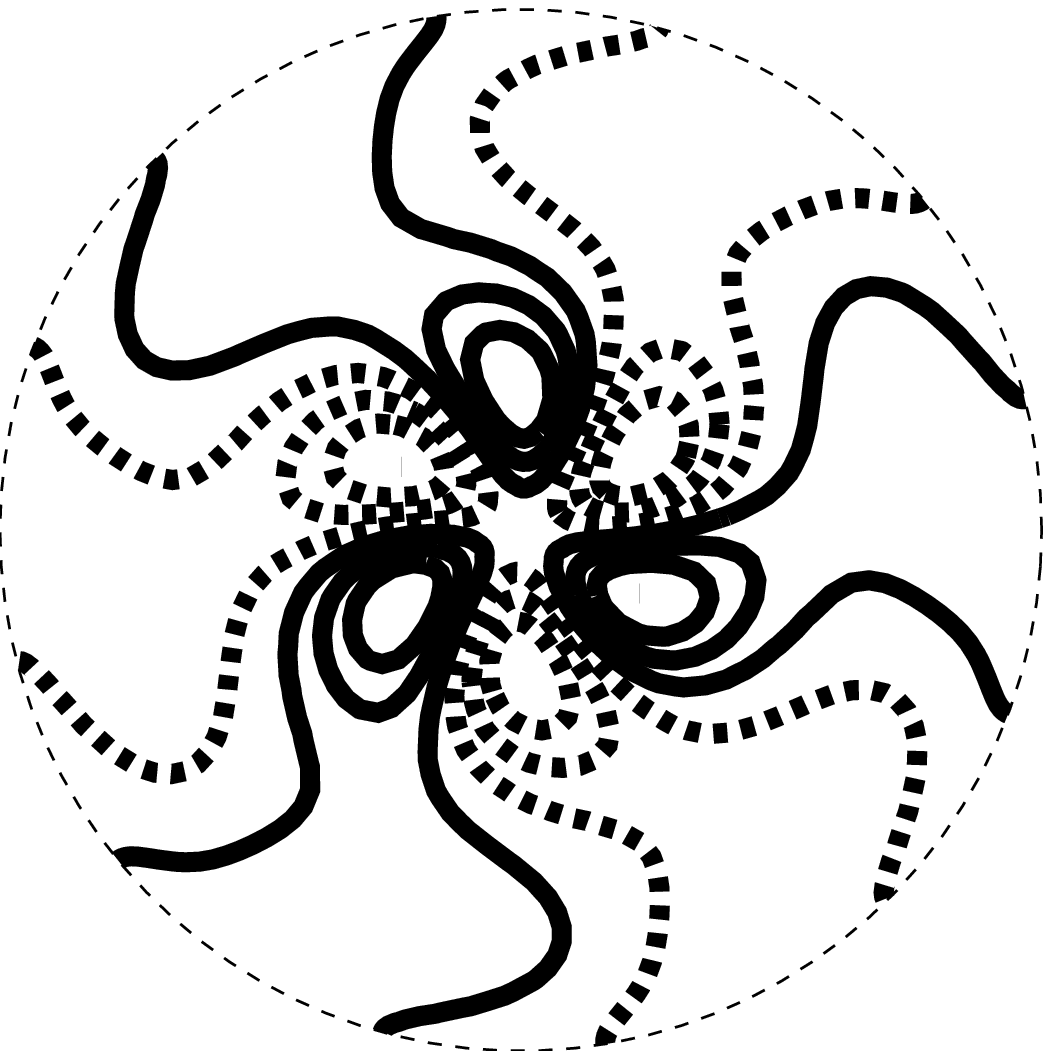}&
\includegraphics[width=2cm]{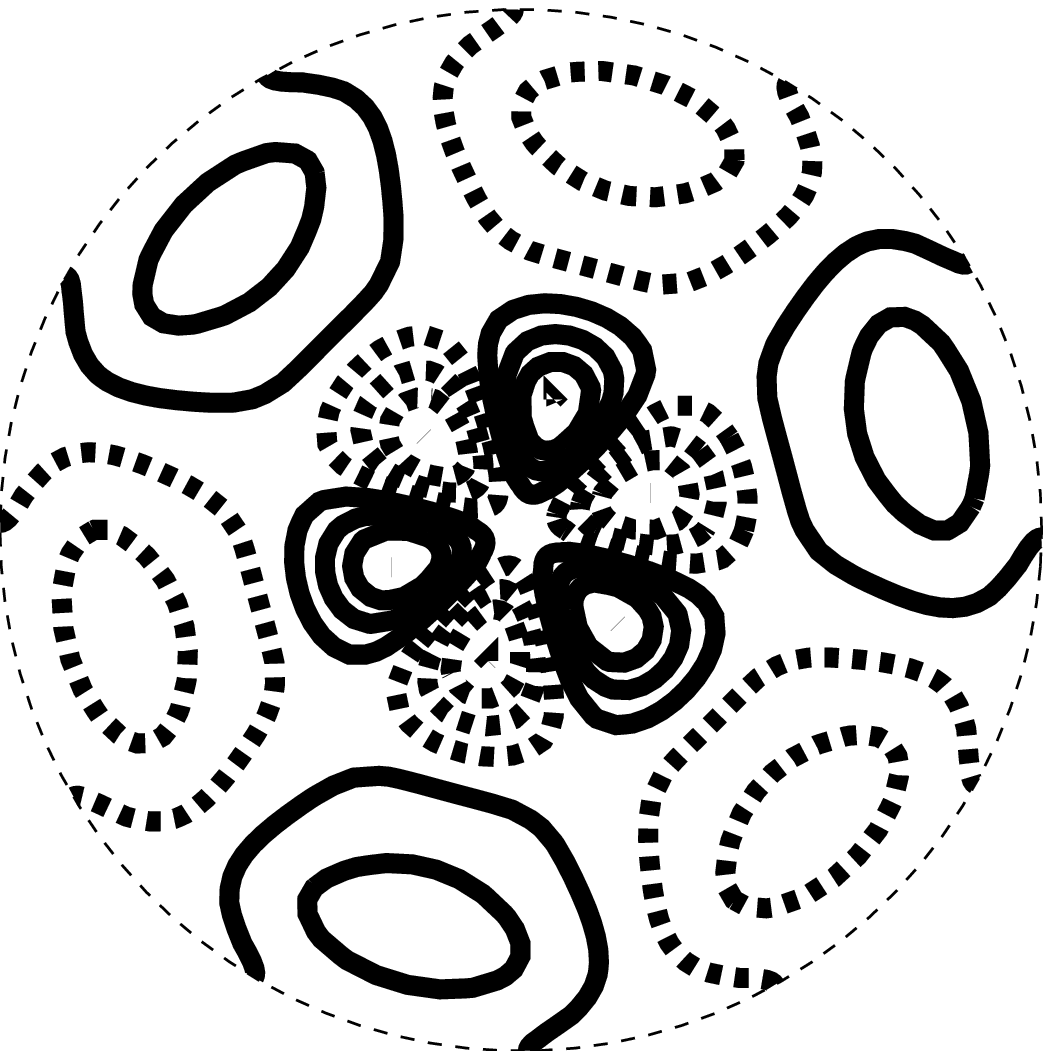}&
\includegraphics[width=2cm]{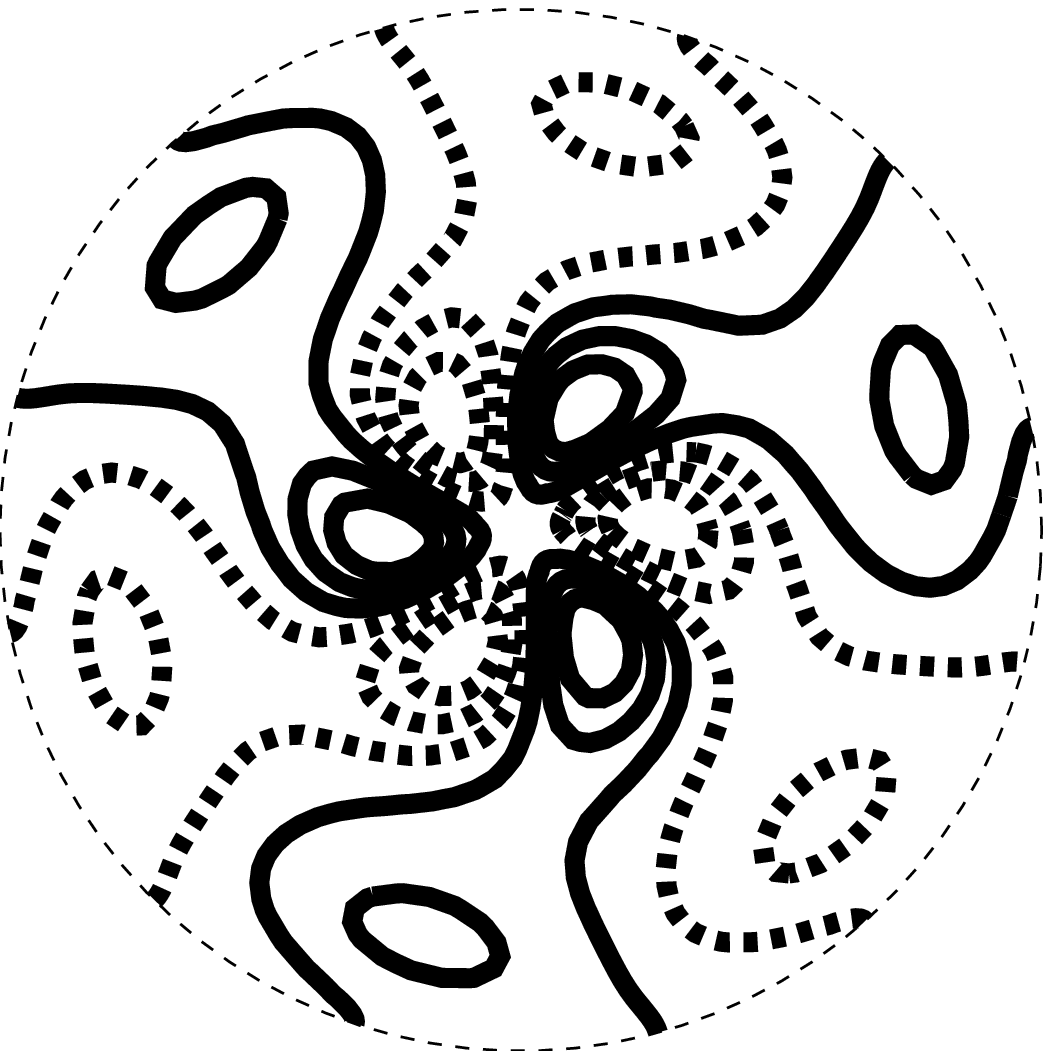} \\
 (\textit{c})  &  (\textit{d})  &   (\textit{e}) & (\textit{f})  &(\textit{g}) 
\end{tabular}
\end{center}
	\caption{Eigenvectors for $\Gamma=1.47$, $Ra=25\,000$ (temperature field contours at $z=0$): 
	(\textit{a}) real part of the critical eigenvector; 
	(\textit{b}) imaginary part of the critical eigenvector; 
	(\textit{c}--\textit{g}) superposition of the two fields via 
$ \hat{h}^R(r,z) \cos(m(\theta-\theta_0)) + \hat{h}^I(r,z) \sin(m(\theta+\theta_0))$,
	with $m\theta_0$ of
	(\textit{c}) $0$, (\textit{d}) $\pi/4$, (\textit{e}) $\pi/2$, (\textit{f}) $3\pi/4$, (\textit{g}) $0.92 \pi$.}
	\label{fig:eigencombs}
\end{figure}

We summarise here the differences between our numerical method 
and that of \cite{WanKuhRat}. 
We linearised a timestepping code in order to, in effect, carry out the
power method (supplemented by an Arnoldi decomposition)
on the exponential $\exp(L \Delta t)$ of the Jacobian.
Wanschura \etal\ constructed the Jacobian matrix $L$
and used inverse iteration to compute its eigenvalues.
Our calculation was restricted to one of the two identical decoupled 
subproblems, corresponding to only one of the invariant subspaces of
the form (\ref{eq:costype}) or (\ref{eq:sintype}).
As a result, the complex eigenfunctions we show in table \ref{tab:lintable}
are all in the eigenspace corresponding to standing waves,
with three axes of reflection symmetry.
Basis vectors for the remainder of the four-dimensional eigenspace can be
found by rotating the eigenvectors of table \ref{tab:lintable},
i.e.~multiplying by $\sin(m\theta)$ instead of $\cos(m\theta)$. 
Wanschura \etal, in contrast,
used the travelling wave form as an initial condition or invariant subspace,
as discussed below.

In figure~\ref{fig:eigencombs}, we show representative elements of
the eigenspace associated with the $m=3$ complex eigenvector
at $Ra=25\,000$.
Figures~\ref{fig:eigencombs}\,(\textit{a,\,b}) show
$\hat{h}^R(r,z) \cos(m\theta)$ and $\hat{h}^I(r,z)\cos(m\theta)$,
while figures~\ref{fig:eigencombs}\,(\textit{c}--\textit{g}) are
generated via 
\begin{equation}
C\left(\hat{h}^R(r,z) \cos(m(\theta-\theta_0))
+\hat{h}^I(r,z) \sin(m(\theta+\theta_0))\right),
\end{equation}
a form equivalent to (\ref{eq:complexevol})
after translation of $\theta$ and of $t$.
Clockwise travelling waves ensue for $m\theta_0=\pi/2$ (\textit{c}),
counterclockwise travelling waves for $m\theta_0=0$ (\textit{e}),
and standing waves at different temporal phases for 
$m\theta_0 = \pm\pi/4$ (\textit{d,f}).
Thus, the angle $m\theta_0$ is similar to that used in figure~\ref{fig:phase}.
An eigenvector which corresponds to neither travelling nor standing waves
is shown in figure~\ref{fig:eigencombs}\,(\textit{g}).
These are all depicted on the slice $z=0$; when we plot the field of
figure \ref{fig:eigencombs}(\textit{c})
at $z=0.3$, we recover the form shown by Wanschura \etal\
We emphasise, however, that the other fields depicted in figure 
\ref{fig:eigencombs} are all equally valid eigenvectors. In particular,
a nonlinear analysis, such as the simulations presented below,
is required to determine whether the resulting nonlinear
flow near onset is a travelling or a standing wave.

\subsection{Weakly unstable standing waves}
\begin{figure}
\begin{center}
\begin{tabular}{cccccccc}
\includegraphics[width=2cm]{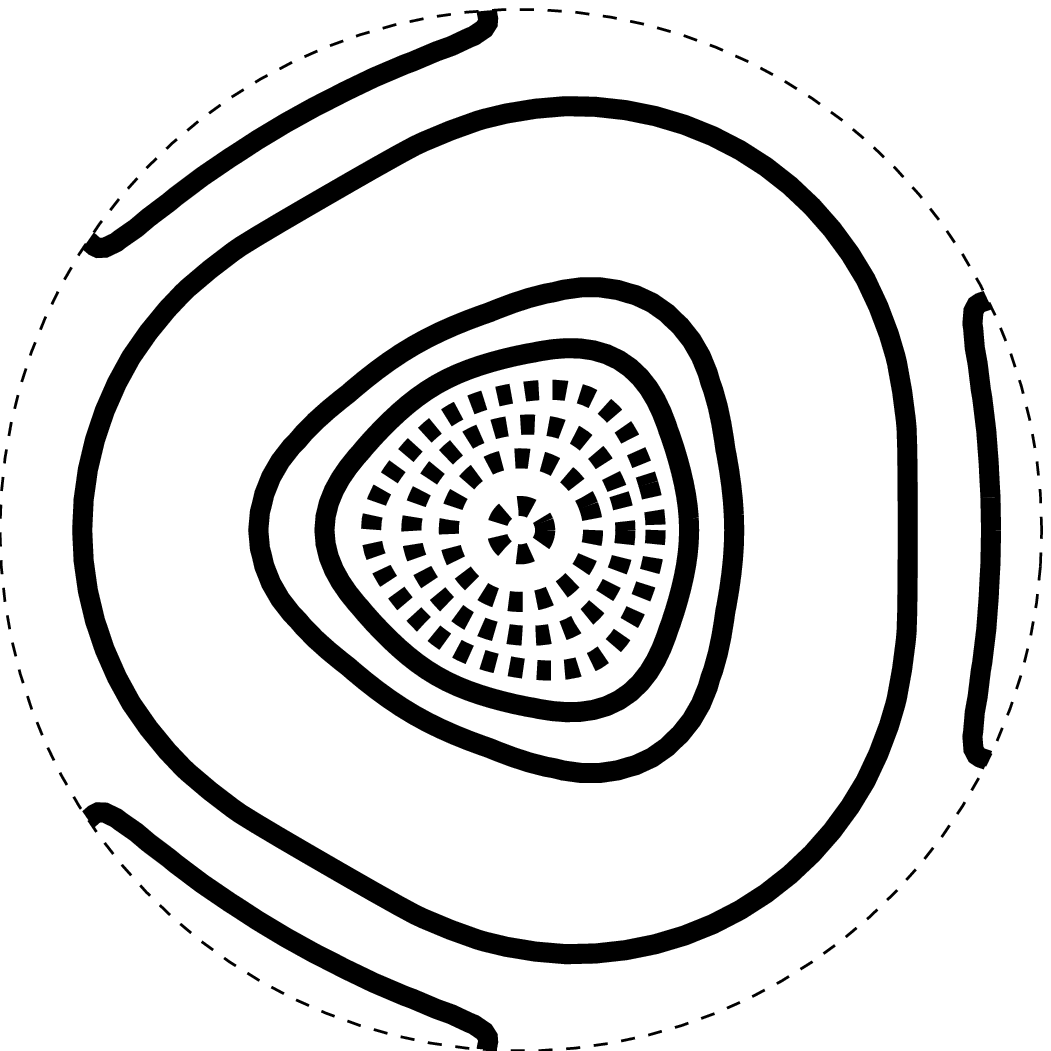}&
\includegraphics[width=2cm]{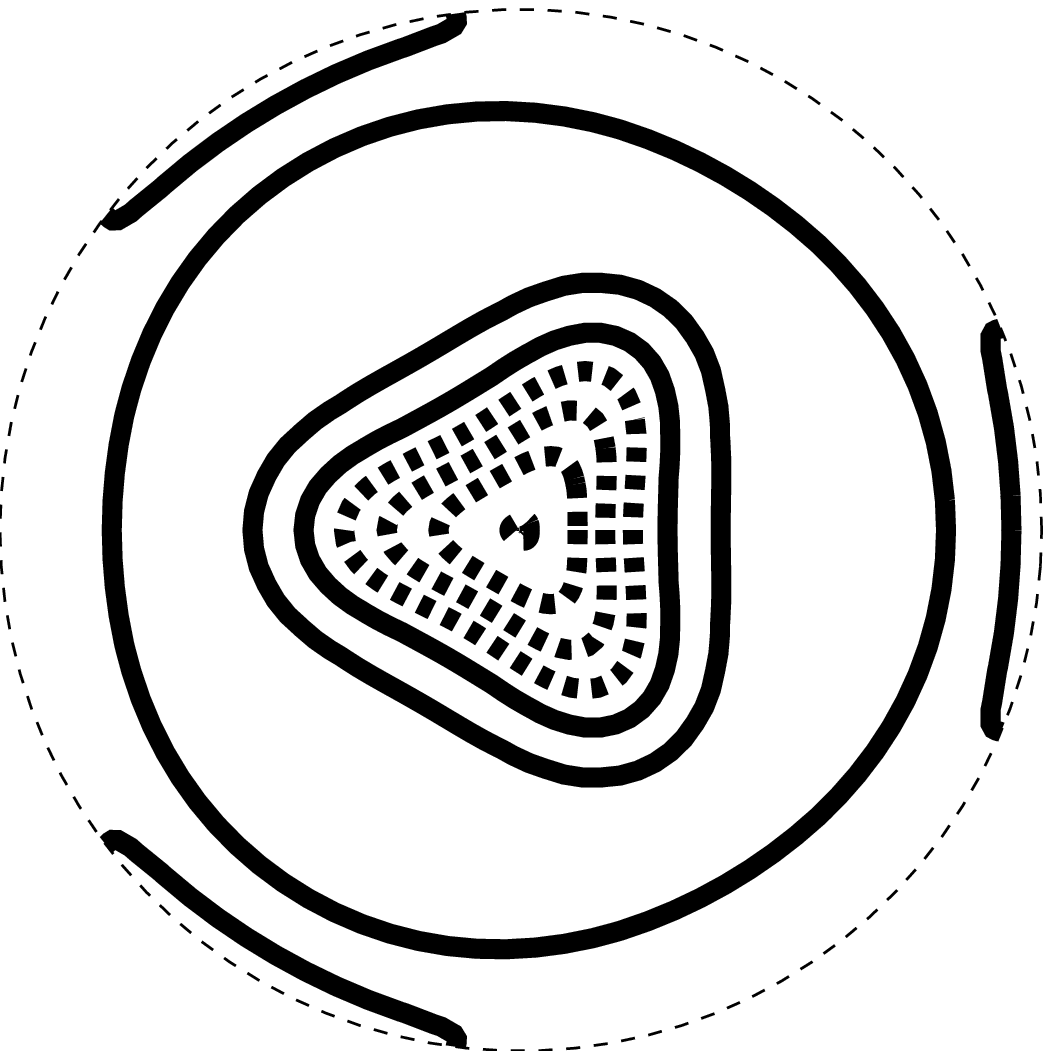}&
\includegraphics[width=2cm]{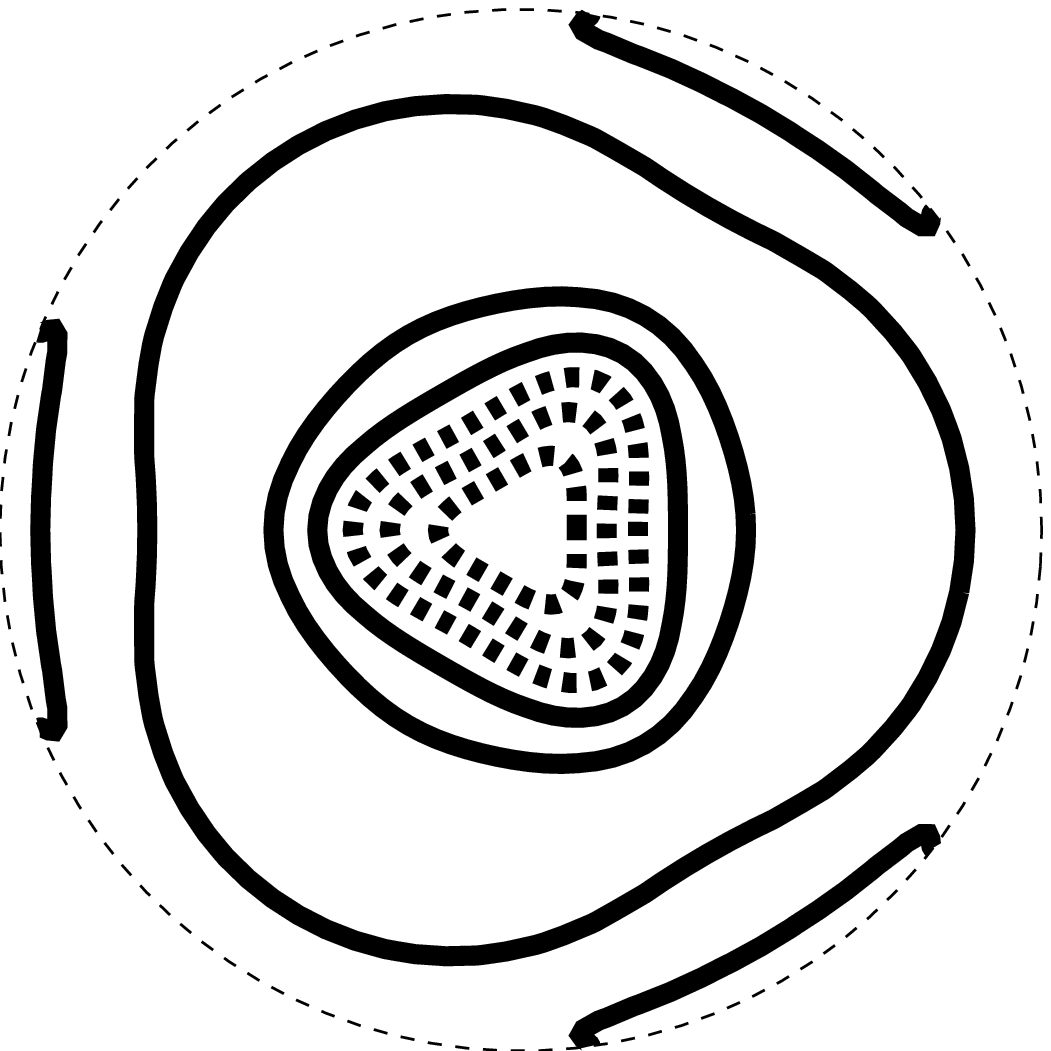}&
\includegraphics[width=2cm]{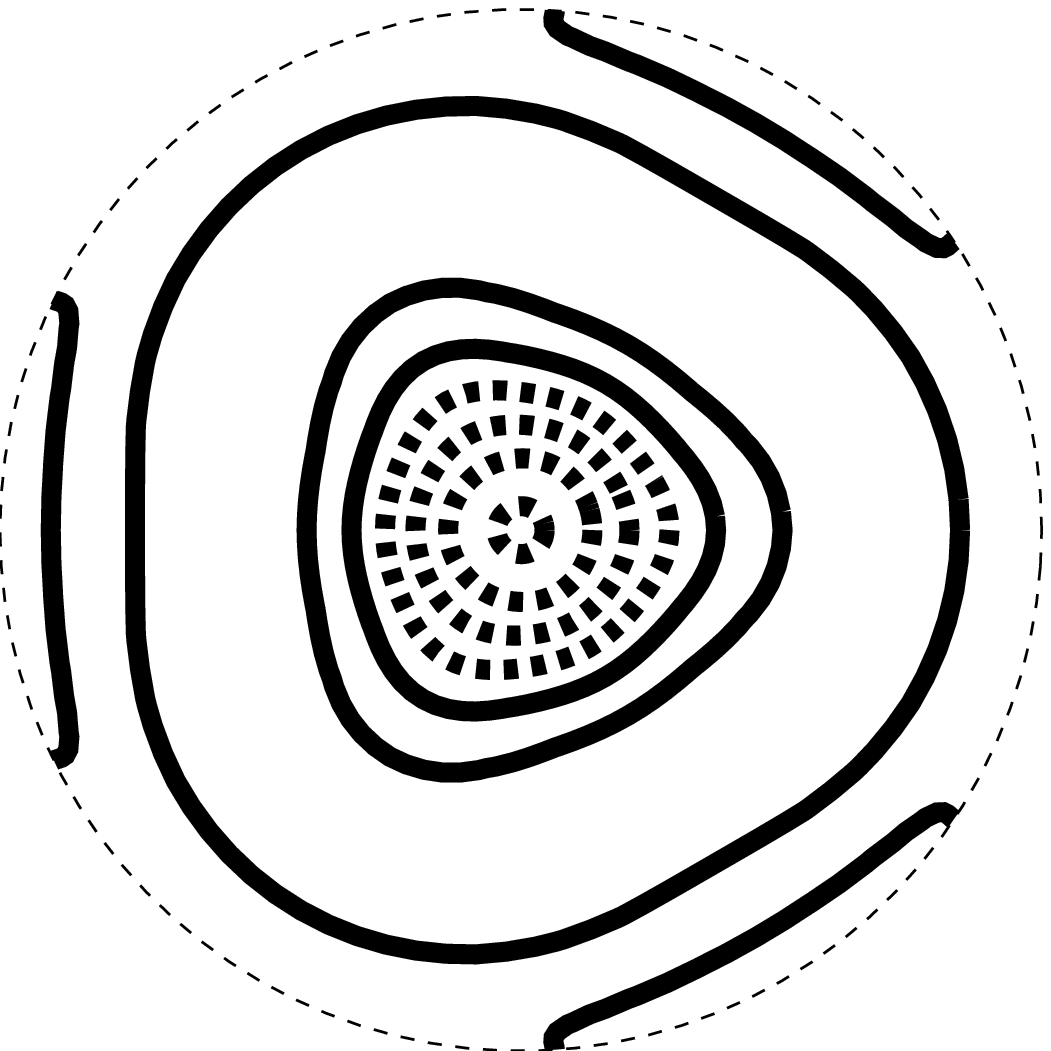}&
\includegraphics[width=2cm]{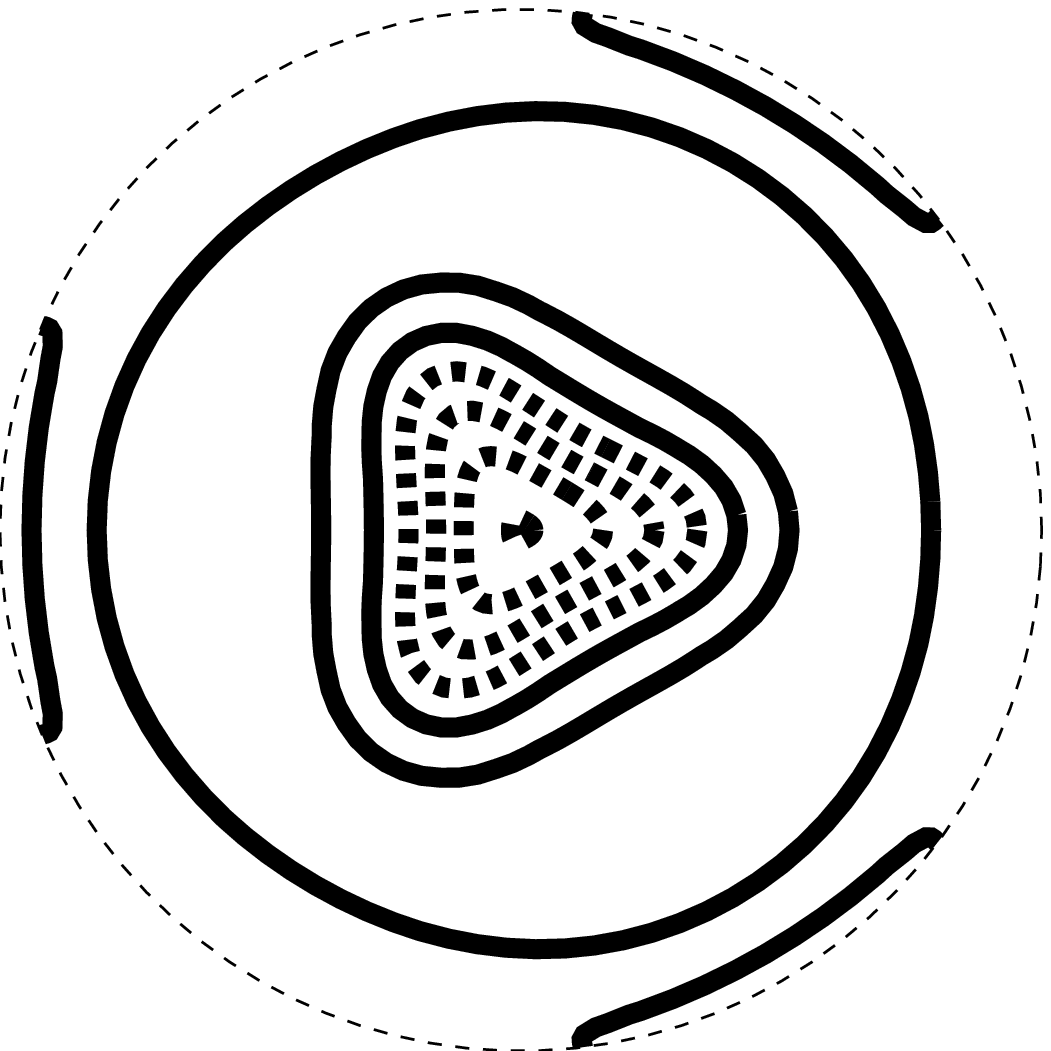}&
\includegraphics[width=2cm]{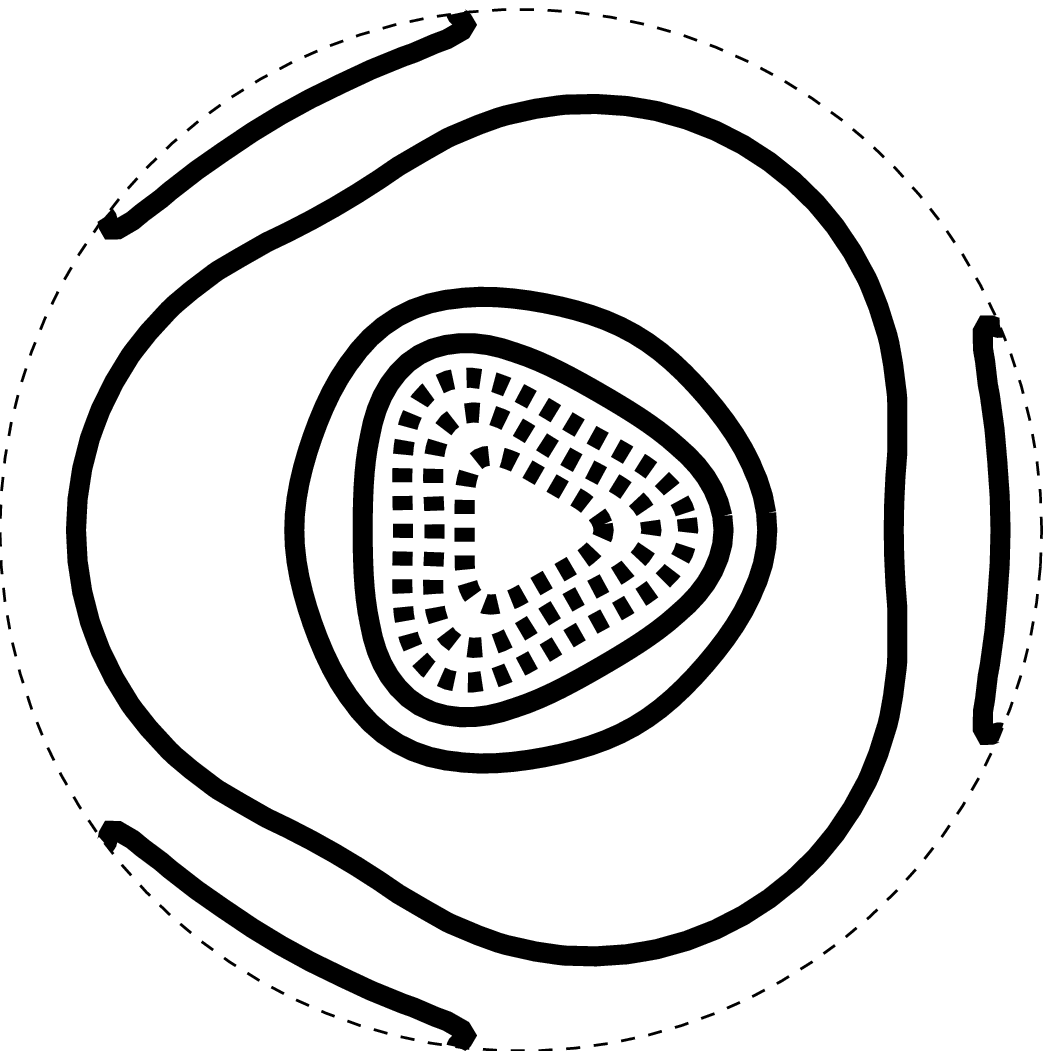}\\
(\textit{a}) & (\textit{b}) & (\textit{c}) & (\textit{d}) & (\textit{e}) & (\textit{f}) 
	\end{tabular}
\end{center}
	\caption{Standing waves at $Ra=26\,000$: temperature contours
        on the midplane at $t=0$, $T/6$, $2T/6$,~$\ldots$}
	\label{fig:SW}
\bigskip
\begin{center}
\begin{tabular}{cccccccc}
\includegraphics[width=2cm]{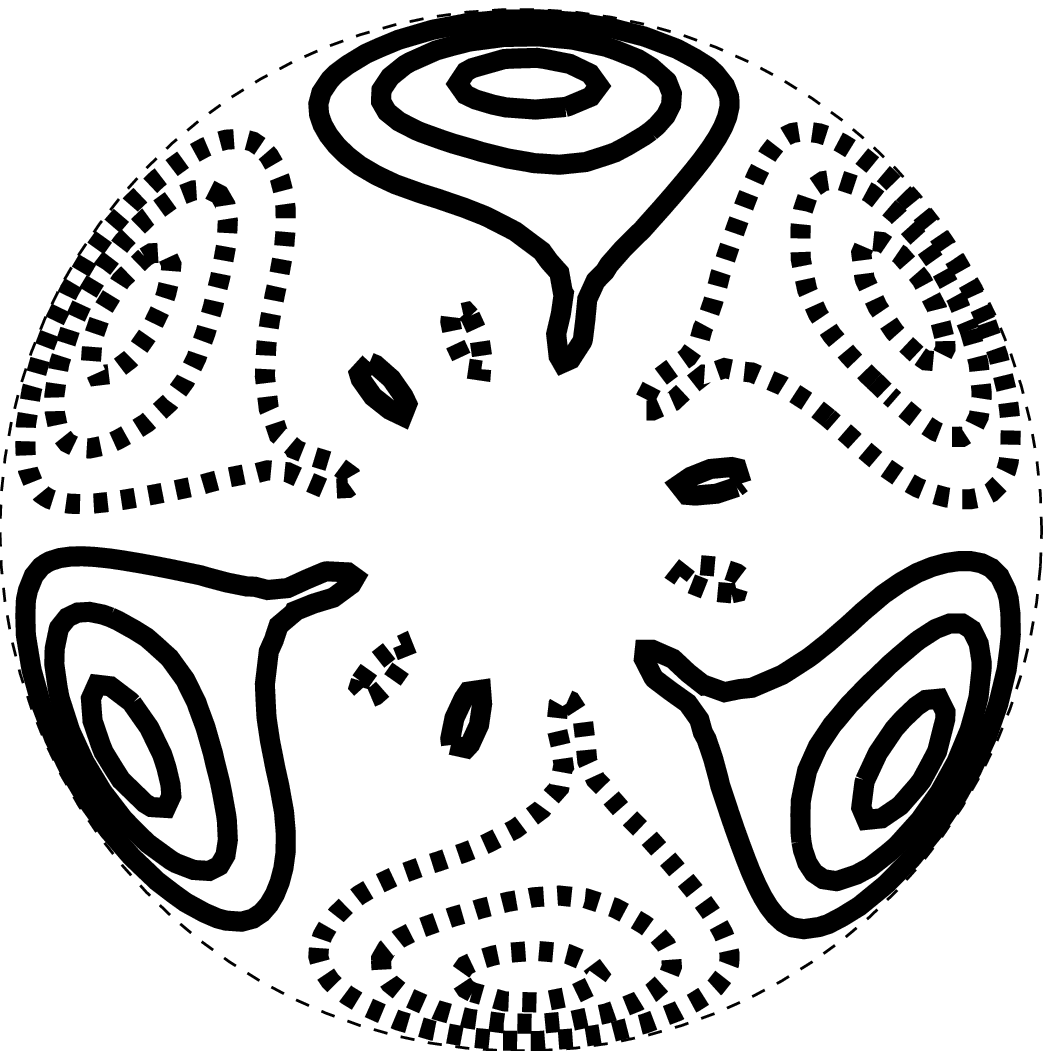}&
\includegraphics[width=2cm]{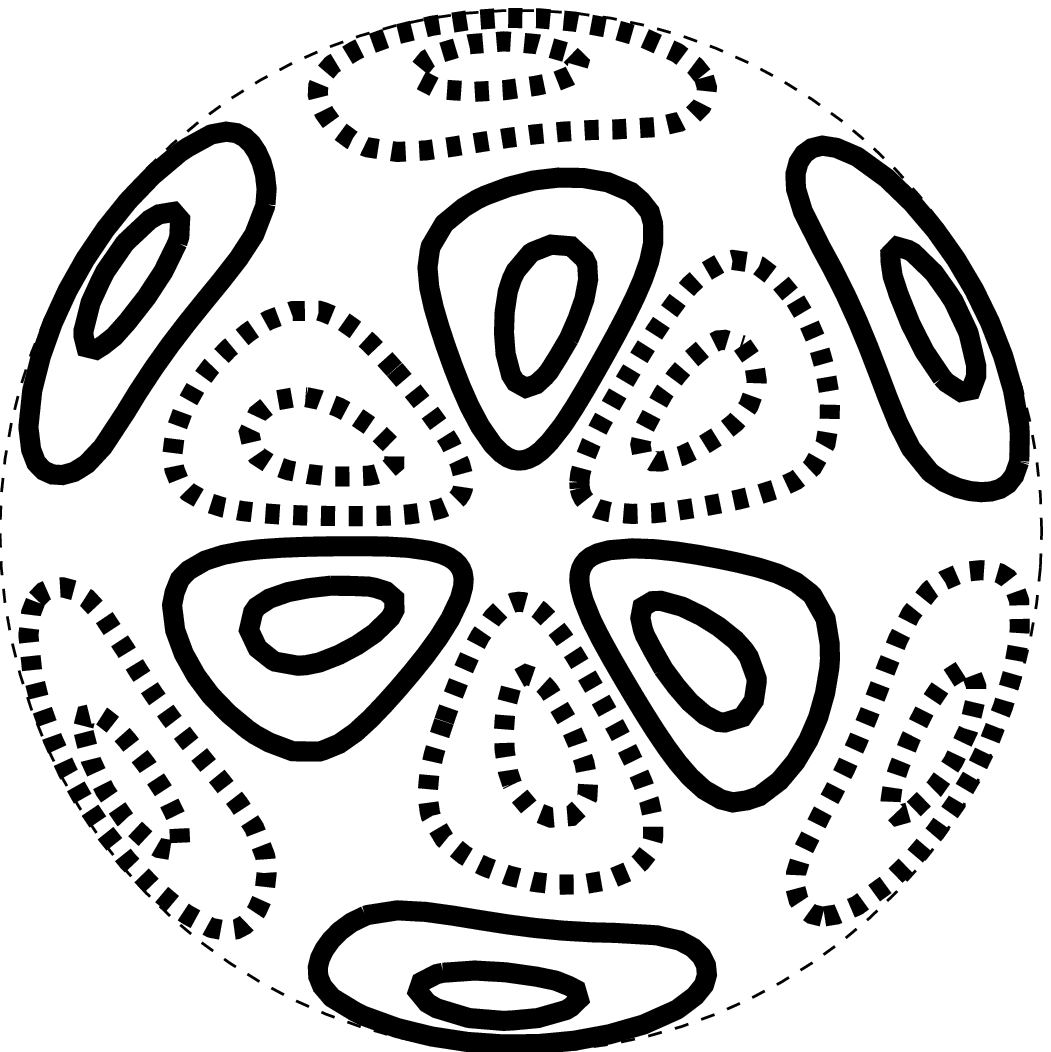}&
\includegraphics[width=2cm]{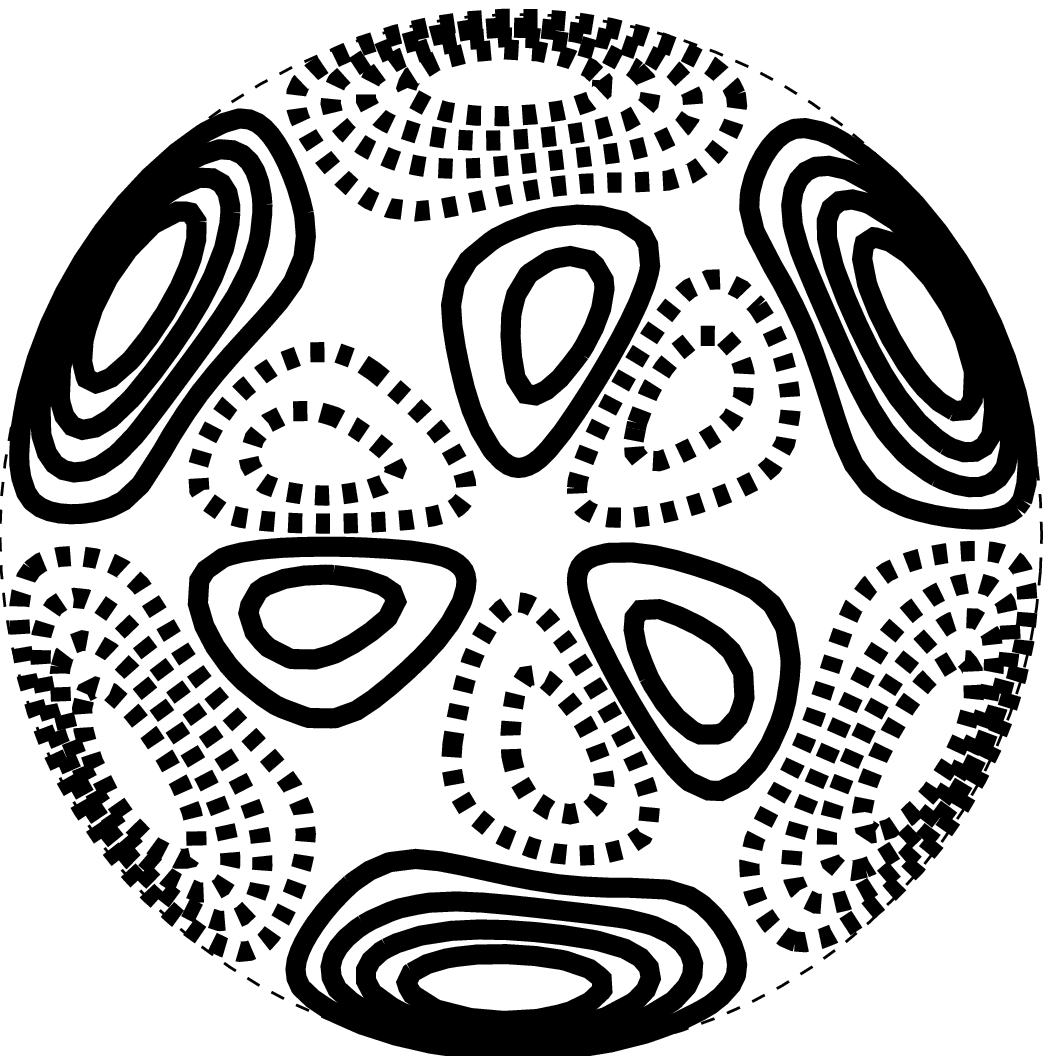}&
\includegraphics[width=2cm]{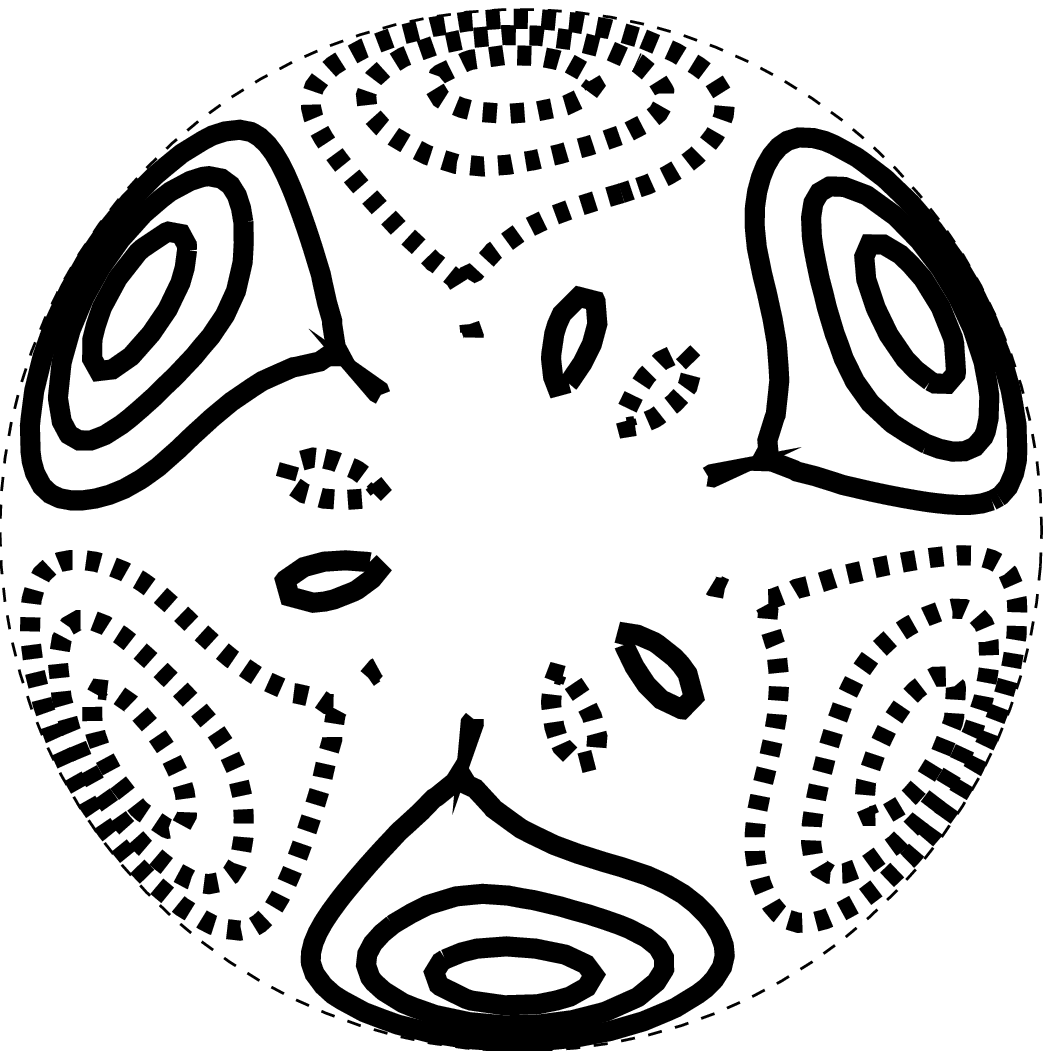}&
\includegraphics[width=2cm]{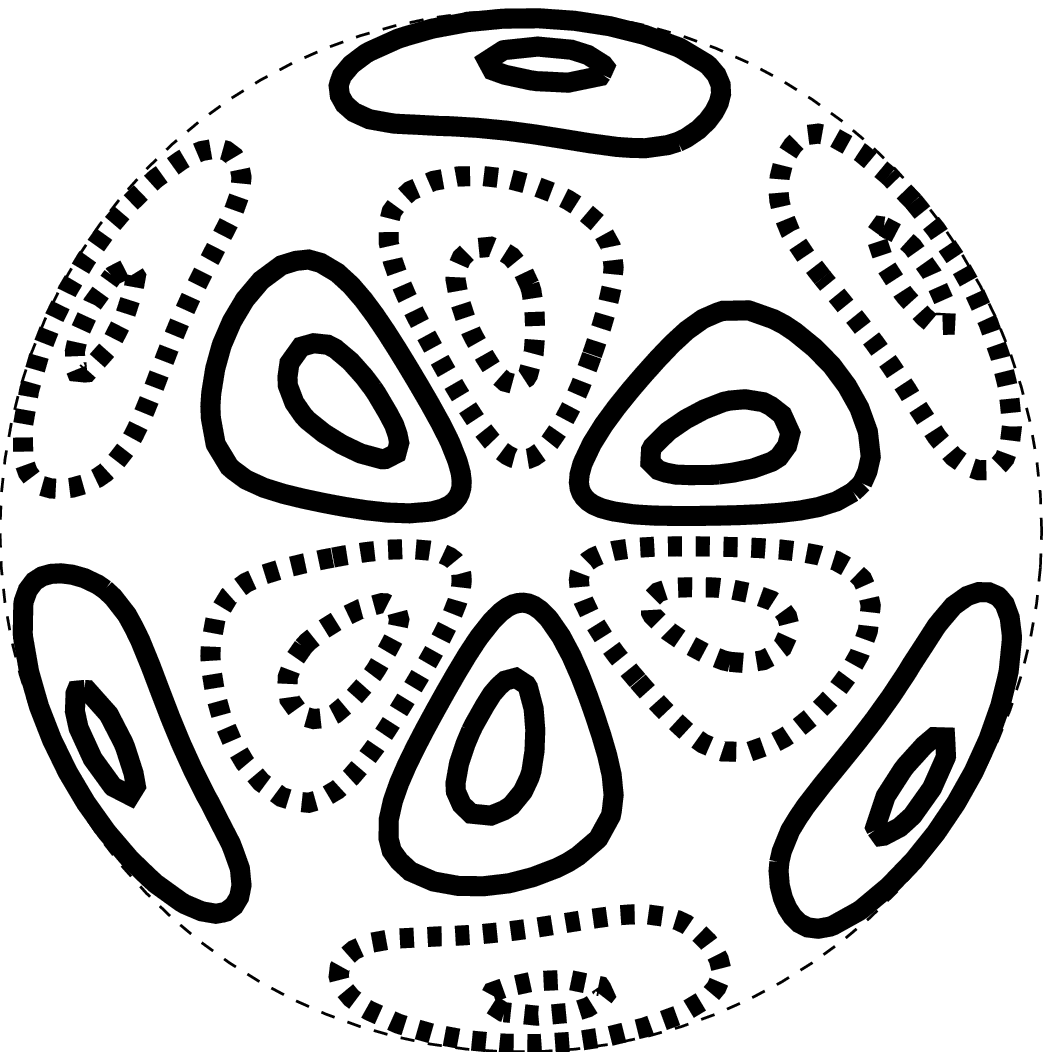}&
\includegraphics[width=2cm]{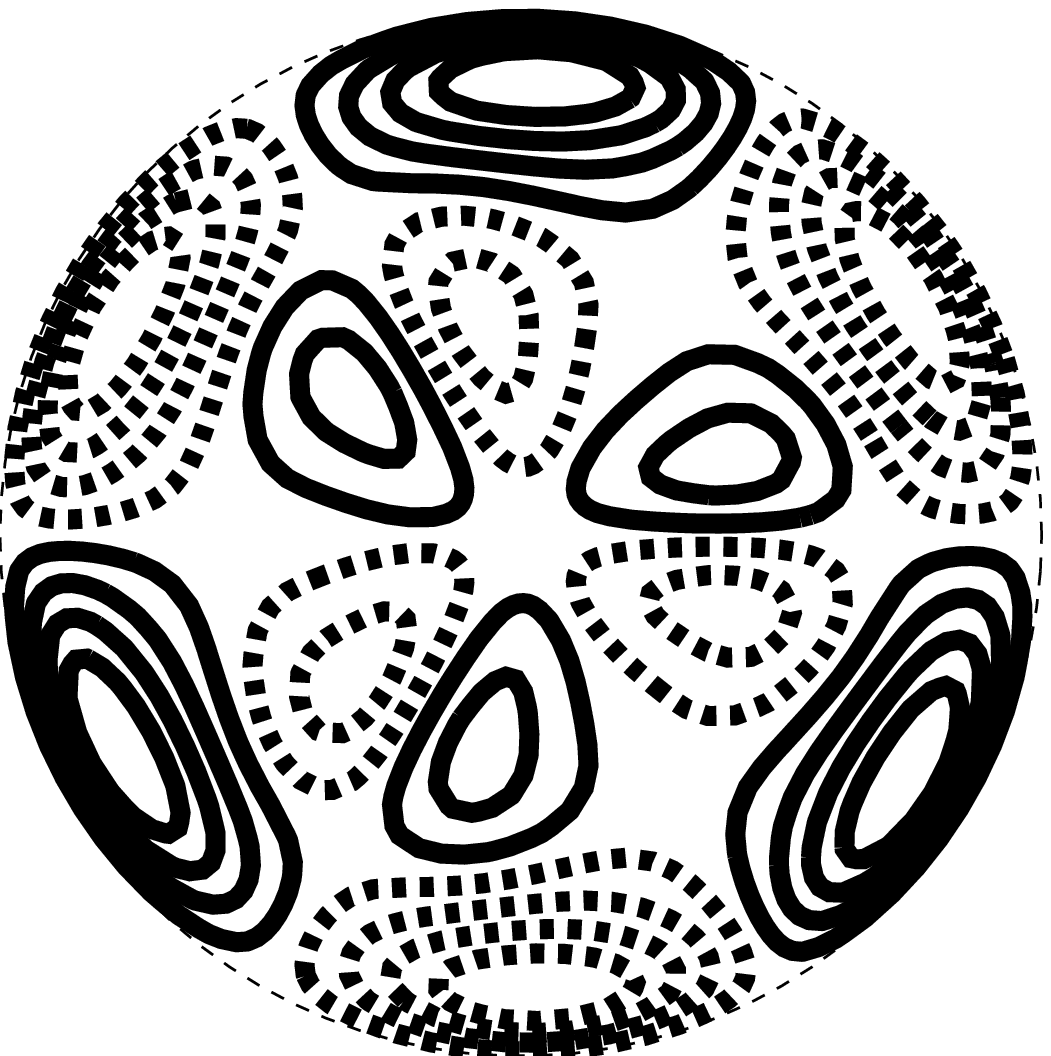}\\
(\textit{a}) & (\textit{b}) & (\textit{c}) & (\textit{d}) & (\textit{e}) & (\textit{f}) 
	\end{tabular}
\end{center}
	\caption{Standing waves at $Ra=26\,000$: contours of azimuthal 
velocity on the midplane
        at $t=0$, $T/6$, $2T/6$,~$\ldots$}
	\label{fig:SW_uth}
\bigskip
\begin{center}
\includegraphics{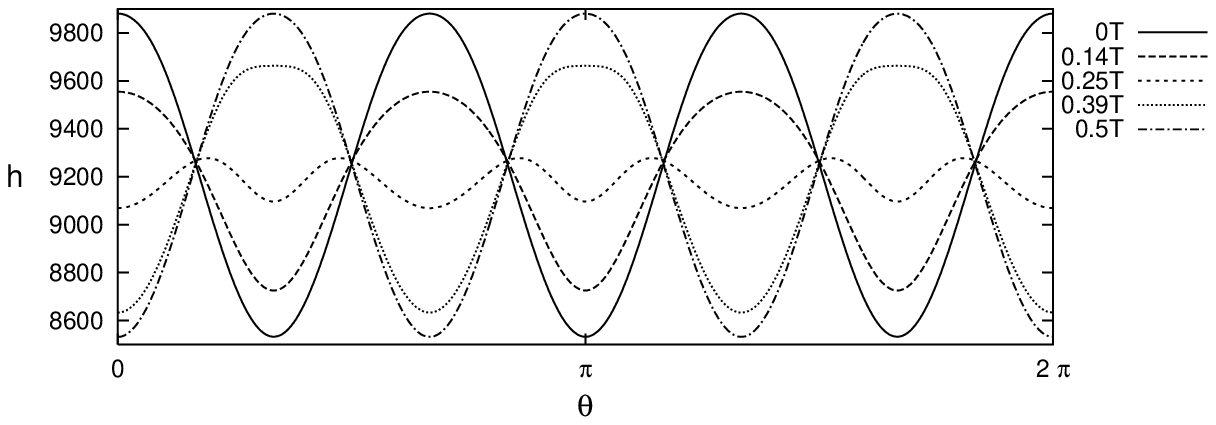}
	\end{center}
	\caption{Standing waves at $Ra=26\,000$: temperature versus $\theta$
	at $(r,z)=(0.7,0.3)$ at five successive times.}
	\label{fig:SW_hvt}
\end{figure}

Above the critical Rayleigh number $Ra_{c2}$, 
a slightly perturbed axisymmetric state evolved in our simulations
towards a three-dimensional time-dependent state, presented in 
figures~\ref{fig:SW}, \ref{fig:SW_uth} and \ref{fig:SW_hvt}. 
Figure~\ref{fig:SW} shows 
temperature contours on the midplane 
at six regularly spaced instants in time within one oscillation period. 
In contrast to the eigenvectors depicted previously,
figure~\ref{fig:SW} displays full nonlinear temperature fields, 
which are dominated by a large axisymmetric component.
There are six pulsing 
extrema, engendering oscillation between two triangular structures
of opposite phases
(figures~\ref{fig:SW}\,\textit{a} and \ref{fig:SW}\,\textit{d}).
At each
instant, the flow is invariant under rotation in $\theta$ by $2\pi/3$. In
addition, this flow is also symmetric with respect to 
three different axes of reflection.  
Figure~\ref{fig:SW_uth} shows contours of azimuthal velocity at
the same times as figure~\ref{fig:SW}.
Figure~\ref{fig:SW_hvt} shows the temperature dependence on the angle $\theta$ for
fixed radius and height at different times.  Six fixed nodes identify this
state as a standing wave with azimuthal wavelength $2\pi / 3$.  

\begin{figure}
\begin{center}
\includegraphics{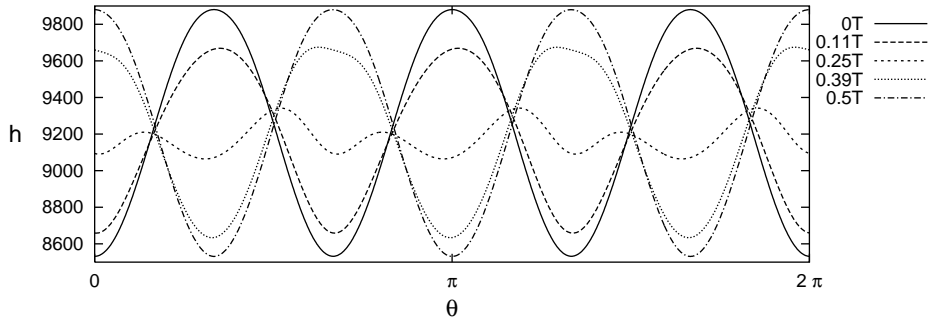}
	\end{center}
	\caption{Standing waves at $Ra=26\,000$ after a time integration
sufficiently long to see the beginning of breaking of reflection symmetry.
Temperature versus $\theta$ at $(r,z)=(0.7,0.3)$ at five successive times.}
	\label{fig:SW2TW_hvt}
\end{figure}

\begin{figure}
\begin{center}
\begin{tabular}{cccccccc}
\includegraphics[width=2cm]{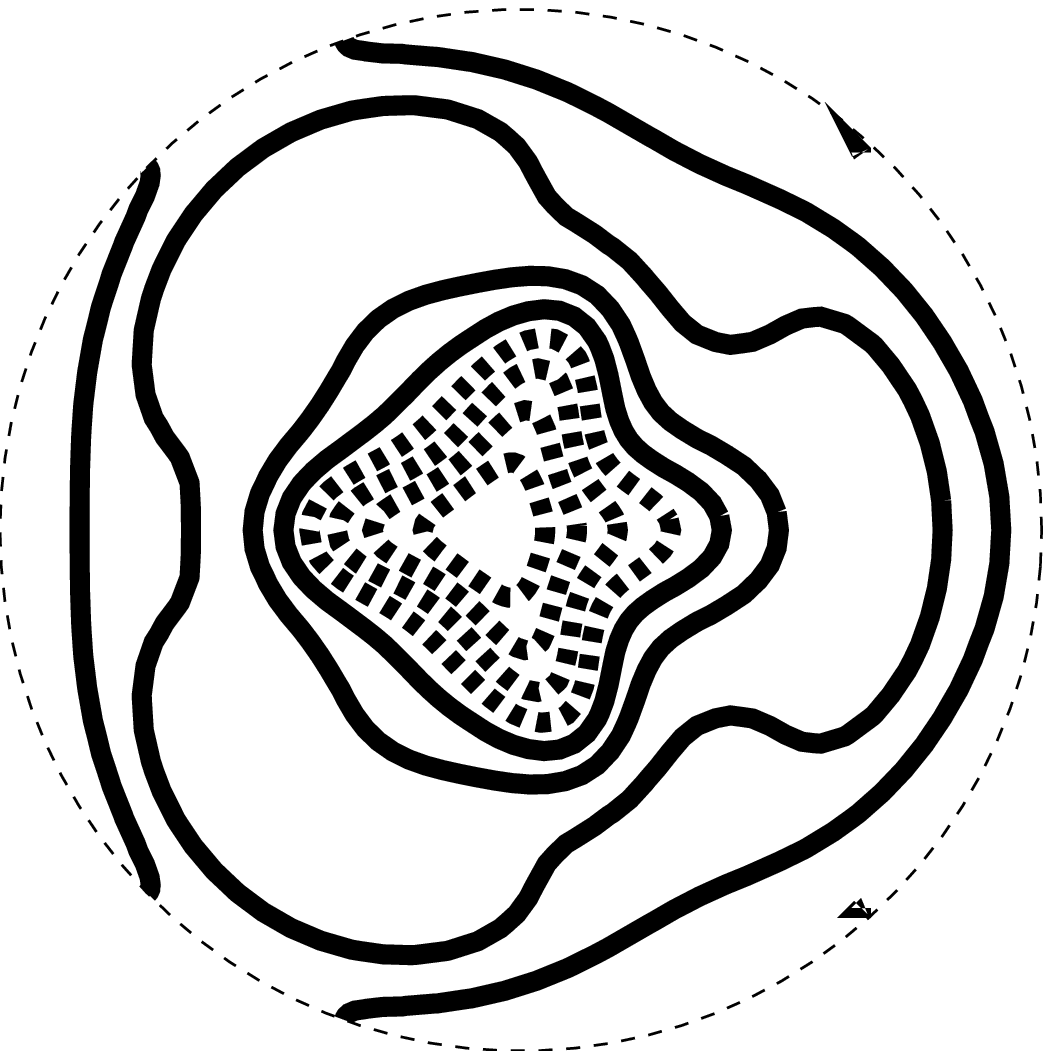}&
\includegraphics[width=2cm]{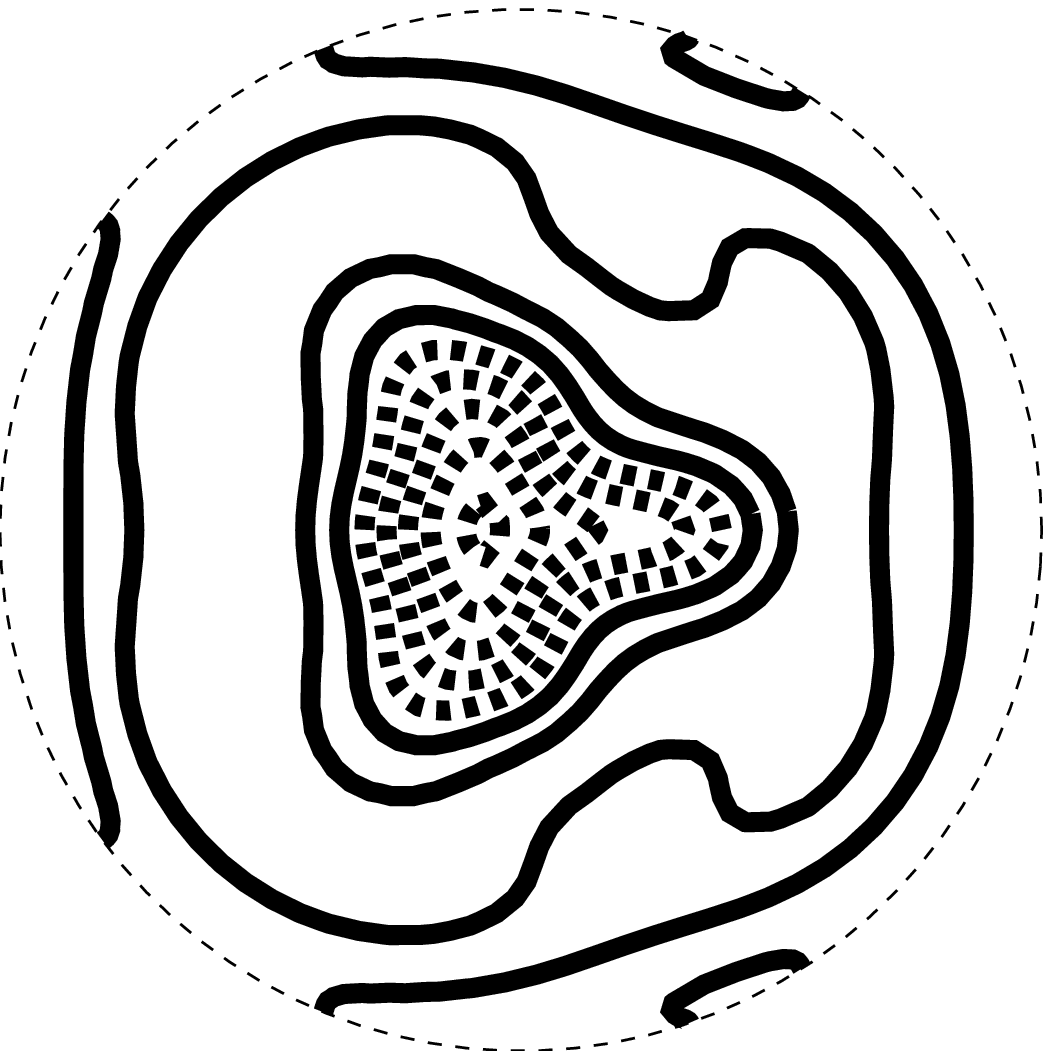}&
\includegraphics[width=2cm]{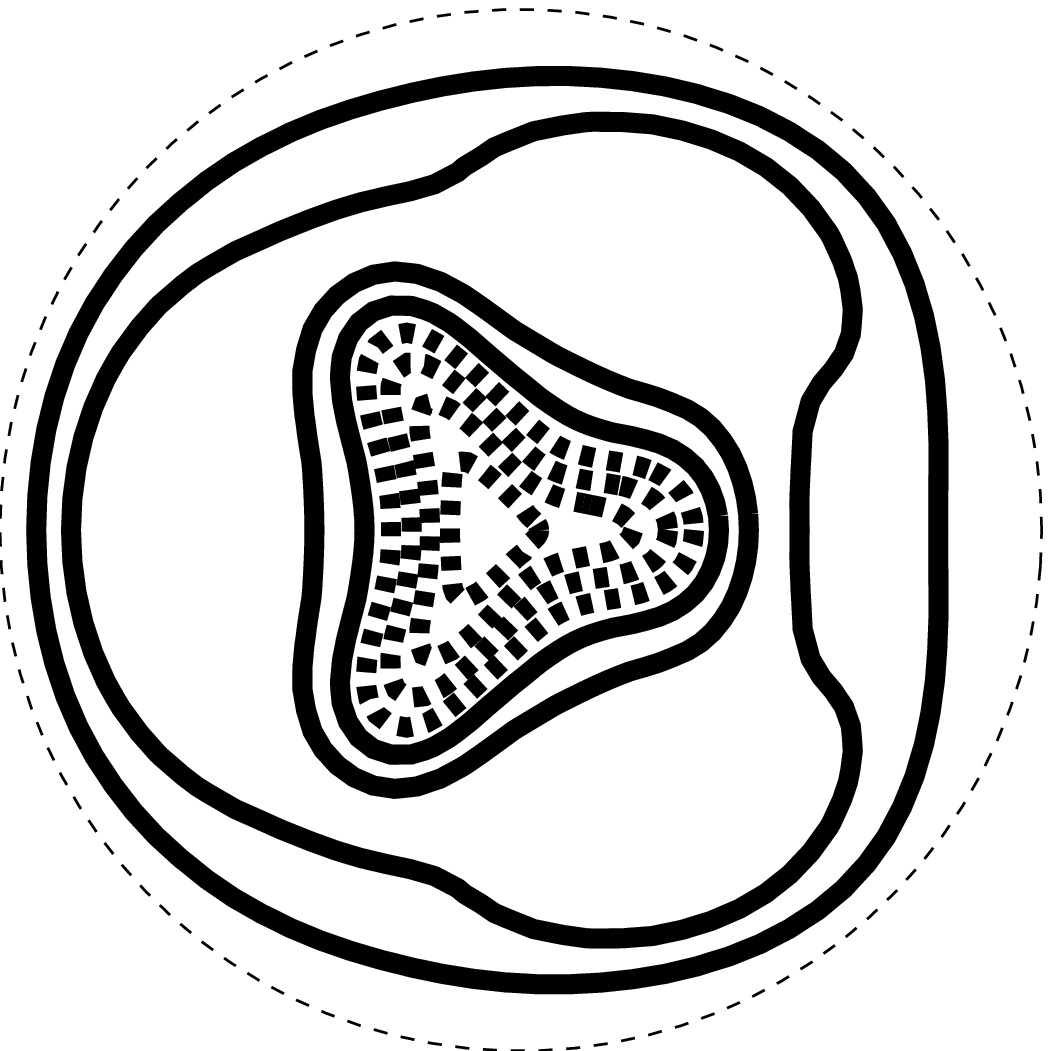}&
\includegraphics[width=2cm]{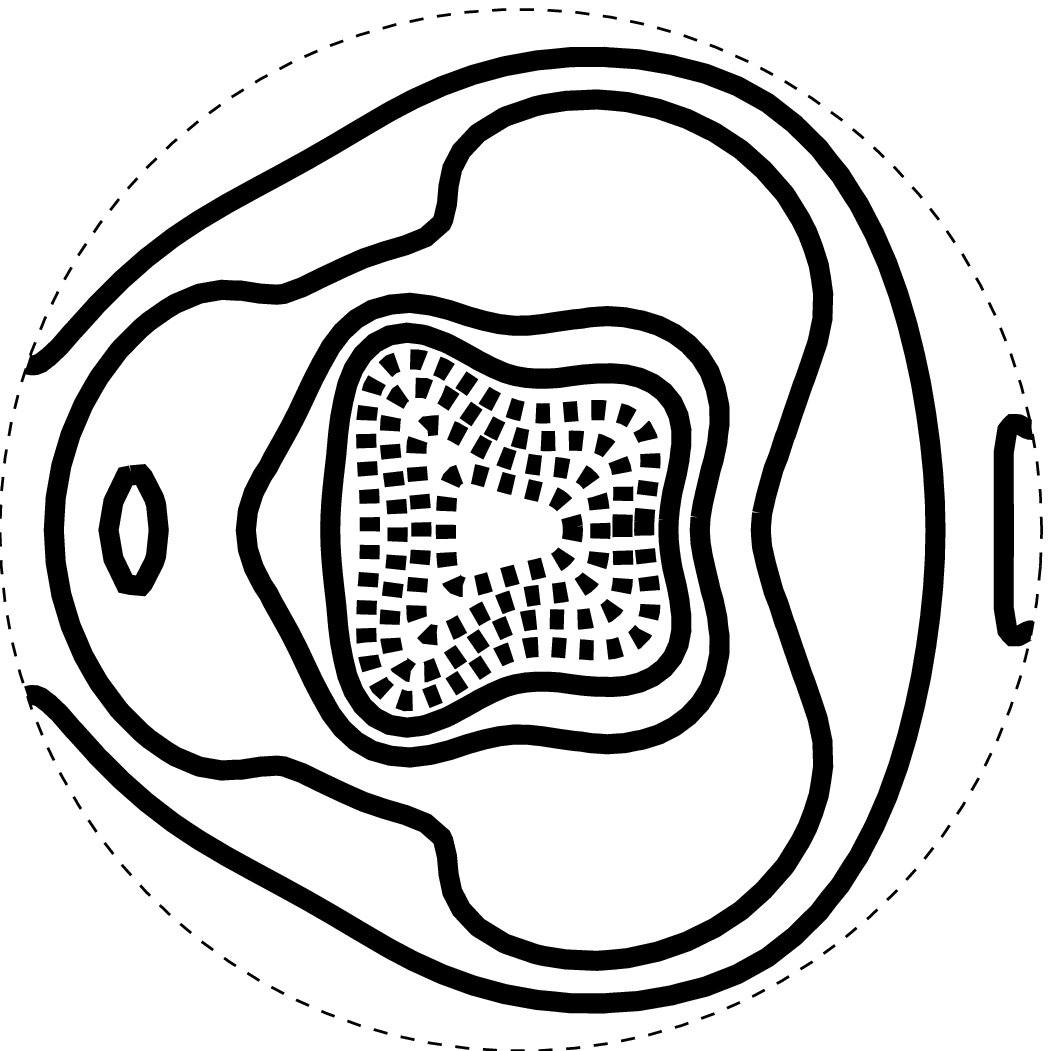}&
\includegraphics[width=2cm]{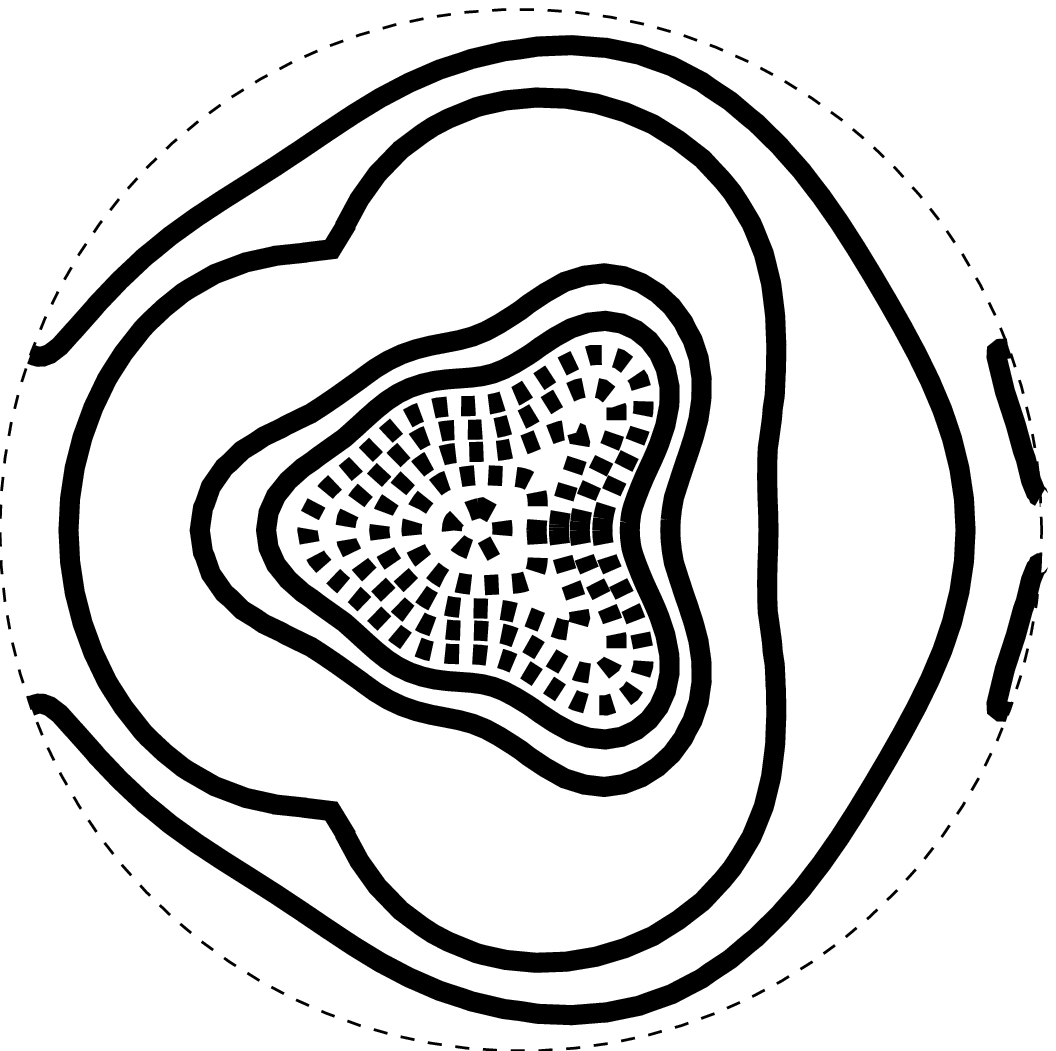}&
\includegraphics[width=2cm]{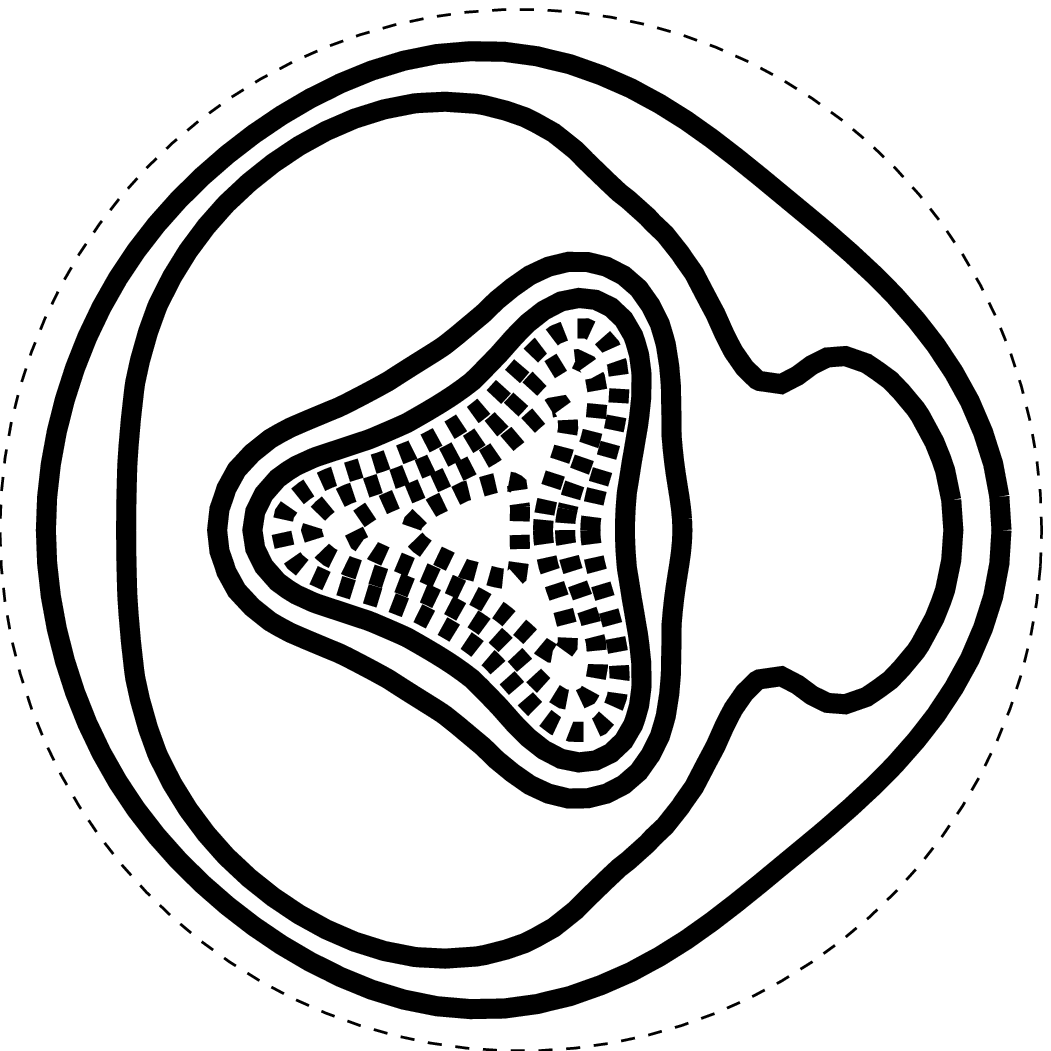}\\
(\textit{a}) & (\textit{b}) & (\textit{c}) & (\textit{d}) & (\textit{e}) & (\textit{f})
	\end{tabular}
\end{center}
	\caption{Oscillatory solution obtained at $Ra=30\,000$ by imposing 
reflection symmetry: temperature contours on the midplane
at $t=0$, $T/6$, $2T/6$,~$\ldots$}
	\label{fig:oc}
\end{figure}

The standing wave state persists
for such a long time that it might seem stable.  However, a small 
reflection-symmetry breaking imperfection develops that eventually 
leads to the transition to travelling waves.  
Figure~\ref{fig:SW2TW_hvt} shows the temperature
dependence on the angle $\theta$ for the same parameters as
figure~\ref{fig:SW_hvt}, but at a later time.  The breaking of reflection
symmetry can be observed when the amplitude of the standing wave is small.
The standing waves can be stabilised by imposing reflection symmetry.
When we did this, above a threshold $Ra_{c3} \approx 27\,000$, 
we discovered a new (unstable) standing-wave solution, displayed in
figure~\ref{fig:oc} for $Ra=30\,000$.

\begin{figure}
\begin{center}
\includegraphics{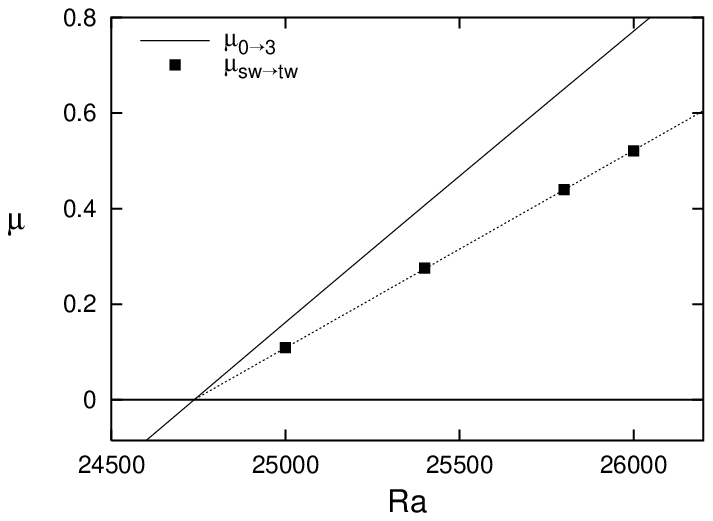}
	\end{center}
	\caption{Growth rates as a function of Rayleigh number.
Solid line: growth rate $\mu_{0\rightarrow 3}$
of $m=3$ eigenvector (either standing or travelling waves)
from the axisymmetric solution (from linear evolution).
Squares: growth rate $\mu_{SW\rightarrow TW}$ of travelling waves
from standing waves (from nonlinear simulation) with linear fit
as dashed line.}
	\label{fig:sigmas_sw2tw}
\end{figure}

In order to study the transition from standing to travelling waves,
we monitored the growth of antisymmetric components.
When the standing wave is still dominant, the amplitude of the
antisymmetric components behaves in time like 
$(A \cos \omega t + B )\exp\left(\mu_{sw\rightarrow tw}\;t\right)$, 
where $\mu_{sw\rightarrow tw}$ is the growth rate
from standing waves to travelling waves.
The growth rate $\mu_{sw\rightarrow tw}$,
shown on figure~\ref{fig:sigmas_sw2tw} as a function of $Ra$,
is about two thirds of $\mu_{0\rightarrow 3}$, 
the growth rate from the axisymmetric state to an $m=3$ flow
(denoted in the previous sections by $\mu$).
The observed lifetime of the standing waves
decreases as the Rayleigh number is increased, since the
growth rate $\mu_{sw\rightarrow tw}$ increases.

\subsection{Stable travelling waves}
\begin{figure}
\begin{center}
\includegraphics[]{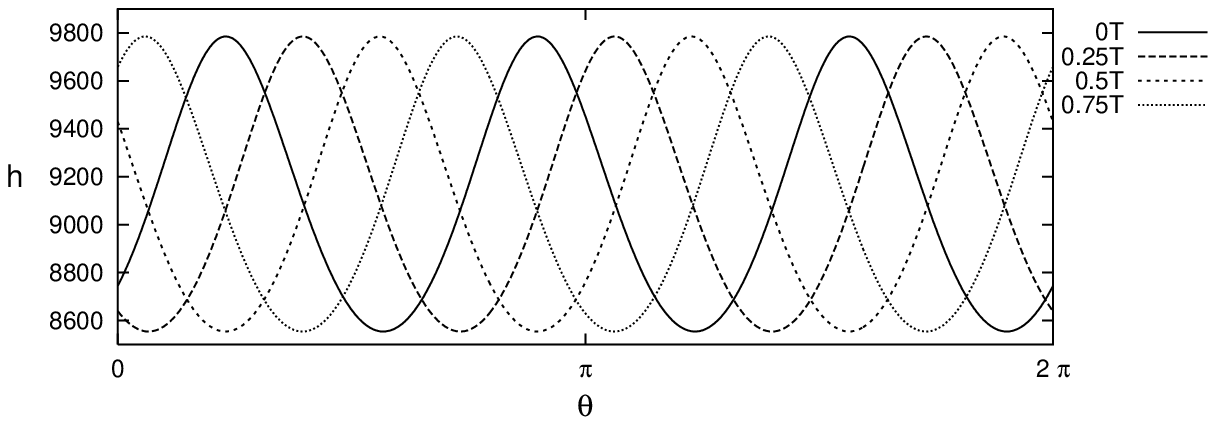}
	\end{center}
	\caption{Travelling waves at $Ra=26\,000$: temperature versus $\theta$ angle,
	for $(r,z)=(0.7,0.3)$, at four different instants 
	during one oscillation period $T$.}
	\label{fig:TW_hvt}
\bigskip
\begin{center}
\begin{tabular}{cccccccc}
\includegraphics[width=2cm]{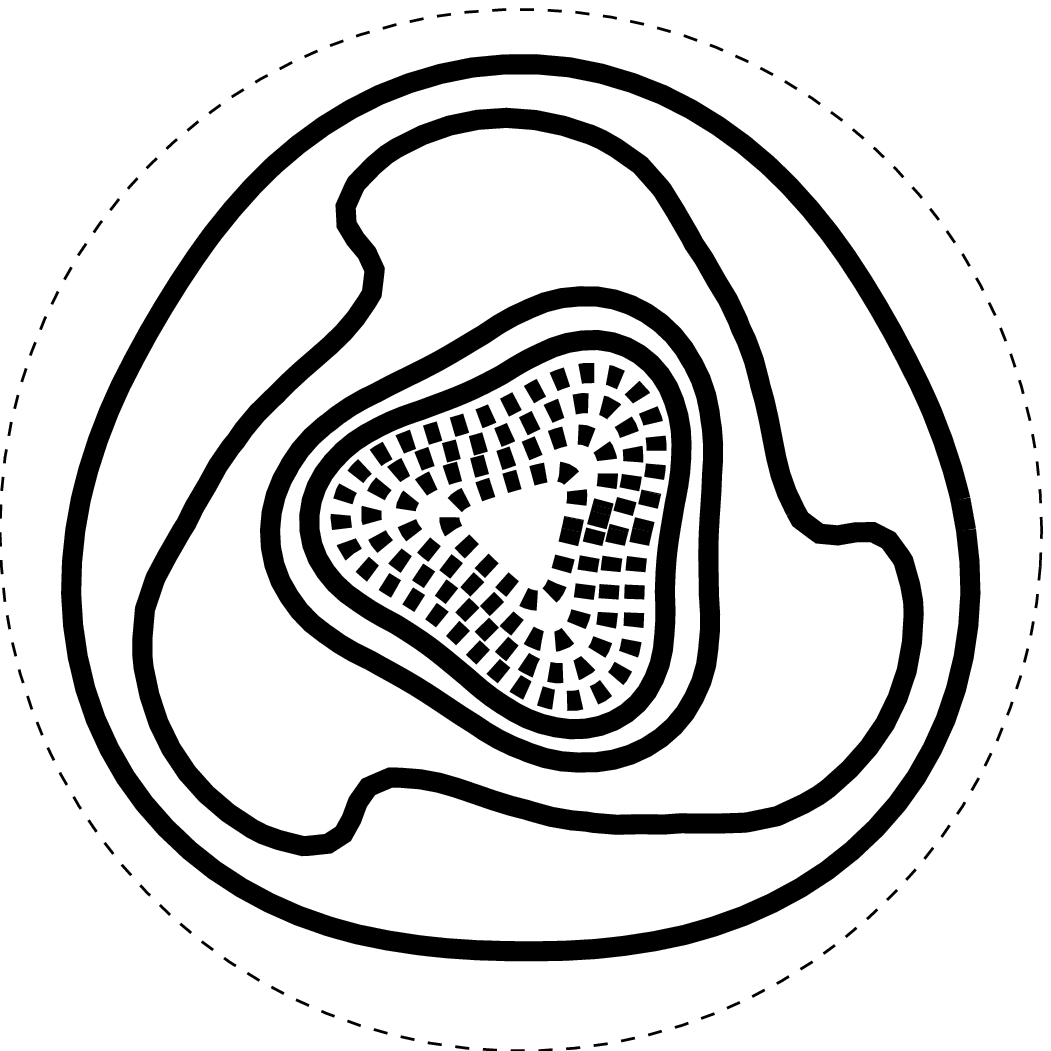}&
\includegraphics[width=2cm]{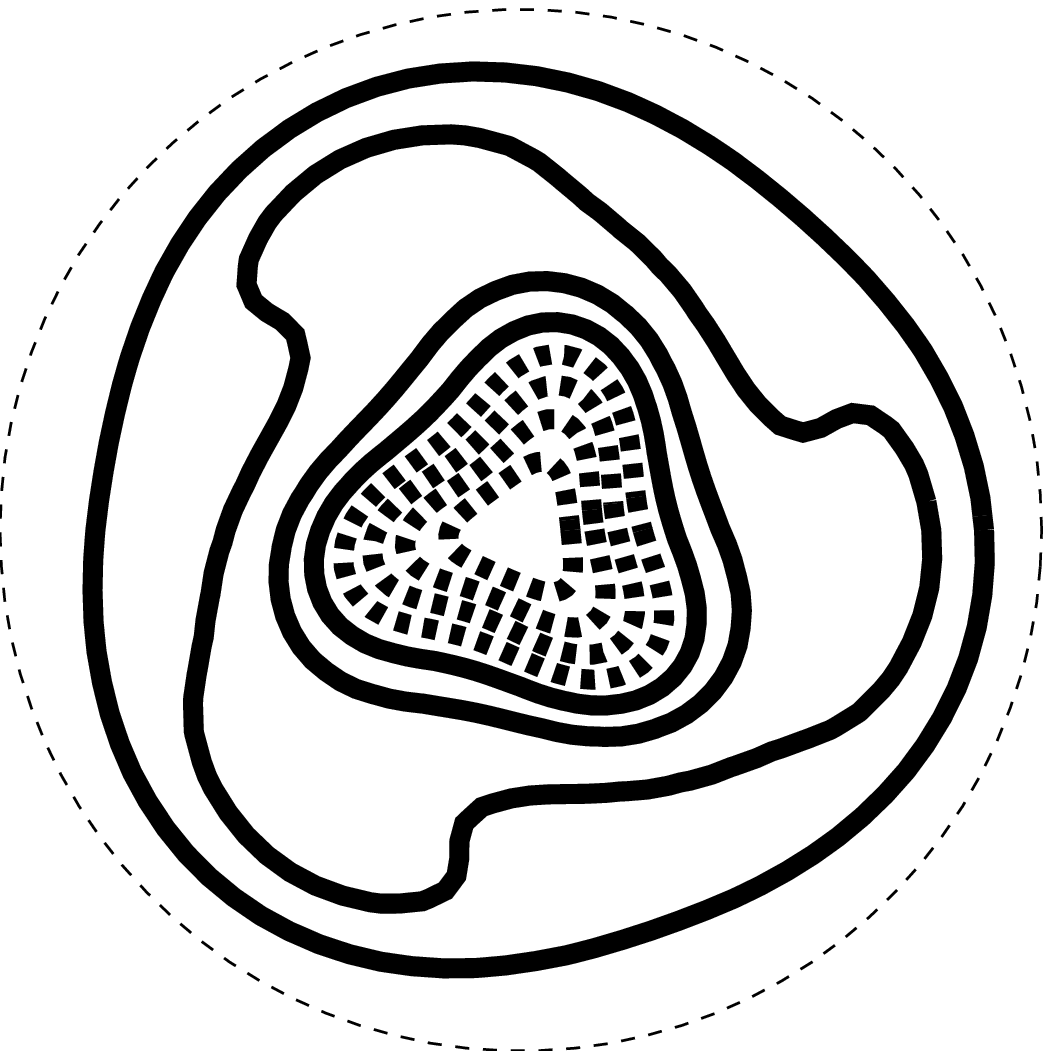}&
\includegraphics[width=2cm]{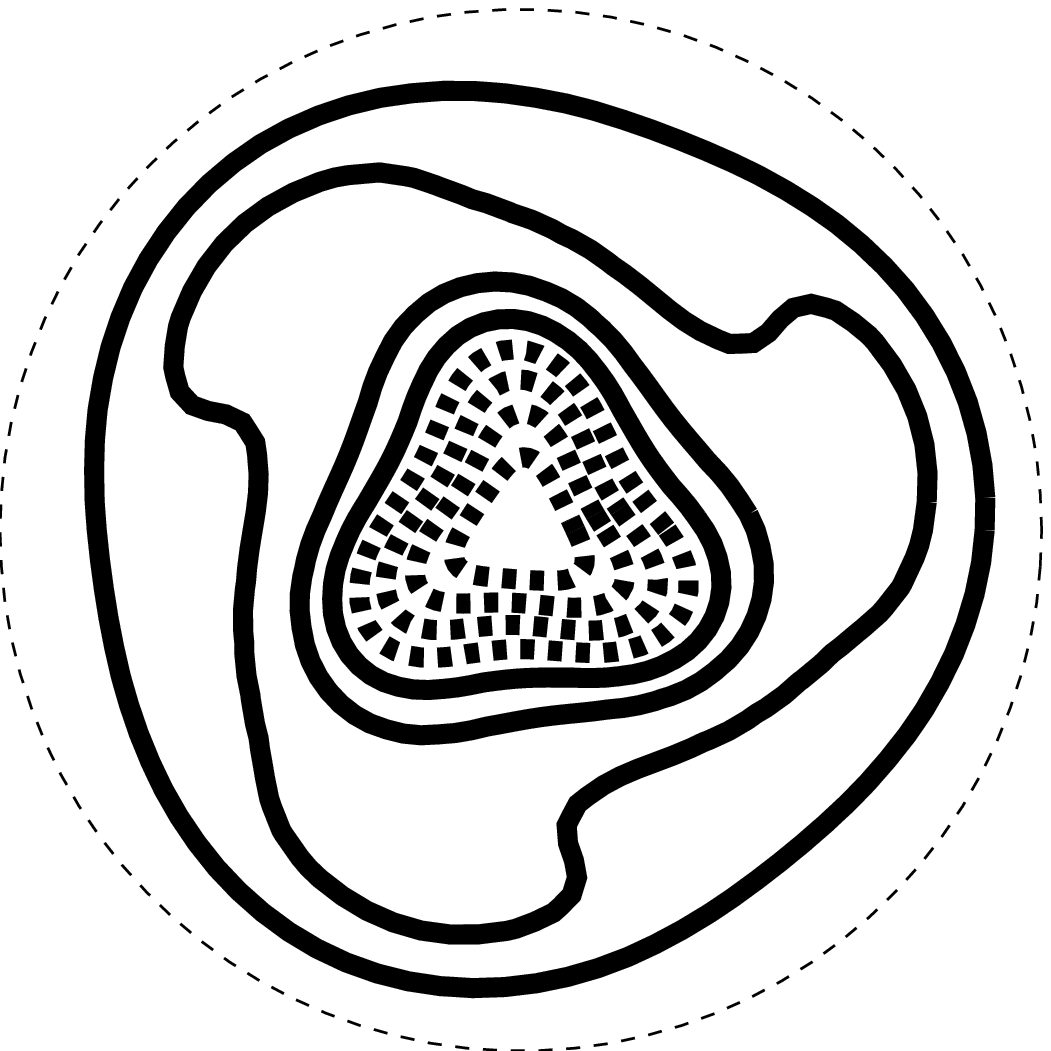}&
\includegraphics[width=2cm]{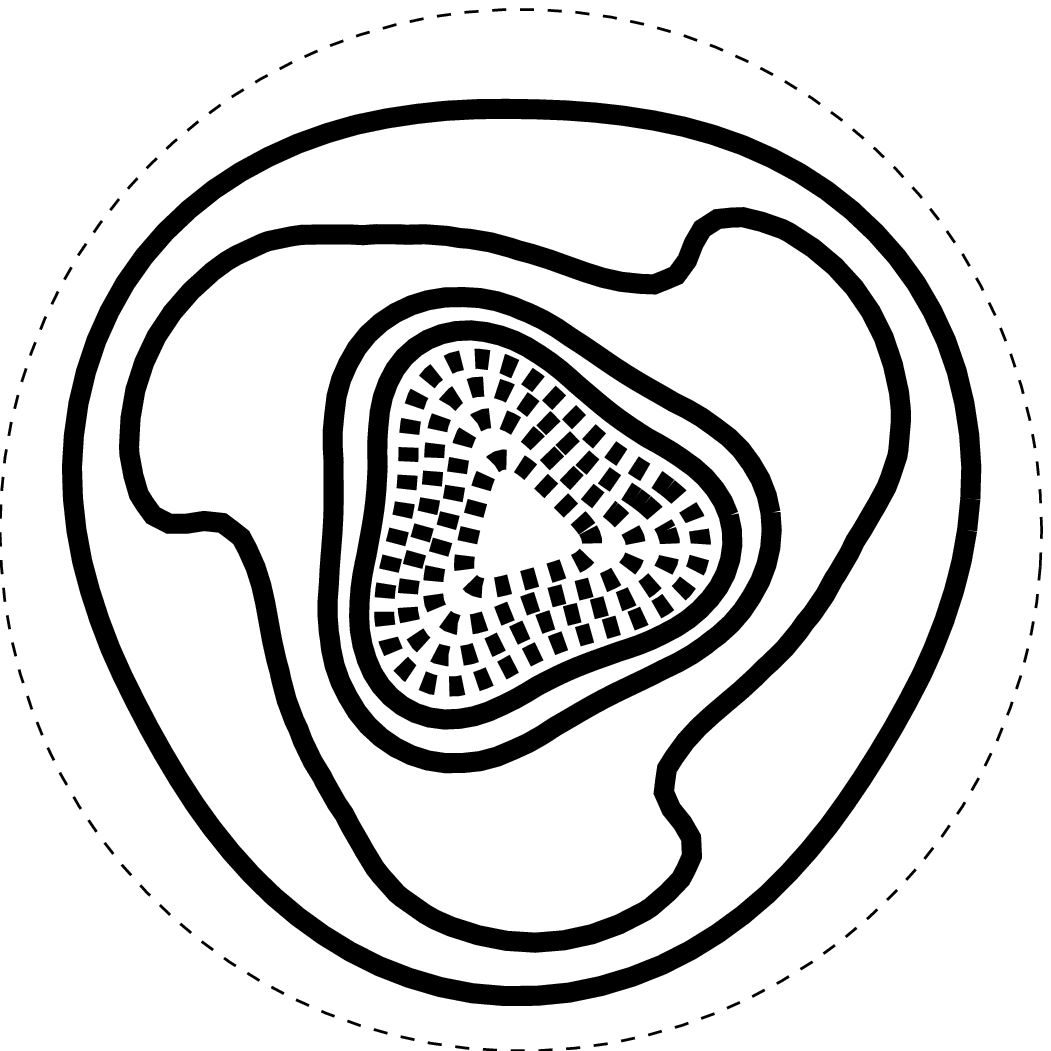}&
\includegraphics[width=2cm]{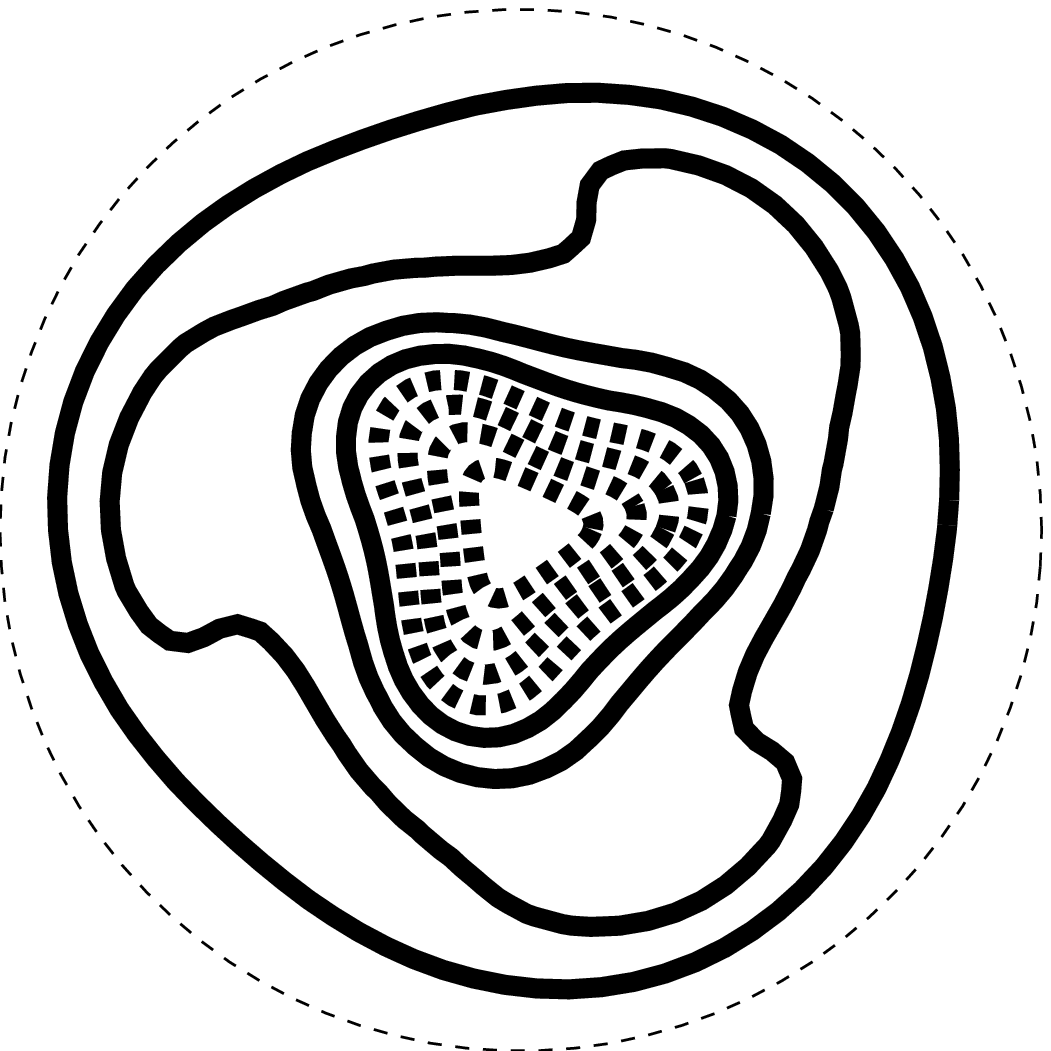}&
\includegraphics[width=2cm]{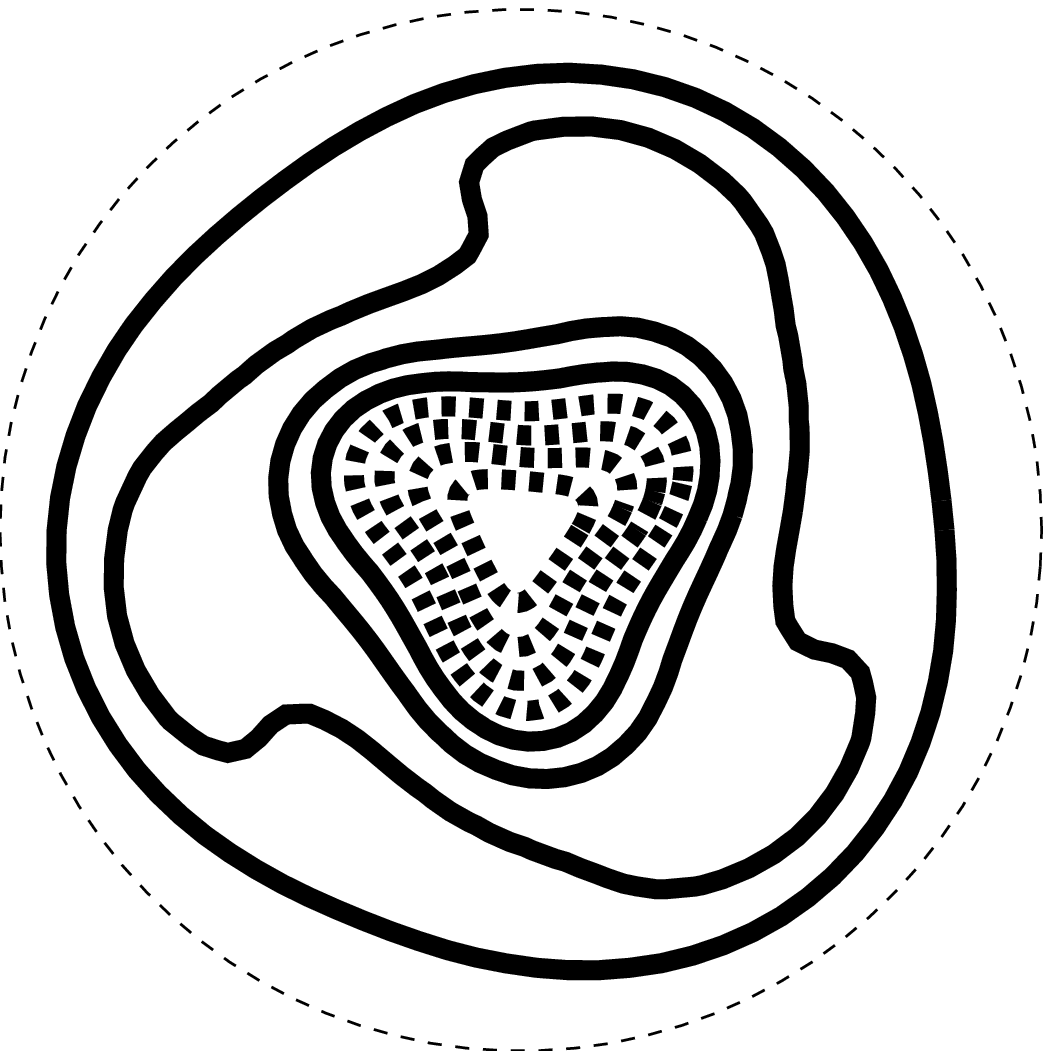}\\
(\textit{a}) & (\textit{b}) & (\textit{c}) & (\textit{d}) & (\textit{e}) & (\textit{f}) 
	\end{tabular}
\end{center}
	\caption{
Counterclockwise
travelling wave  at $Ra=26\,000$:
	temperature contours on the midplane at $t=0$, $T/6$, $2T/6$,~$\ldots$}
	\label{fig:TW}
\bigskip
\begin{center}
\begin{tabular}{cccccccc}
\includegraphics[width=2cm]{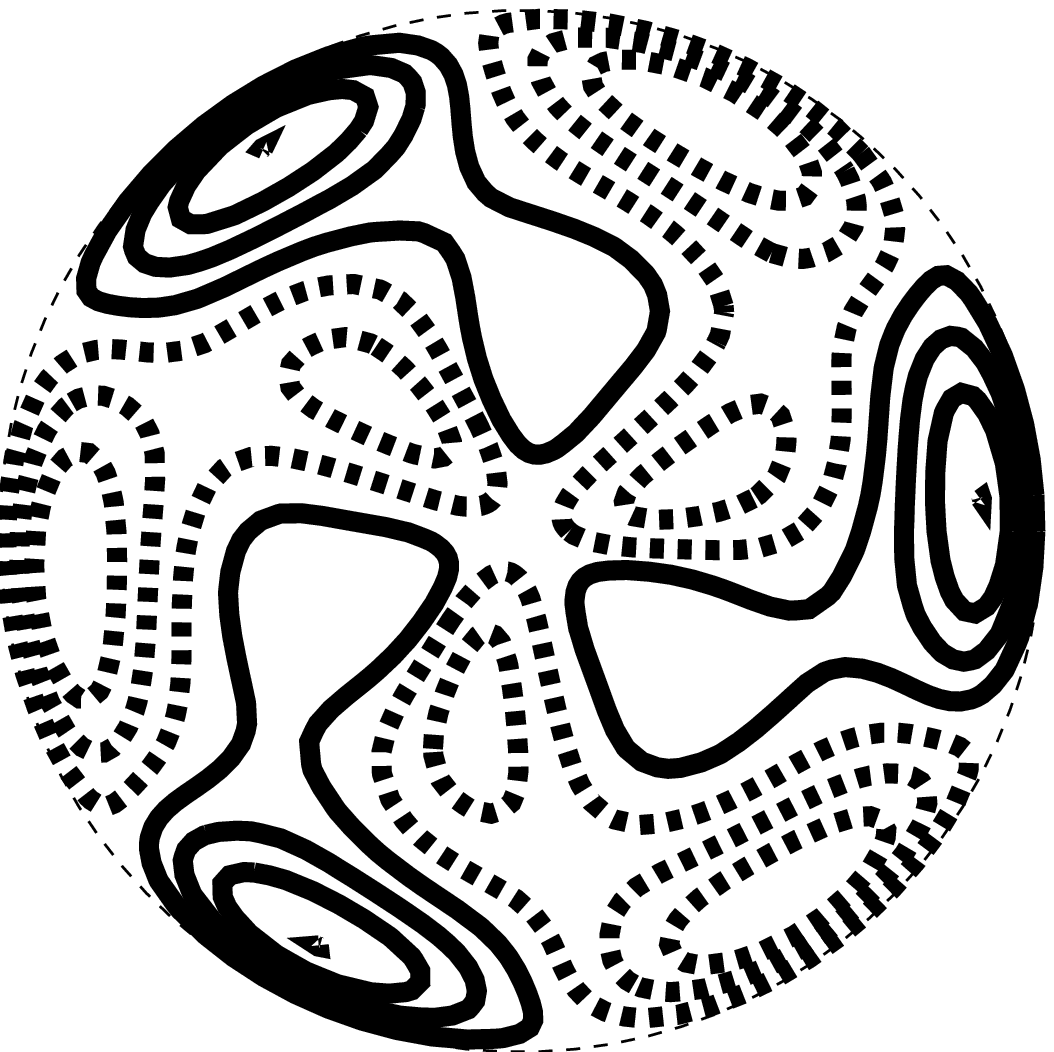}&
\includegraphics[width=2cm]{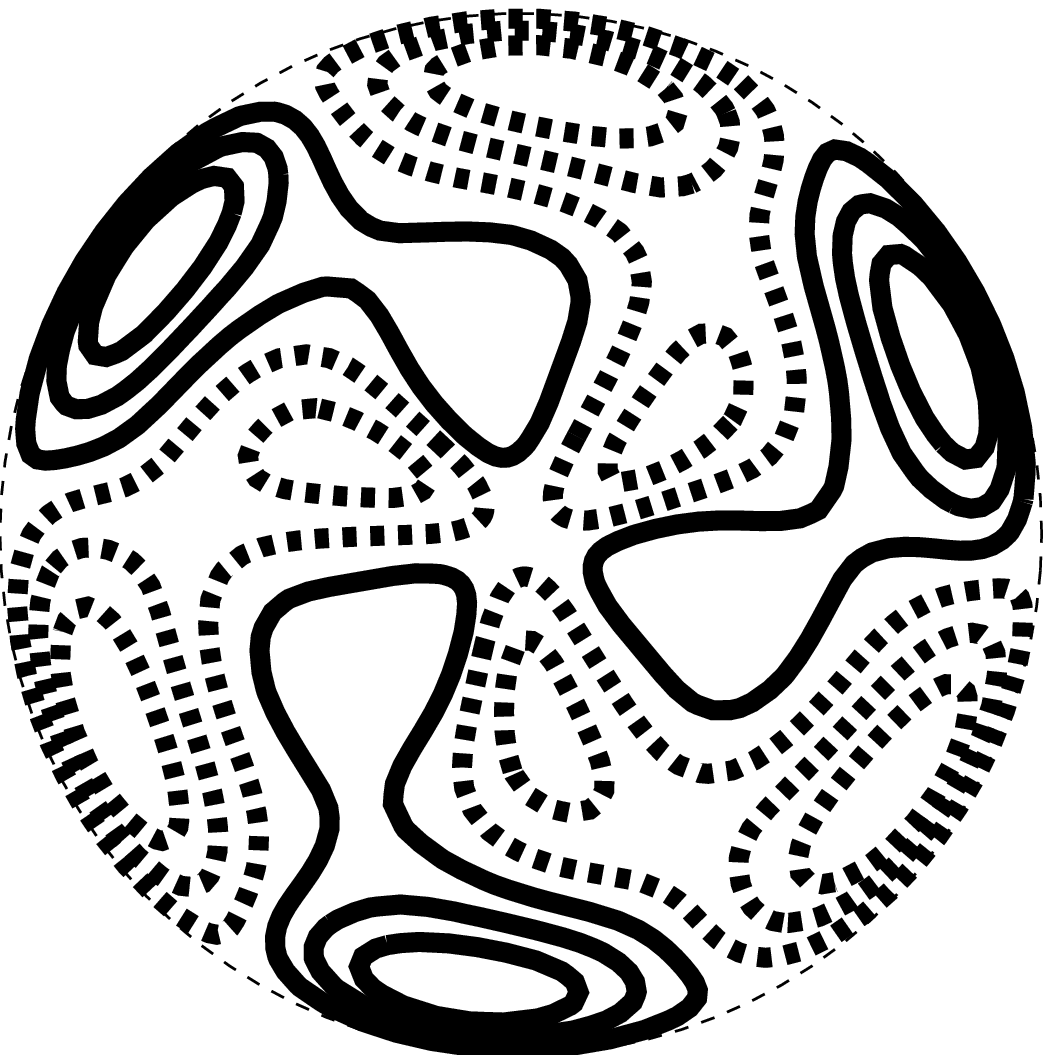}&
\includegraphics[width=2cm]{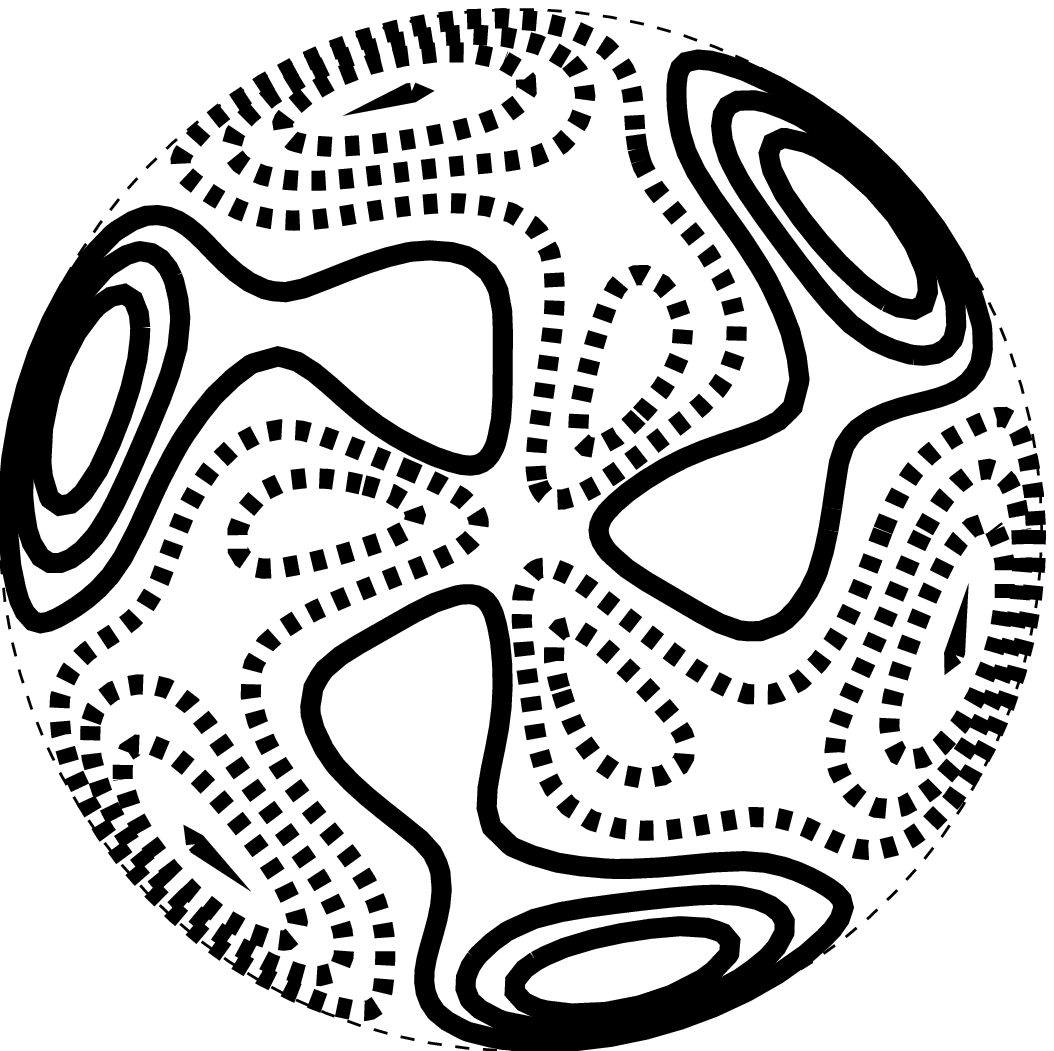}&
\includegraphics[width=2cm]{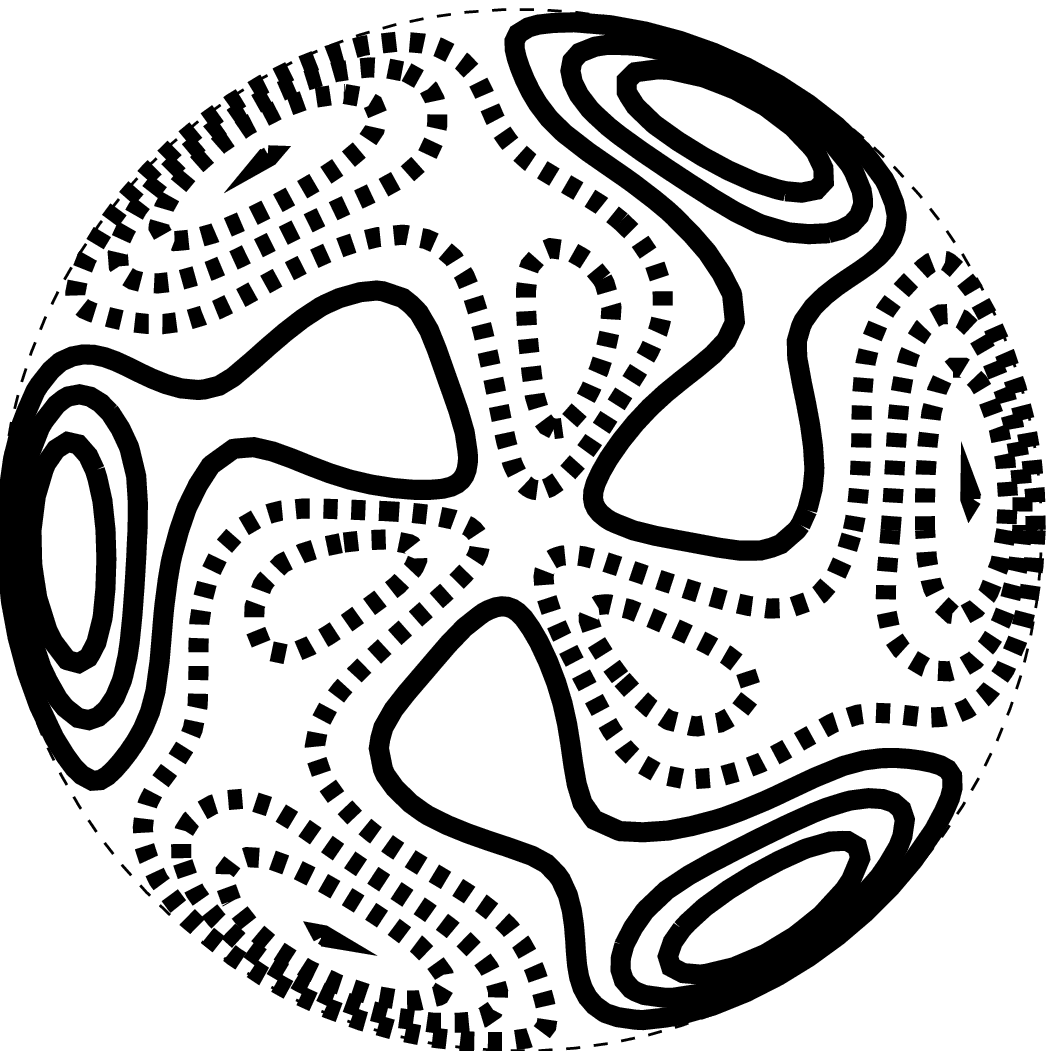}&
\includegraphics[width=2cm]{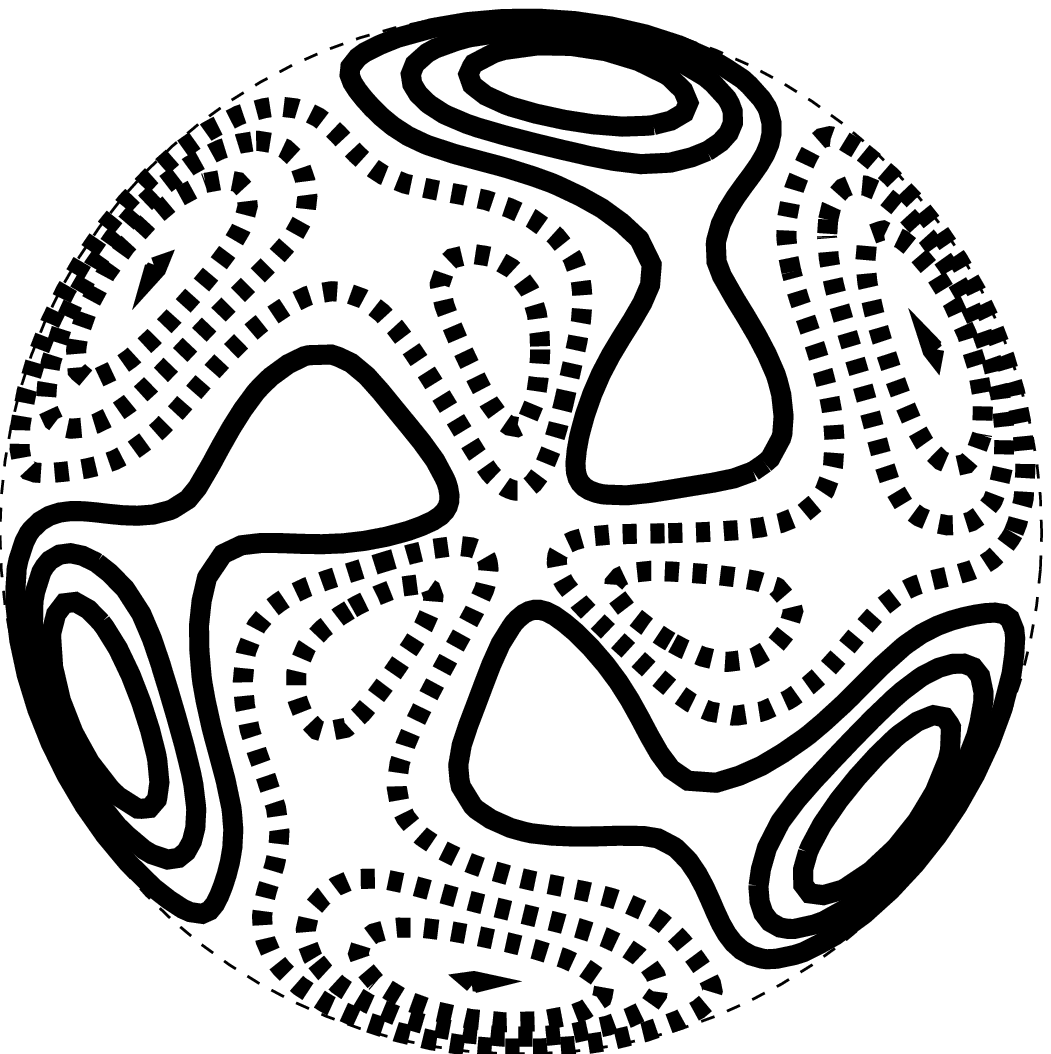}&
\includegraphics[width=2cm]{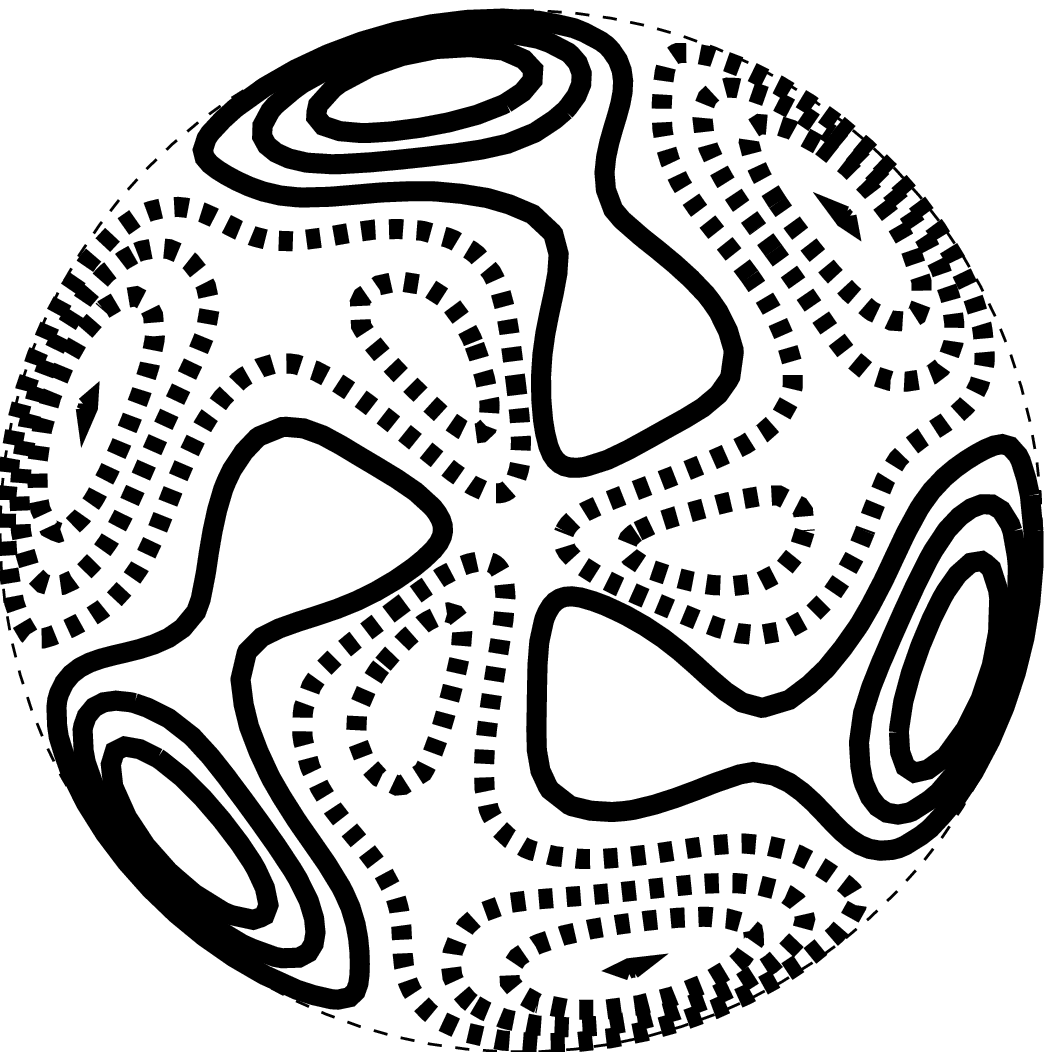}\\
(\textit{a}) & (\textit{b}) & (\textit{c}) & (\textit{d}) & (\textit{e}) & (\textit{f}) 
	\end{tabular}
\end{center}
	\caption{
Counterclockwise travelling wave at $Ra=26\,000$:
	contours of azimuthal velocity on the midplane
at $t=0$, $T/6$, $2T/6$,~$\ldots$}
	\label{fig:TW_uth}
\end{figure}

After the pattern has evolved sufficiently from the standing wave state,
the fixed antinodes abruptly begin to rotate about the cylinder axis.
The six pulsing spots change
into three rotating  spots, as the standing waves become travelling waves with
the same azimuthal wavelength. 
Figure~\ref{fig:TW_hvt}, \ref{fig:TW} and \ref{fig:TW_uth}
depict temperature profiles and  contours of the temperature 
and the azimuthal velocity of the travelling waves at different times. The
travelling waves, like the standing waves, have three-fold rotational symmetry,
but do not have reflection symmetry.

Travelling waves are the final state of the time evolution.
The reason for which we obtained standing waves before
travelling waves in our simulations is that our initial conditions
were reflection symmetric and our numerical procedures introduce
antisymmetric perturbations at a low rate. (This is also seen in the
simulations of thermocapillary flow by \cite{Leypoldt}.)
When the Rayleigh number is decreased, travelling waves persist until
$Ra$ reaches $Ra_{c2}$. 

We conducted simulations for several values of $\Gamma$ in 
the range $1.45 \leq \Gamma < 1.53$ and observed
weakly unstable standing waves and stable travelling
waves for all of them.
We believe that the same scenario also occurs for 
$1.53 \leq \Gamma \leq 1.57$, 
but with azimuthal wavenumber $m=4$ instead of $m=3$.

\subsection{Amplitudes and frequencies}
\label{Amplitudes and frequencies}
 
We calculated the energy $E$ of both types of waves by first defining a norm 
whose square is
\begin{equation}
\frac{1}{Ra} \left(\frac{\langle{\bf u},{\bf u}\rangle}{Pr} + \frac{\langle h,h \rangle}{Ra}\right),\label{definenorm}
\end{equation}
where $\langle , \rangle$ denotes spatial integration; 
(\ref{definenorm}) is one of many possible choices for this system.
We then simulated the nonlinear evolution equations and calculated
$({\bf u},h)$ as the difference between the 
three-dimensional and the axisymmetric solution. We define $E$ to be
the integral of (\ref{definenorm}) over one oscillation period.

\begin{figure}
\begin{center}
	\begin{tabular}{cc}
	\includegraphics[]{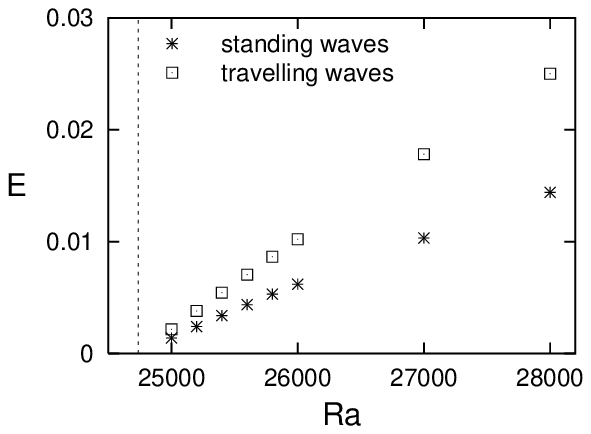} &
	\includegraphics[]{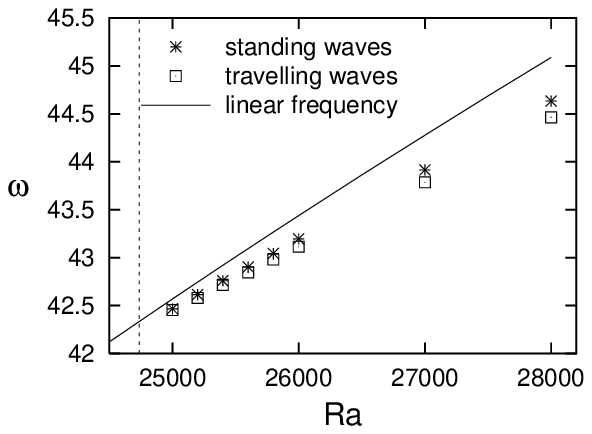}\\
	(\textit{a}) & (\textit{b})
	\end{tabular}
	\parindent 0pt
\end{center}	
\caption{Dependence
of energy and frequency on Rayleigh number for standing and
	travelling waves. Vertical dashed line indicates
the critical Rayleigh number $Ra_{c2}$ for onset of the waves. }
	\label{fig:amplfreq}
\end{figure}

The energies $E_{sw}$, $E_{tw}$ and frequencies $\omega_{sw}$, $\omega_{tw}$
as a function of $Ra$ are shown in 
figure~\ref{fig:amplfreq}.
The energies and frequencies for the two types of waves are quite close.
The frequency $\omega_{0\rightarrow 3}$ obtained from linear stability 
analysis is also reproduced from figure~\ref{fig:eigenvalues}\,(\textit{b})
for comparison.
For both types of waves, the frequencies near the threshold are close to the Hopf frequency and the energy satisfies $ E \propto (Ra-Ra_{c2})$.
These are hallmarks of a supercritical Hopf bifurcation.

\subsection{Normal form coefficients}
Using the growth rates, amplitudes and frequencies of the standing and
travelling waves that we have presented in 
sections \ref{Eigenvalues and eigenvectors} and \ref{Amplitudes and frequencies},
it is possible to calculate the coefficients 
of the normal form (\ref{nf}) for our particular case.
The bifurcation parameter $\mu=\mu_{0\rightarrow 3}$ and frequency $\omega=\omega_{0\rightarrow 3}$
vary linearly with $Ra-Ra_{c2}$, 
while the other coefficients $a_r$, $b_r$, $a_i$, $b_i$ are constants.

From the data in figures
~\ref{fig:eigenvalues}\,(\textit{a},\textit{b}), we extract the fits
\begin{subequations}
\begin{eqnarray}
\mu_{0\rightarrow 3}= 14.98\;\frac{Ra-Ra_{c2}}{Ra_{c2}}, \label{fitmu}\\
\omega_{0\rightarrow 3}= 42.33+21.21\;\frac{Ra-Ra_{c2}}{Ra_{c2}}. \label{fitom}
\end{eqnarray}
\end{subequations}
From the data in figure~\ref{fig:amplfreq} we extract the fits
\begin{subequations}
\label{fitfour}
\begin{eqnarray}
E_{tw}=A_{tw}^2 = \rho_+^2 = \frac{-\mu}{b_r}=0.2037
 \;\frac{Ra-Ra_{c2}}{Ra_{c2}}, \label{fitbr}\\
E_{sw}=A_{sw}^2 = \rho_+^2 + \rho_-^2 = 2\frac{-\mu}{a_r+2b_r} 
=0.13 \;\frac{Ra-Ra_{c2}}{Ra_{c2}}, \\
\omega_{tw}=\omega_{0\rightarrow 3}-\frac{b_i}{b_r}\;\mu = 42.33+ 16.26 \;\frac{Ra-Ra_{c2}}{Ra_{c2}}, \label{fitbi}\\
\omega_{sw}=\omega_{0\rightarrow 3}
-\frac{a_i+2b_i}{a_r+2b_r}\;\mu=42.33+17.29 \;\frac{Ra-Ra_{c2}}{Ra_{c2}}. \label{fitai}
\end{eqnarray}
\end{subequations}
Equations (\ref{fitfour}) are used to determine the nonlinear coefficients as
\begin{subequations}
\begin{eqnarray}
b_r 
= -73.5,\\
a_r
= -83.6, \label{fitar}\\
b_i 
= -24.3,\\
a_i
=11.7.
\end{eqnarray}
\end{subequations}
An additional equation is provided by the data in 
figure~\ref{fig:sigmas_sw2tw} showing the growth rate 
$\mu_{sw\rightarrow tw}$ from standing to travelling waves:
\begin{equation}
\mu_{sw\rightarrow tw}=\frac{2a_r}{a_r+2b_r}\mu = 
10.23 \;\frac{Ra-Ra_{c2}}{Ra_{c2}} \label{fitar2} .\\
\end{equation}
and provides a second determination of $a_r$
\begin{equation}
a_r = \frac{-\mu_{sw\rightarrow tw}}{A_{sw}^2} =-78.8.
\end{equation}
which differs by 6\% from (\ref{fitar}).

\section{Conclusion}
We have used both nonlinear simulations and linear stability analysis
to elucidate the behaviour of Rayleigh--B\'enard convection in the
parameter region of $1.45\leq\Gamma\leq 1.57$, $Pr=1$ first studied by 
\cite{WanKuhRat}.
In this regime, the primary axisymmetric convective state
loses stability to an $m=3$ perturbation via a Hopf bifurcation whose
critical eigenspace is four-dimensional. We calculated representative
eigenvectors and explained how these relate to those 
computed by Wanschura \etal\ 
The bifurcation scenario guarantees that branches of
standing waves and of travelling waves are created at the bifurcation,
but that at most one of these branches is stable.
Our nonlinear simulations showed 
a supercritical bifurcation leading to
long-lived standing waves which
were eventually succeeded by travelling waves, both as time
progressed and as the Rayleigh number was increased.
We explained this by showing that the rate of transition from
standing waves to travelling waves, while positive, is nevertheless small.
In the absence of long-time integration and of these analyses,
it would be easy to conclude that the standing waves were stable.
This underlines the importance of calculating growth rates,
in addition to carrying out nonlinear simulations, and of
using established bifurcation scenarios to interpret physical phenomena.

The numerical and theoretical techniques we have used can
be generally applied to study transitions in hydrodynamic problems.
Our main tool was direct numerical simulation of the governing
Boussinesq equations using a pseudo-spectral semi-implicit timestepping code. 
We complemented this approach with several other techniques.
To carry out stability analysis, we first linearised the code.
This requires very little modification of the existing code,
but yields results which are far more precise and robust
than restricting integration to the time interval during which
perturbations to the basic state are small.
Integrating the linearised equations is, in effect, an implementation
of the power method for finding the fastest growing eigenvalues 
and corresponding eigenvectors.
Eigenvectors with different azimuthal wavenumbers can be found simultaneously,
since the linearised evolution of each Fourier mode is independent 
of the others.
For a single wavenumber, this use of the power method is rendered
more accurate and more general by postprocessing the results
of linearised time integration with the Arnoldi decomposition
to extract several, possibly complex, eigenvectors.
We also interpreted our results in light of known results 
concerning axisymmetry-breaking Hopf bifurcations in systems
with $O(2)$ symmetry. This framework allows us to generate
the four-dimensional eigenspace by combining eigenvectors
with different symmetries. Traditionally, eigenvectors corresponding
to clockwise and counterclockwise travelling waves are combined
to form standing waves; we used a complementary,
but equivalent, approach of combining standing waves
of different spatial phases to form travelling waves.
Finally, we interpreted our results in terms of the 
four ordinary differential equations comprising the 
normal form for Hopf bifurcations in systems with $O(2)$ symmetry.
Using our nonlinear simulations of the governing Boussinesq equations,
we were able to calculate the various coefficients in the normal
form equations.

We have not sought to determine the limits of the range of this
phenomenon, in aspect ratio and Prandtl number.
As these ranges were given by Wanschura \etal\ only for $Pr=1$, a future
direction would be to determine the whole zone in the parameter space where
the Hopf bifurcation occurs.
It would be interesting also to examine more closely 
the pulsing pattern found by \cite{HofLucMul} at $Ra=33\,000$, 
$\Gamma=2$, $Pr=6.7$,
in order to determine whether this state, evolving from axisymmetric flow,
is the result of a bifurcation similar to that described in the present paper.

\begin{acknowledgments}
The computations were performed on the NEC SX5 of the 
IDRIS (Institut du D\'eveloppement et des Ressources
en Informatique Scientifique) supercomputer center of the 
CNRS (Centre National pour la Recherche Scientifique)
under project 1119.
\end{acknowledgments}

\bibliographystyle{jfm}
\bibliography{wholebib}
\end{document}